\baselineskip 18 pt plus 2 pt minus 1 pt



%
%
%
\magnification = \magstep 1

\baselineskip=15pt plus 1pt minus 1pt
\font\secfnt=cmss10

\font\titlefnt=cmssbx10 scaled \magstep 1
\def\title#1{\centerline{{\titlefnt #1}}\medskip}

\def\author#1{\smallskip\centerline{#1}\medskip}
\def\address#1{\centerline{#1}}
\def\date#1{\smallskip\centerline{{\it #1}}\smallskip}

\def\abstract#1{\par\vskip\normalbaselineskip\par
    {\baselineskip=\normalbaselineskip
    \parindent=0 pt
    {\hfill\vbox{\hsize= 11 cm  #1  }\hfill}}
    \bigskip}

\def\section#1{\bigskip\centerline{{\secfnt #1}}\medskip}


\newcount\eqncnt
\eqncnt=0
\def\eqprefix{}
\def\eqn{\global\advance\eqncnt by 1 {\rm(\eqprefix\the\eqncnt)}}
\def\eqname#1{\eqn\xdef#1{\eqprefix\the\eqncnt}}


\newcount\refcnt
\refcnt=0
\def\ref#1.#2\par{\global\advance\refcnt by 1\xdef#1{\the\refcnt}}


\newcount\figcnt
\figcnt=0
\def\fig#1.#2\par{\global\advance\figcnt by 1\xdef#1{\the\figcnt}}

\def\figures{
\par\vfill\eject
\def\fig##1.##2\par{\item{{\secfnt Fig.\ ##1.}}##2}
\frenchspacing
\parskip 6pt plus 1pt minus 1pt
\parindent 0 pt
\par\section{Figures}
\def\temp{1.34}%
\let\tempp=\relax
\expandafter\ifx\csname psboxversion\endcsname\relax
  \message{PSBOX(\temp) loading}%
\else
    \ifdim\temp cm>\psboxversion cm
      \message{PSBOX(\temp) loading}%
    \else
      \message{PSBOX(\psboxversion) is already loaded: I won't load
        PSBOX(\temp)!}%
      \let\temp=\psboxversion
      \let\tempp= 
    \fi
\fi
\tempp
\let\psboxversion=\temp
\catcode`\@=11
%
%
\def\psfortextures{
\def\PSspeci@l##1##2{%
\special{illustration ##1\space scaled ##2}%
}}%
\def\psfordvitops{
\def\PSspeci@l##1##2{%
\special{dvitops: import ##1\space \the\drawingwd \the\drawinght}%
}}%
\def\psfordvips{
\def\PSspeci@l##1##2{%
\d@my=0.1bp \d@mx=\drawingwd \divide\d@mx by\d@my
\includegraphics{##1\space}}}%
\def\psforoztex{
\def\PSspeci@l##1##2{%
\special{##1 \space
      ##2 1000 div dup scale
      \number-\psllx\space \number-\pslly\space translate
}}}%
\def\psfordvitps{
\def\psdimt@n@sp##1{\d@mx=##1\relax\edef\psn@sp{\number\d@mx}}
\def\PSspeci@l##1##2{%
\special{dvitps: Include0 "psfig.psr"}
\psdimt@n@sp{\drawingwd}
\special{dvitps: Literal "\psn@sp\space"}
\psdimt@n@sp{\drawinght}
\special{dvitps: Literal "\psn@sp\space"}
\psdimt@n@sp{\psllx bp}
\special{dvitps: Literal "\psn@sp\space"}
\psdimt@n@sp{\pslly bp}
\special{dvitps: Literal "\psn@sp\space"}
\psdimt@n@sp{\psurx bp}
\special{dvitps: Literal "\psn@sp\space"}
\psdimt@n@sp{\psury bp}
\special{dvitps: Literal "\psn@sp\space startTexFig\space"}
\special{dvitps: Include1 "##1"}
\special{dvitps: Literal "endTexFig\space"}
}}%
\def\psfordvialw{
\def\PSspeci@l##1##2{
\special{language "PostScript",
position = "bottom left",
literal "  \psllx\space \pslly\space translate
  ##2 1000 div dup scale
  -\psllx\space -\pslly\space translate",
include "##1"}
}}%
\def\psforptips{
\def\PSspeci@l##1##2{{
\d@mx=\psurx bp
\advance \d@mx by -\psllx bp
\divide \d@mx by 1000\multiply\d@mx by \xscale
\incm{\d@mx}
\let\tmpx\dimincm
\d@my=\psury bp
\advance \d@my by -\pslly bp
\divide \d@my by 1000\multiply\d@my by \xscale
\incm{\d@my}
\let\tmpy\dimincm
\d@mx=-\psllx bp
\divide \d@mx by 1000\multiply\d@mx by \xscale
\d@my=-\pslly bp
\divide \d@my by 1000\multiply\d@my by \xscale
\at(\d@mx;\d@my){\special{ps:##1 x=\tmpx, y=\tmpy}}
}}}%
\def\psonlyboxes{
\def\PSspeci@l##1##2{%
\at(0cm;0cm){\boxit{\vbox to\drawinght
  {\vss\hbox to\drawingwd{\at(0cm;0cm){\hbox{({\tt##1})}}\hss}}}}
}}%
\def\psloc@lerr#1{%
\let\savedPSspeci@l=\PSspeci@l%
\def\PSspeci@l##1##2{%
\at(0cm;0cm){\boxit{\vbox to\drawinght
  {\vss\hbox to\drawingwd{\at(0cm;0cm){\hbox{({\tt##1}) #1}}\hss}}}}
\let\PSspeci@l=\savedPSspeci@l
}}%
%
%
\newread\pst@mpin
\newdimen\drawinght\newdimen\drawingwd
\newdimen\psxoffset\newdimen\psyoffset
\newbox\drawingBox
\newcount\xscale \newcount\yscale \newdimen\pscm\pscm=1cm
\newdimen\d@mx \newdimen\d@my
\newdimen\pswdincr \newdimen\pshtincr
\let\ps@nnotation=\relax
{\catcode`\|=0 |catcode`|\=12 |catcode`|
|catcode`#=12 |catcode`*=14
|xdef|backslashother{\}*
|xdef|percentother{
|xdef|tildeother{~}*
|xdef|sharpother{#}*
}%
\def\R@moveMeaningHeader#1:->{}%
\def\uncatcode#1{%
\edef#1{\expandafter\R@moveMeaningHeader\meaning#1}}%
\def\execute#1{#1}
\def\psm@keother#1{\catcode`#112\relax}
\def\executeinspecs#1{%
\execute{\begingroup\let\do\psm@keother\dospecials\catcode`\^^M=9#1\endgroup}}%
\def\@mpty{}%
\def\matchexpin#1#2{
  \fi%
  \edef\tmpb{{#2}}%
  \expandafter\makem@tchtmp\tmpb%
  \edef\tmpa{#1}\edef\tmpb{#2}%
  \expandafter\expandafter\expandafter\m@tchtmp\expandafter\tmpa\tmpb\endm@tch%
  \if\match%
}%
\def\matchin#1#2{%
  \fi%
  \makem@tchtmp{#2}%
  \m@tchtmp#1#2\endm@tch%
  \if\match%
}%
\def\makem@tchtmp#1{\def\m@tchtmp##1#1##2\endm@tch{%
  \def\tmpa{##1}\def\tmpb{##2}\let\m@tchtmp=\relax%
  \ifx\tmpb\@mpty\def\match{YN}%
  \else\def\match{YY}\fi%
}}%
\def\incm#1{{\psxoffset=1cm\d@my=#1
 \d@mx=\d@my
  \divide\d@mx by \psxoffset
  \xdef\dimincm{\number\d@mx.}
  \advance\d@my by -\number\d@mx cm
  \multiply\d@my by 100
 \d@mx=\d@my
  \divide\d@mx by \psxoffset
  \edef\dimincm{\dimincm\number\d@mx}
  \advance\d@my by -\number\d@mx cm
  \multiply\d@my by 100
 \d@mx=\d@my
  \divide\d@mx by \psxoffset
  \xdef\dimincm{\dimincm\number\d@mx}
}}%
%
\newif\ifNotB@undingBox
\newhelp\PShelp{Proceed: you'll have a 5cm square blank box instead of
your graphics (Jean Orloff).}%
\def\s@tsize#1 #2 #3 #4\@ndsize{
  \def\psllx{#1}\def\pslly{#2}%
  \def\psurx{#3}\def\psury{#4}
  \ifx\psurx\@mpty\NotB@undingBoxtrue
  \else
    \drawinght=#4bp\advance\drawinght by-#2bp
    \drawingwd=#3bp\advance\drawingwd by-#1bp
  \fi
  }%
\def\sc@nBBline#1:#2\@ndBBline{\edef\p@rameter{#1}\edef\v@lue{#2}}%
\def\g@bblefirstblank#1#2:{\ifx#1 \else#1\fi#2}%
{\catcode`\%=12
\xdef\B@undingBox{
\def\ReadPSize#1{
 \readfilename#1\relax
 \let\PSfilename=\lastreadfilename
 \openin\pst@mpin=#1\relax
 \ifeof\pst@mpin \errhelp=\PShelp
   \errmessage{I haven't found your postscript file (\PSfilename)}%
   \psloc@lerr{was not found}%
   \s@tsize 0 0 142 142\@ndsize
   \closein\pst@mpin
 \else
   \if\matchexpin{\GlobalInputList}{, \lastreadfilename}%
   \else\xdef\GlobalInputList{\GlobalInputList, \lastreadfilename}%
     \immediate\write\psbj@inaux{\lastreadfilename,}%
   \fi%
   \loop
     \executeinspecs{\catcode`\ =10\global\read\pst@mpin to\n@xtline}%
     \ifeof\pst@mpin
       \errhelp=\PShelp
       \errmessage{(\PSfilename) is not an Encapsulated PostScript File:
           I could not find any \B@undingBox: line.}%
       \edef\v@lue{0 0 142 142:}%
       \psloc@lerr{is not an EPSFile}%
       \NotB@undingBoxfalse
     \else
       \expandafter\sc@nBBline\n@xtline:\@ndBBline
       \ifx\p@rameter\B@undingBox\NotB@undingBoxfalse
         \edef\t@mp{%
           \expandafter\g@bblefirstblank\v@lue\space\space\space}%
         \expandafter\s@tsize\t@mp\@ndsize
       \else\NotB@undingBoxtrue
       \fi
     \fi
   \ifNotB@undingBox\repeat
   \closein\pst@mpin
 \fi
\message{#1}%
}%
%
%
\def\psboxto(#1;#2)#3{\vbox{
   \ReadPSize{#3}%
   \divide\drawingwd by 1000
   \divide\drawinght by 1000
   \d@mx=#1
   \ifdim\d@mx=0pt\xscale=1000
         \else \xscale=\d@mx \divide \xscale by \drawingwd\fi
   \d@my=#2
   \ifdim\d@my=0pt\yscale=1000
         \else \yscale=\d@my \divide \yscale by \drawinght\fi
   \ifnum\yscale=1000
         \else\ifnum\xscale=1000\xscale=\yscale
                    \else\ifnum\yscale<\xscale\xscale=\yscale\fi
              \fi
   \fi
   \divide\pswdincr by 1000 \multiply\pswdincr by \xscale
   \divide\pshtincr by 1000 \multiply\pshtincr by \xscale
   \divide\psxoffset by1000 \multiply\psxoffset by\xscale
   \divide\psyoffset by1000 \multiply\psyoffset by\xscale
   \global\divide\pscm by 1000
   \global\multiply\pscm by\xscale
   \multiply\drawingwd by\xscale \multiply\drawinght by\xscale
   \ifdim\d@mx=0pt\d@mx=\drawingwd\fi
   \ifdim\d@my=0pt\d@my=\drawinght\fi
   \message{scaled \the\xscale}%
 \hbox to\d@mx{\hss\vbox to\d@my{\vss
   \global\setbox\drawingBox=\hbox to 0pt{\kern\psxoffset\vbox to 0pt{
      \kern-\psyoffset
      \PSspeci@l{\PSfilename}{\the\xscale}%
      \vss}\hss\ps@nnotation}%
   \advance\pswdincr by \drawingwd
   \advance\pshtincr by \drawinght
   \global\wd\drawingBox=\the\pswdincr
   \global\ht\drawingBox=\the\pshtincr
   \baselineskip=0pt
   \copy\drawingBox
 \vss}\hss}%
  \global\psxoffset=0pt
  \global\psyoffset=0pt
  \global\pswdincr=0pt
  \global\pshtincr=0pt 
  \global\pscm=1cm 
  \global\drawingwd=\drawingwd
  \global\drawinght=\drawinght
}}%
%
%
\def\psboxscaled#1#2{\vbox{
  \ReadPSize{#2}%
  \xscale=#1
  \message{scaled \the\xscale}%
  \advance\drawingwd by\pswdincr\advance\drawinght by\pshtincr
  \divide\pswdincr by 1000 \multiply\pswdincr by \xscale
  \divide\pshtincr by 1000 \multiply\pshtincr by \xscale
  \divide\psxoffset by1000 \multiply\psxoffset by\xscale
  \divide\psyoffset by1000 \multiply\psyoffset by\xscale
  \divide\drawingwd by1000 \multiply\drawingwd by\xscale
  \divide\drawinght by1000 \multiply\drawinght by\xscale
  \global\divide\pscm by 1000
  \global\multiply\pscm by\xscale
  \global\setbox\drawingBox=\hbox to 0pt{\kern\psxoffset\vbox to 0pt{
     \kern-\psyoffset
     \PSspeci@l{\PSfilename}{\the\xscale}%
     \vss}\hss\ps@nnotation}%
  \advance\pswdincr by \drawingwd
  \advance\pshtincr by \drawinght
  \global\wd\drawingBox=\the\pswdincr
  \global\ht\drawingBox=\the\pshtincr
  \baselineskip=0pt
  \copy\drawingBox
  \global\psxoffset=0pt
  \global\psyoffset=0pt
  \global\pswdincr=0pt
  \global\pshtincr=0pt 
  \global\pscm=1cm
  \global\drawingwd=\drawingwd
  \global\drawinght=\drawinght
}}%
%
\def\psbox#1{\psboxscaled{1000}{#1}}%
\newif\ifn@teof\n@teoftrue
\newif\ifc@ntrolline
\newif\ifmatch
\newread\j@insplitin
\newwrite\j@insplitout
\newwrite\psbj@inaux
\immediate\openout\psbj@inaux=psbjoin.aux
\immediate\write\psbj@inaux{\string\joinfiles}%
\immediate\write\psbj@inaux{\jobname,}%
%
%
\def\toother#1{\ifcat\relax#1\else\expandafter%
  \toother@ux\meaning#1\endtoother@ux\fi}%
\def\toother@ux#1 #2#3\endtoother@ux{\def\tmp{#3}%
  \ifx\tmp\@mpty\def\tmp{#2}\let\next=\relax%
  \else\def\next{\toother@ux#2#3\endtoother@ux}\fi%
\next}%
%
%
\let\readfilenamehook=\relax
\def\re@d{\expandafter\re@daux}
\def\re@daux{\futurelet\nextchar\stopre@dtest}%
\def\re@dnext{\xdef\lastreadfilename{\lastreadfilename\nextchar}%
  \afterassignment\re@d\let\nextchar}%
\def\stopre@d{\egroup\readfilenamehook}%
\def\stopre@dtest{%
  \ifcat\nextchar\relax\let\nextread\stopre@d
  \else
    \ifcat\nextchar\space\def\nextread{%
      \afterassignment\stopre@d\chardef\nextchar=`}%
    \else\let\nextread=\re@dnext
      \toother\nextchar
      \edef\nextchar{\tmp}%
    \fi
  \fi\nextread}%
\def\readfilename{\vbox\bgroup%
  \let\\=\backslashother \let\%=\percentother \let\~=\tildeother
  \let\#=\sharpother \xdef\lastreadfilename{}%
  \re@d}%
%
%
\xdef\GlobalInputList{\jobname}%
\def\psnewinput{%
  \def\readfilenamehook{
    \if\matchexpin{\GlobalInputList}{, \lastreadfilename}%
    \else\xdef\GlobalInputList{\GlobalInputList, \lastreadfilename}%
      \immediate\write\psbj@inaux{\lastreadfilename,}%
    \fi%
    \ps@ldinput\lastreadfilename\relax%
    \let\readfilenamehook=\relax%
  }\readfilename%
}%
\expandafter\ifx\csname @@input\endcsname\relax    
  \immediate\let\ps@ldinput=\input\def\input{\psnewinput}%
\else
  \immediate\let\ps@ldinput=\@@input
  \def\@@input{\psnewinput}%
\fi%
\def\nowarnopenout{%
 \def\warnopenout##1##2{%
   \readfilename##2\relax
   \message{\lastreadfilename}%
   \immediate\openout##1=\lastreadfilename\relax}}%
\def\warnopenout#1#2{%
 \readfilename#2\relax
 \def\t@mp{TrashMe,psbjoin.aux,psbjoint.tex,}\uncatcode\t@mp
 \if\matchexpin{\t@mp}{\lastreadfilename,}%
 \else
   \immediate\openin\pst@mpin=\lastreadfilename\relax
   \ifeof\pst@mpin
     \else
     \errhelp{If the content of this file is so precious to you, abort (ie
press x or e) and rename it before retrying.}%
     \errmessage{I'm just about to replace your file named \lastreadfilename}%
   \fi
   \immediate\closein\pst@mpin
 \fi
 \message{\lastreadfilename}%
 \immediate\openout#1=\lastreadfilename\relax}%
{\catcode`\%=12\catcode`\*=14
\gdef\splitfile#1{*
 \readfilename#1\relax
 \immediate\openin\j@insplitin=\lastreadfilename\relax
 \ifeof\j@insplitin
   \message{! I couldn't find and split \lastreadfilename!}*
 \else
   \immediate\openout\j@insplitout=TrashMe
   \message{< Splitting \lastreadfilename\space into}*
   \loop
     \ifeof\j@insplitin
       \immediate\closein\j@insplitin\n@teoffalse
     \else
       \n@teoftrue
       \executeinspecs{\global\read\j@insplitin to\spl@tinline\expandafter
         \ch@ckbeginnewfile\spl@tinline
       \ifc@ntrolline
       \else
         \toks0=\expandafter{\spl@tinline}*
         \immediate\write\j@insplitout{\the\toks0}*
       \fi
     \fi
   \ifn@teof\repeat
   \immediate\closeout\j@insplitout
 \fi\message{>}*
}*
\gdef\ch@ckbeginnewfile#1
 \def\t@mp{#1}*
 \ifx\@mpty\t@mp
   \def\t@mp{#3}*
   \ifx\@mpty\t@mp
     \global\c@ntrollinefalse
   \else
     \immediate\closeout\j@insplitout
     \warnopenout\j@insplitout{#2}*
     \global\c@ntrollinetrue
   \fi
 \else
   \global\c@ntrollinefalse
 \fi}*
\gdef\joinfiles#1\into#2{*
 \message{< Joining following files into}*
 \warnopenout\j@insplitout{#2}*
 \message{:}*
 {*
 \edef\w@##1{\immediate\write\j@insplitout{##1}}*
\w@{
\w@{
\w@{
\w@{
\w@{
\w@{
\w@{
\w@{
\w@{
\w@{
\w@{\string\input\space psbox.tex}*
\w@{\string\splitfile{\string\jobname}}*
\w@{\string\let\string\autojoin=\string\relax}*
}*
 \expandafter\tre@tfilelist#1, \endtre@t
 \immediate\closeout\j@insplitout
 \message{>}*
}*
\gdef\tre@tfilelist#1, #2\endtre@t{*
 \readfilename#1\relax
 \ifx\@mpty\lastreadfilename
 \else
   \immediate\openin\j@insplitin=\lastreadfilename\relax
   \ifeof\j@insplitin
     \errmessage{I couldn't find file \lastreadfilename}*
   \else
     \message{\lastreadfilename}*
     \immediate\write\j@insplitout{
     \executeinspecs{\global\read\j@insplitin to\oldj@ininline}*
     \loop
       \ifeof\j@insplitin\immediate\closein\j@insplitin\n@teoffalse
       \else\n@teoftrue
         \executeinspecs{\global\read\j@insplitin to\j@ininline}*
         \toks0=\expandafter{\oldj@ininline}*
         \let\oldj@ininline=\j@ininline
         \immediate\write\j@insplitout{\the\toks0}*
       \fi
     \ifn@teof
     \repeat
   \immediate\closein\j@insplitin
   \fi
   \tre@tfilelist#2, \endtre@t
 \fi}*
}%
\def\autojoin{%
 \immediate\write\psbj@inaux{\string\into{psbjoint.tex}}%
 \immediate\closeout\psbj@inaux
 \expandafter\joinfiles\GlobalInputList\into{psbjoint.tex}%
}%
%
%
%
\def\centinsert#1{\midinsert\line{\hss#1\hss}\endinsert}%
\def\psannotate#1#2{\vbox{%
  \def\ps@nnotation{#2\global\let\ps@nnotation=\relax}#1}}%
\def\pscaption#1#2{\vbox{%
   \setbox\drawingBox=#1
   \copy\drawingBox
   \vskip\baselineskip
   \vbox{\hsize=\wd\drawingBox\setbox0=\hbox{#2}%
     \ifdim\wd0>\hsize
       \noindent\unhbox0\tolerance=5000
    \else\centerline{\box0}%
    \fi
}}}%
%
\def\at(#1;#2)#3{\setbox0=\hbox{#3}\ht0=0pt\dp0=0pt
  \rlap{\kern#1\vbox to0pt{\kern-#2\box0\vss}}}%
%
\newdimen\gridht \newdimen\gridwd
\def\gridfill(#1;#2){%
  \setbox0=\hbox to 1\pscm
  {\vrule height1\pscm width.4pt\leaders\hrule\hfill}%
  \gridht=#1
  \divide\gridht by \ht0
  \multiply\gridht by \ht0
  \gridwd=#2
  \divide\gridwd by \wd0
  \multiply\gridwd by \wd0
  \advance \gridwd by \wd0
  \vbox to \gridht{\leaders\hbox to\gridwd{\leaders\box0\hfill}\vfill}}%
%
\def\fillinggrid{\at(0cm;0cm){\vbox{%
  \gridfill(\drawinght;\drawingwd)}}}%
%
%
\def\textleftof#1:{%
  \setbox1=#1
  \setbox0=\vbox\bgroup
    \advance\hsize by -\wd1 \advance\hsize by -2em}%
\def\textrightof#1:{%
  \setbox0=#1
  \setbox1=\vbox\bgroup
    \advance\hsize by -\wd0 \advance\hsize by -2em}%
\def\endtext{%
  \egroup
  \hbox to \hsize{\valign{\vfil##\vfil\cr%
\box0\cr%
\noalign{\hss}\box1\cr}}}%
%
\def\frameit#1#2#3{\hbox{\vrule width#1\vbox{%
  \hrule height#1\vskip#2\hbox{\hskip#2\vbox{#3}\hskip#2}%
        \vskip#2\hrule height#1}\vrule width#1}}%
\def\boxit#1{\frameit{0.4pt}{0pt}{#1}}%
\catcode`\@=12 
%
 \psfordvips   

\dofigure{
$$\psboxto(15 true cm;0cm){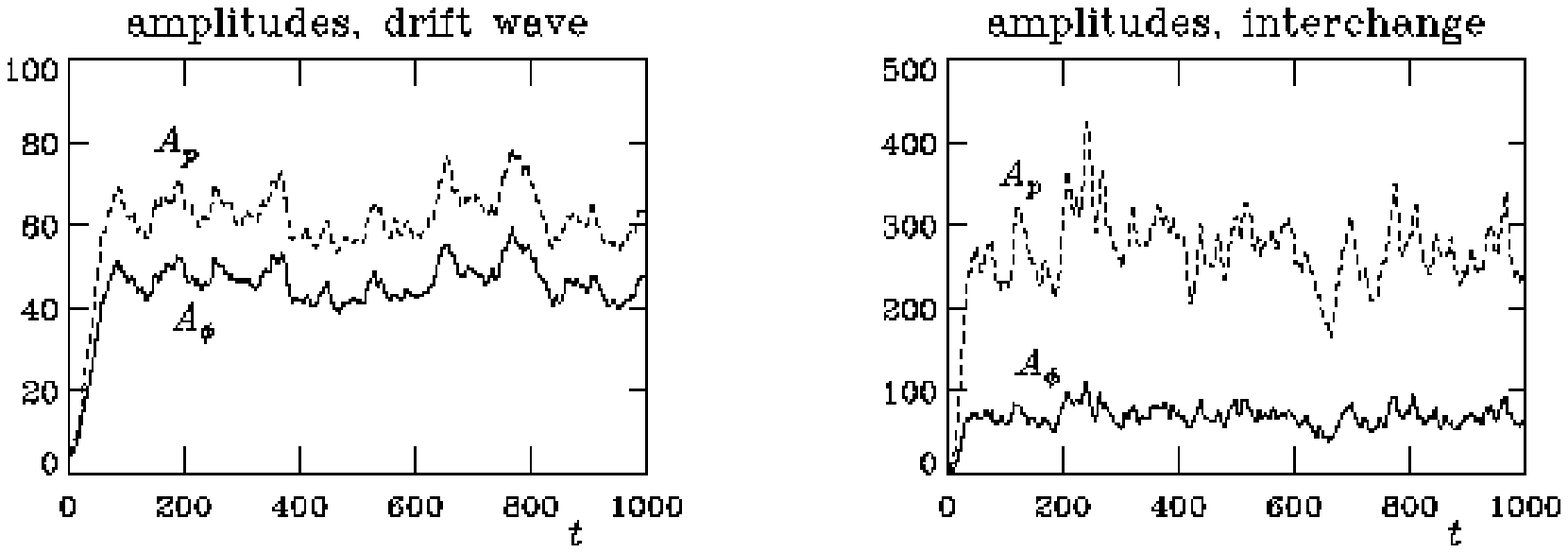}$$
}
{Figure \figctrldgdw.}{ (left) Time traces of the half squared
amplitudes of $\pefl$ and $\phifl$, respectively labelled by
$A_p$ and $A_\phi$, in the drift wave model (right) and the
interchange model (right), showing saturation.  Only in the drift wave
model are the time traces for $A_p$ and $A_\phi$ similar.  
}

\dofigure{
$$\psboxto(15 true cm;0cm){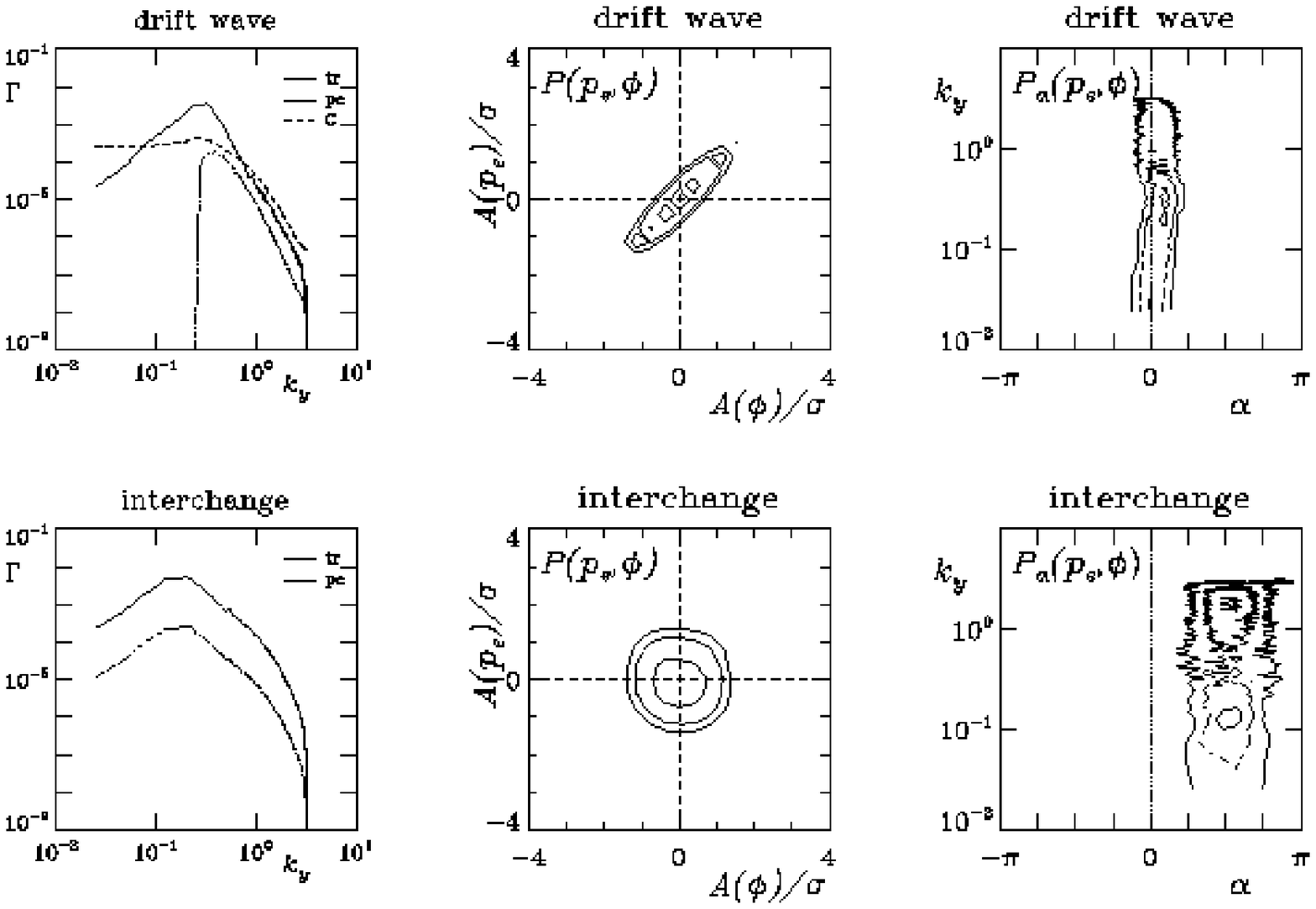}$$
}
{Figure \figcontrol.}{  (left to right) Drive and transfer
spectra, and cross coherence and phase shifts between $\pefl$ and
$\phifl$,  for the drift wave (top row) and interchange (bottom row)
models.
In the drift wave case the sink spectrum (labelled `C') is relatively
flat, and the transfer (`tr') is also due to $\dpl\Jfl$ and is positive
at short wavelength and negative at long wavelength, while in the
interchange case it is due to $\kappacv(\pefl)$ which follows the source
spectrum (`pe'). 
The drift wave case shows strong cross
coherence and a narrow phase shift distribution closer to zero than to
$\pi/2$.  The interchange case shows dominance by the longest
wavelengths, no cross coherence, and phase shifts near $\pi/2$ due to
the strong driving and weak coupling.  
The phase shift distributions contours are $0.3$,
$0.5$, and $0.8$ times the maximum, and for the cross coherence they are 
$0.37$, $0.5$, and $0.8$ times the maximum.
}

\dofigure{
$$\psboxto(15 true cm;0cm){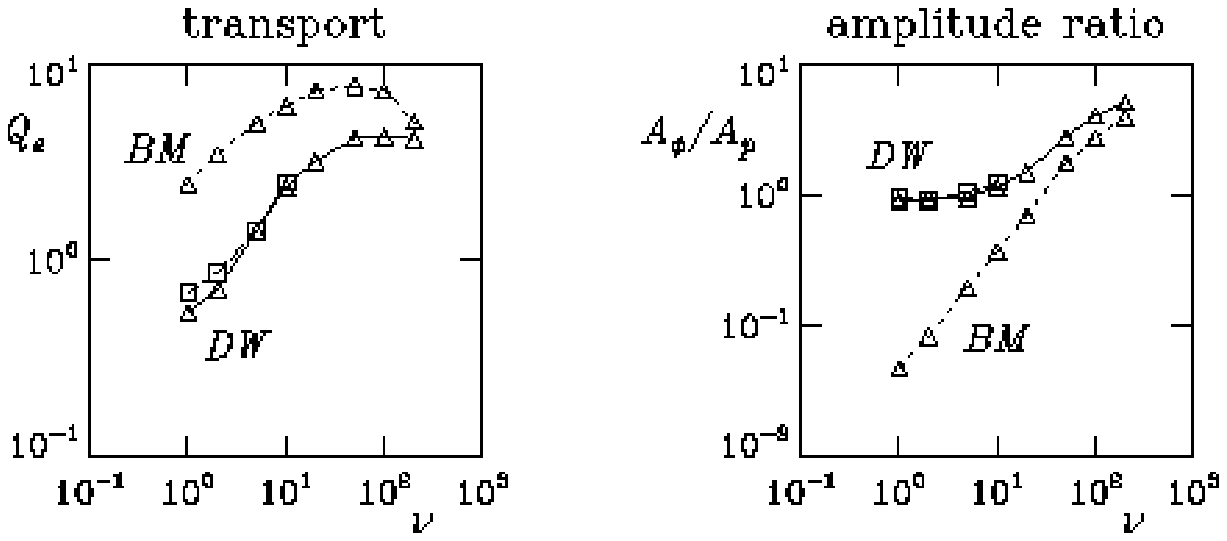}$$
}
{Figure \figbdmode.}{  (left) Transport scaling of drift wave and
interchange turbulence in toroidal geometry, from the DALF3 and reduced
resistive MHD models labelled `DW' and `BM', respectively.
The drift wave cases show the clearest scaling with 
collisionality at low $C=2.55\nu$.  At asymptotically large $C$
the trends will merge, but that limit is not reached.  (right) Half
squared amplitude ratio (including only $k_y\ne 0$)
for the two sets of cases.  Due to the adiabatic response,
$\phifl$ tracks $\pefl$ for drift wave turbulence, but in the MHD
model of the interchange cases $\pefl$ is unaffected by the Alfv\'en
dynamics, and so instead of $\phifl$ being forced towards $\pefl$ it is
forced towards zero.  The extra points marked with squares for DW
($\nu=1,2,5,10$) are with double resolution in the drift plane.
}

\dofigure{
$$\psboxto(15 true cm;0cm){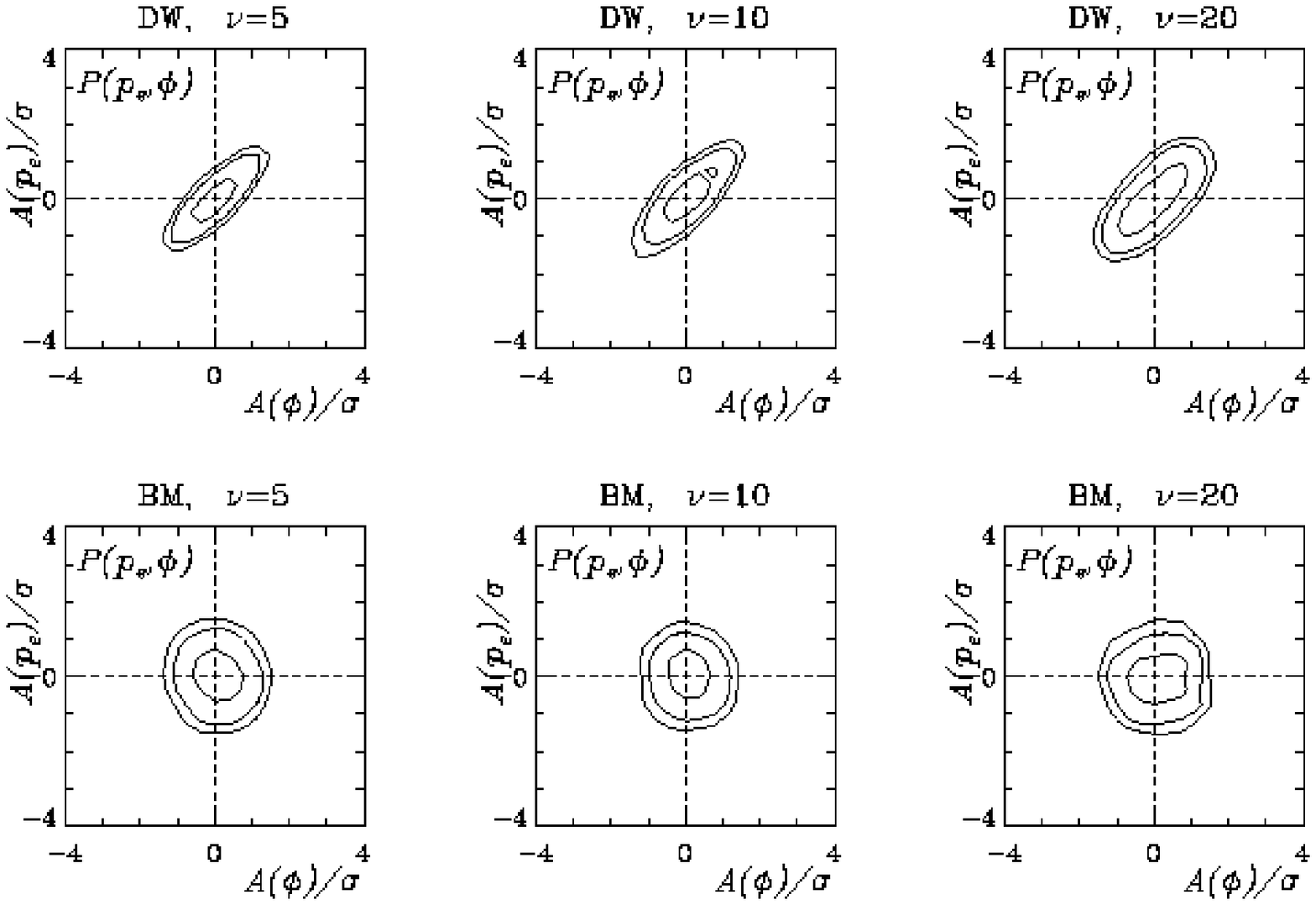}$$
}
{Figure \figplcoher.}{  Cross coherence between $\pefl$ and
$\phifl$, for drift wave (top row) and interchange (bottom row)
turbulence, for 
$\nu=5$, $10$, and $20$ (left to right), where $C=2.55\nu$ and
$\nu_B=C\wcv$ as defined in Eq.~(\eqresbal).  Compare
with the results in Section~\seccontrol\ for drift wave and interchange
turbulence.  The DALF3 model results in drift 
wave mode structure even for larger $\nu_B$, while the MHD model
always shows interchange mode structure. 
Contours are as in Fig.~\figcontrol.
}

\dofigure{
$$\psboxto(15 true cm;0cm){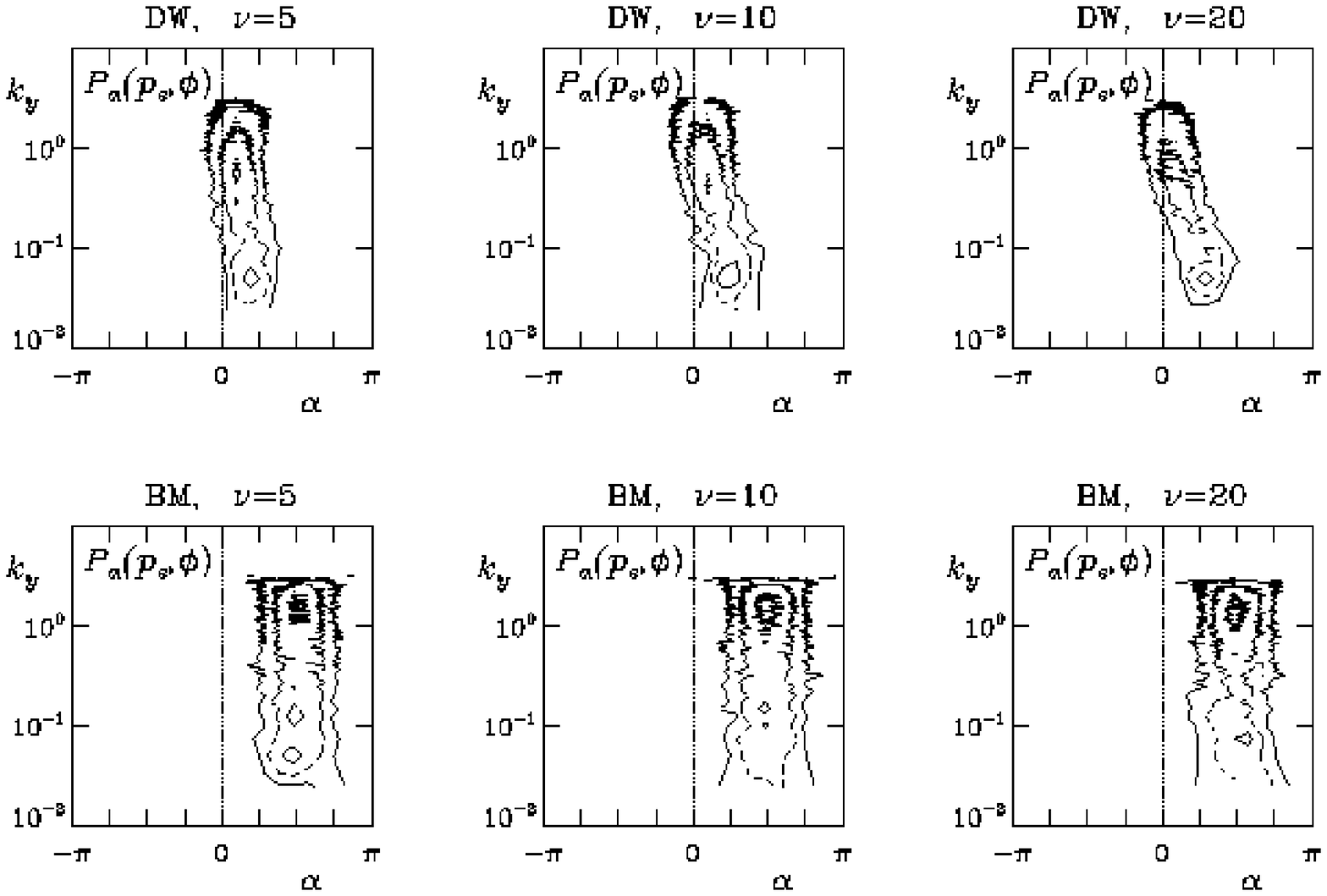}$$
}
{Figure \figphasdist.}{  Phase shift distributions of $\pefl$
ahead of $\phifl$ at each $k_y$, 
for drift wave (top row) and interchange (bottom row)
turbulence, for 
$\nu=5$, $10$, and $20$ (left to right), where $C=2.55\nu$ and
$\nu_B=C\wcv$ as defined in Eq.~(\eqresbal).  Compare
with the results in Section~\seccontrol\ for drift wave and interchange
turbulence.  The DALF3 model results in drift 
wave mode structure for $\nu<10$, while the MHD model
always shows interchange mode structure.  
The transition to resistive ballooning turbulence in the DALF3 model
occurs in the longest wavelengths, $k_y\rs<0.1$.  
Contours are as in Fig.~\figcontrol.  }

\dofigure{
$$\psboxto(15 true cm;0cm){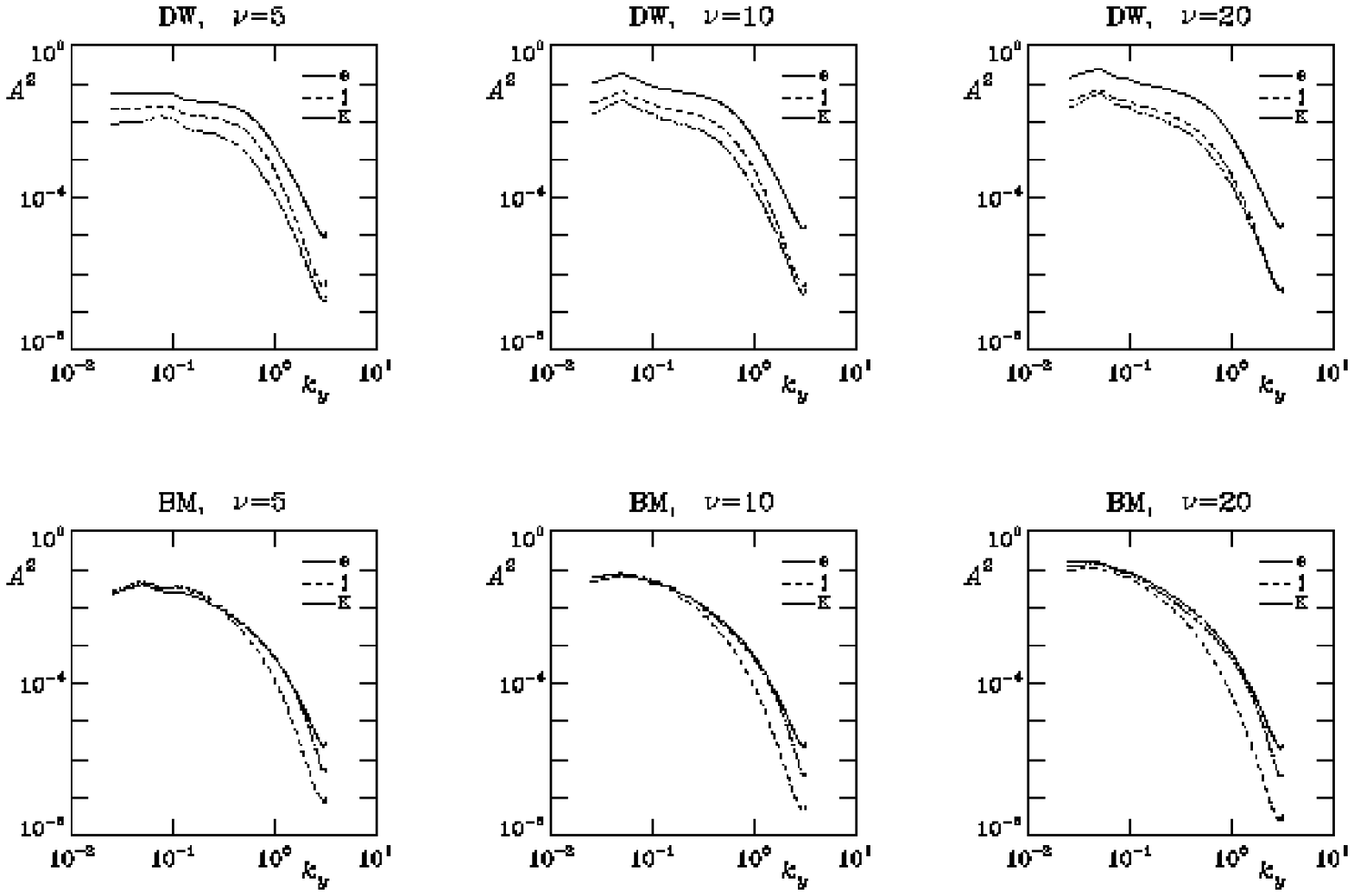}$$
}
{Figure \figwtrans.}{  Spectra of the rms transfer dynamical
levels for each $k_y$ in the spectrum, comparing the
sizes of $\phifl\dpl\Jfl$ (`j'), $\phifl\kappacv(\pefl)$ (`k'), and 
$\phifl\vedl\vorfl$ (`e'),
for drift wave (top row) and interchange (bottom row) turbulence, for
$\nu=5$, $10$, and $20$ (left to right), where $C=2.55\nu$ and
$\nu_B=C\wcv$.
In drift wave turbulence the transfer through the current is
too large to be accounted for by the curvature, and at all wavelengths
the vorticity nonlinearity is balanced only by the linear time
derivative.  Nonlinear vorticity dynamics is generally much stronger
when the adiabatic coupling mechanism $\pefl\fromto\Jfl$ is present,
leading to a self consistent situation in which both mechanisms catalyse
each other. 
}

\dofigure{
$$\psboxto(15 true cm;0cm){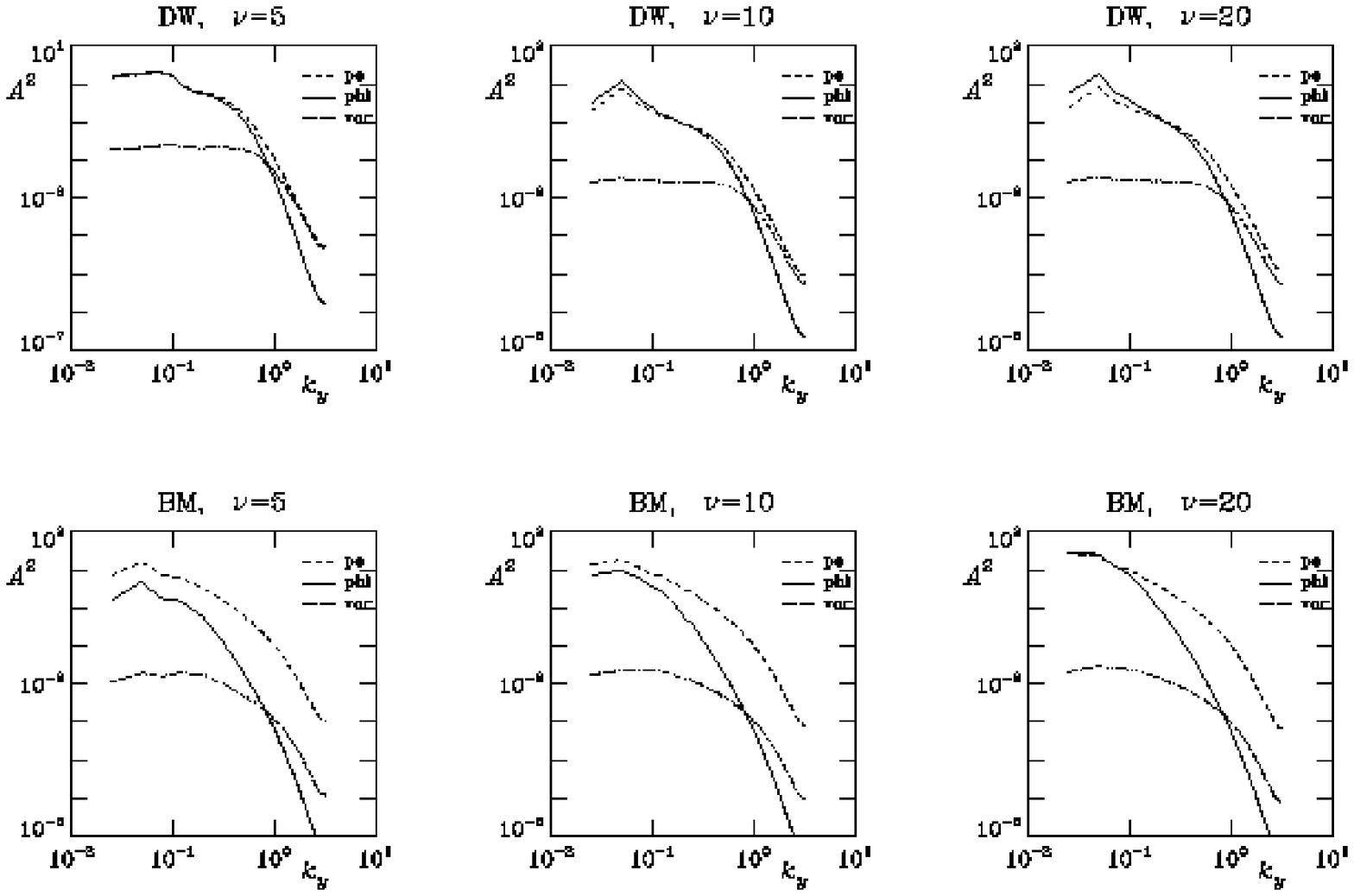}$$
}
{Figure \figpldwa.}{  Amplitude spectra of $\pefl$,
$\phifl$,and  $\vorfl$, respectively labelled by `pe', `phi', and `vor',
for drift wave (top row) and interchange (bottom row) turbulence, for
$\nu=5$, $10$, and $20$ (left to right), where $C=2.55\nu$ and
$\nu_B=C\wcv$. 
With only the density present
one cannot distinguish the mode structure or dynamics, but if $\phifl$
is also present the amplitude ratio, shown in Fig.~\figbdmode, is
decisive, as the spectrum of $\phifl$ follows that of $\pefl$ only for
drift wave turbulence.
The transition to resistive ballooning turbulence in the DALF3 model
($\phifl\gg\pefl$) occurs in the longest wavelengths, $k_y\rs<0.1$.  
}

\dofigure{
$$\psboxto(15 true cm;0cm){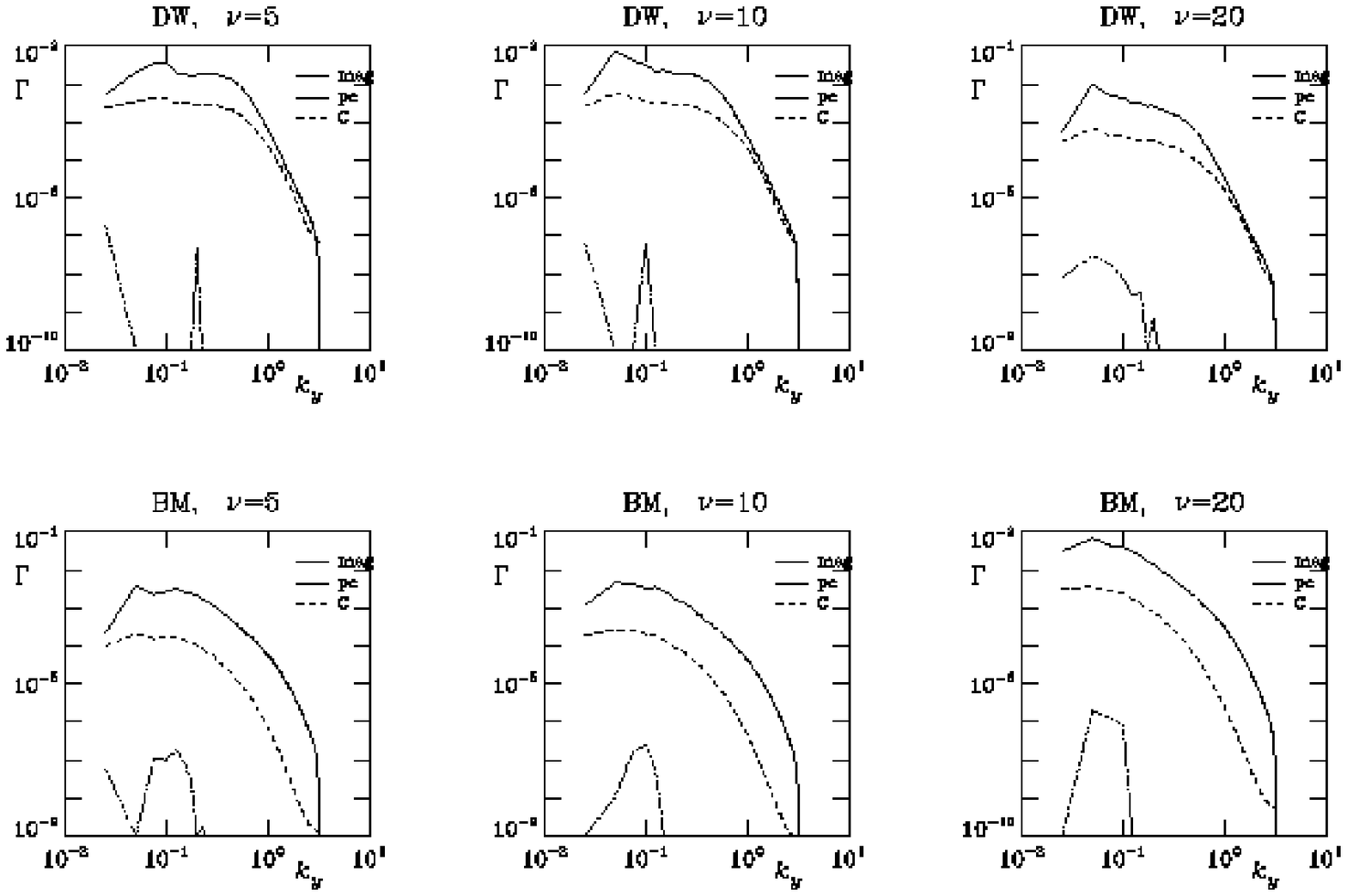}$$
}
{Figure \figpldwb.}{  Spectra of the ExB gradient drive (`pe'),
dissipation (`C'), and the magnetic flutter drive (`mag'),
for drift wave (top row) and interchange (bottom row) turbulence, for
$\nu=5$, $10$, and $20$ (left to right), where $C=2.55\nu$ and
$\nu_B=C\wcv$. 
For drift wave turbulence the shorter wavelengths contribute more to the
energetic drive and hence the ExB transport.  Compare the positions of
the energy containing range (Fig.~\figpldwa), the energy producing range
(this figure) and the vorticity catalysing range (Fig.~\figwtrans).
These show that for drift wave turbulence the entire spectrum acts as a
single unit.  
}

\dofigure{
$$\psboxto(15 true cm;0cm){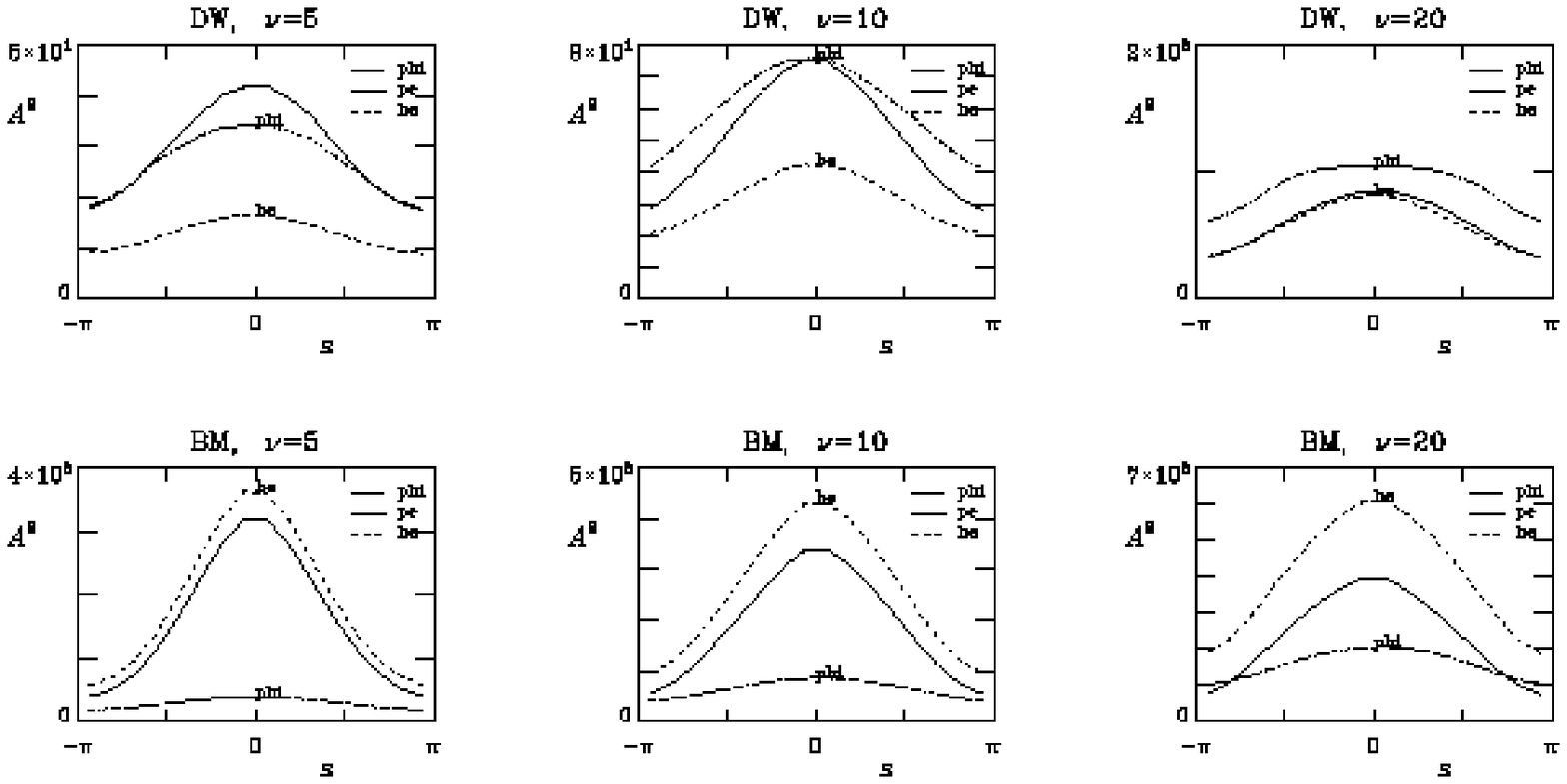}$$
}
{Figure \figplfls.}{  Mean squared amplitude envelopes 
($k_y\ne 0$ only)
showing parallel structure of $\phifl$, $\pefl$,
and $\hefl=\pefl-\phifl$, respectively labelled by `phi', `pe', and  `he',
for drift wave (top row) and interchange (bottom row) turbulence, 
for $\nu=5$, $10$, and $20$ (left to right), where $C=2.55\nu$ and
$\nu_B=C\wcv$.
Drift wave mode structure is
exemplary for $\nu<10$ in the DALF3 model, with $\hefl$ flatter and
smaller than either $\pefl$ or $\phifl$. 
}

\dofigure{
$$\psboxto(15 true cm;0cm){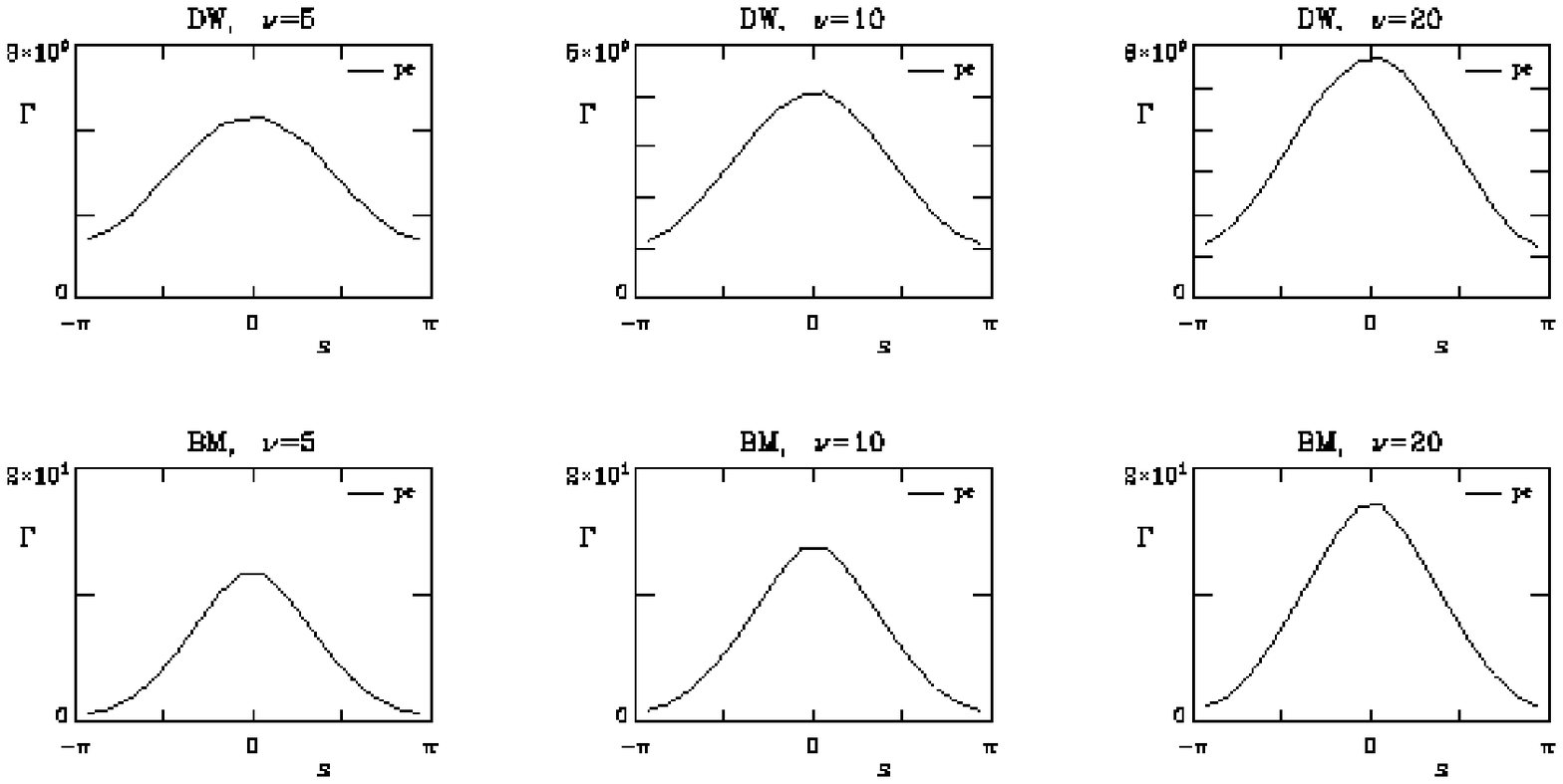}$$
}
{Figure \figplbal.}{  Amplitude envelope of the ExB transport
(`pe'), 
for drift wave (top row) and interchange (bottom row) turbulence, for
$\nu=5$, $10$, and $20$ (left to right), where $C=2.55\nu$ and
$\nu_B=C\wcv$.
The magnetic flutter transport is negligible on this scale.
The ballooning in the
transport becomes somewhat more pronounced in the transition to interchange
character for $\nu>10$, but for all cases the ballooning is much
stronger for the MHD model than for DALF3.  
}


\dofigure{
$$\psboxto(15 true cm;0cm){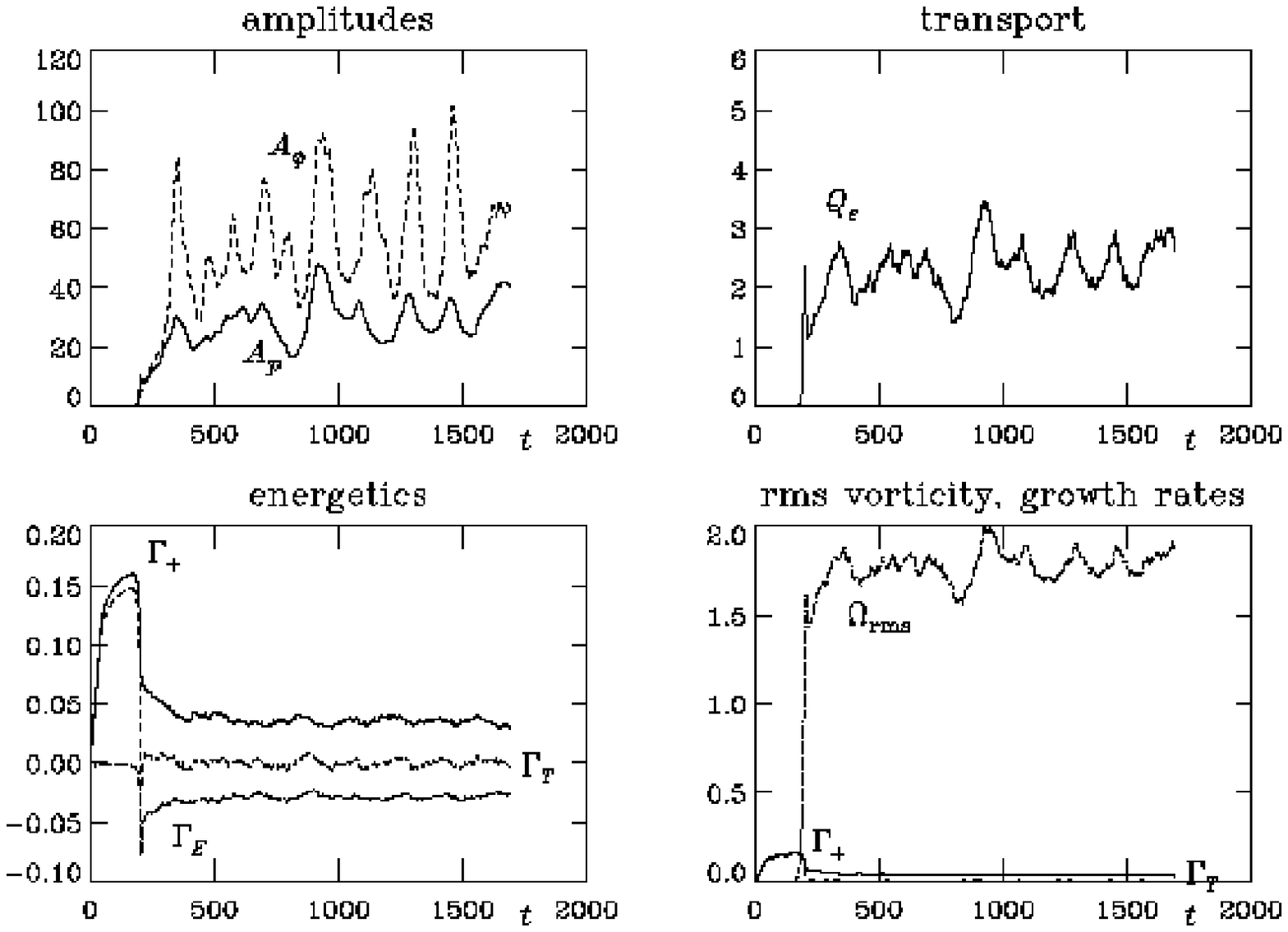}$$
}
{Figure \figdgdwl.}{ Time evolution of the DALF3 case with
$\nu=10$ (hence $C=25.5$ and $\nu_B=1.25$) out of a linear initial state
at small amplitude, run to $t=1582$.
(top left) Half squared amplitudes of $\phifl$ (denoted $A_\phi$) and 
$\pefl$ ($A_p$).  The larger amplitude
departures of $A_\phi$ from $A_p$ reflect zonal flow activity in the fully
developed turbulence.
(top right) The transport caused by the turbulence.
(bottom left) Growth rate ($\Gamma_T$), gradient drive rate ($\Gamma_+$), 
and total dissipation rate ($\Gamma_E$). 
(bottom right) Vorticity (rms) compared to $\Gamma_+$ and $\Gamma_T$.
The linear mode leads to overshoot,
initial saturation ($t\approx 200$)
is reflected in the first drop of $\Gamma_T$ to zero,
and then the nonlinear mode structure takes over shortly thereafter,
with full development with robust transport and zonal flow activity
established after $t=400$.  The turbulence not only saturates but
changes character when the rms vorticity overcomes the linear growth
rate ($\gamma_L$, equal to $\Gamma_T$ in the linear stage).
Nonlinear saturation of the linear instability would obtain if
$\Omega_{{\rm rms}}$ were comparable to $\gamma_L$, but the situation with 
$\Omega_{{\rm rms}}\gg\gamma_L$ indicates complete supersession of the
instability by turbulence which has its own dynamics.
}

\dofigure{
$$\psboxto(15 true cm;0cm){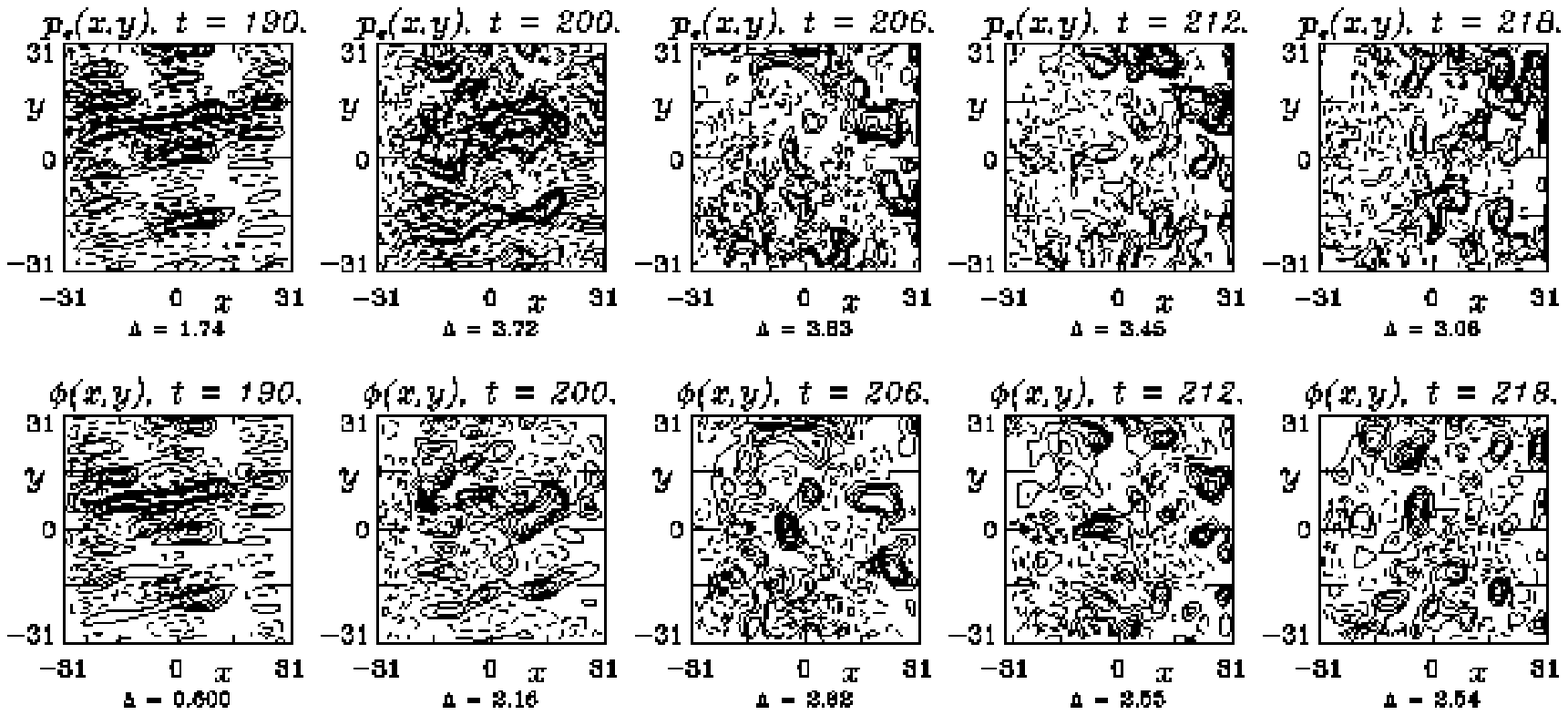}$$
}
{Figure \figplctrl.}{  Saturation and initial transition to
turbulence,
as seen in the spatial morphology of $\phifl$ and $\pefl$
in the linear growth stage
to $t=190$, and the initial saturation stage after $t=200$.
The transition between linear and nonlinear mode structure is most clear
in the disappearance of $x$-direction flows remniscent of bouyant plumes.
This represents a transition away from interchange dominated dynamics to
a more isotropic turbulence as the vorticity nonlinearity replaces the
interchange forcing as the principal mechanism supporting finite
parallel currents (nonadiabatic electron dynamics).
(Positive/negative values are indicated by solid/dashed lines.
Only $1/4$ of the computational domain in the $y$-direction is shown.
The exact moment of saturation is $t=201$, and the growth rate becomes
positive again at $t=212$.)
}

\dofigure{
$$\psboxto(15 true cm;0cm){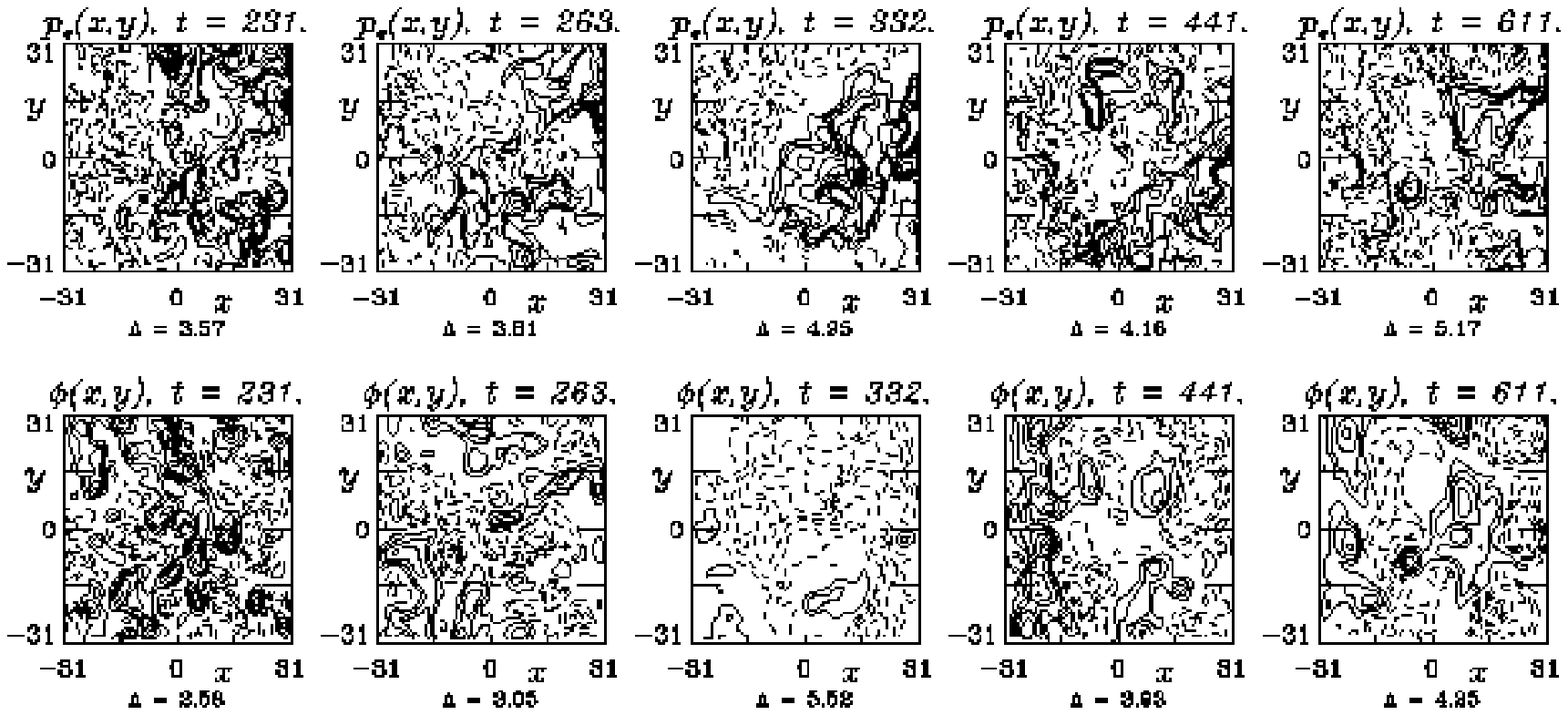}$$
}
{Figure \figplctrlb.}{  Transition to fully developed turbulence,
as seen in the spatial morphology of $\phifl$ and $\pefl$ through the
stage of nonlinear structure adjustment, and then in the stage of
statistical saturation in which the nonlinear growth rate fluctuates
near zero.  As the saturated state finds itself, the scale of motion
increases, and until $t=345$ the nonlinear growth rate is positive.
Zonal flows emerge after about $t=400$ and reach statistical equilibrium
after about $t=600$.  The zonal flows are part of the nonlinear mode
structure, but by the time they emerge the interchange driven flows of
the linear stage are long gone.  The correlation time is about 6 in
these units.
(Positive/negative values are indicated by solid/dashed lines.
Only $1/4$ of the computational domain in the $y$-direction is shown.
The exact moment of saturation is $t=201$, and the growth rate becomes
positive again at $t=212$, and negative again at $t=345$.)
}

\dofigure{
$$\psboxto(15 true cm;0cm){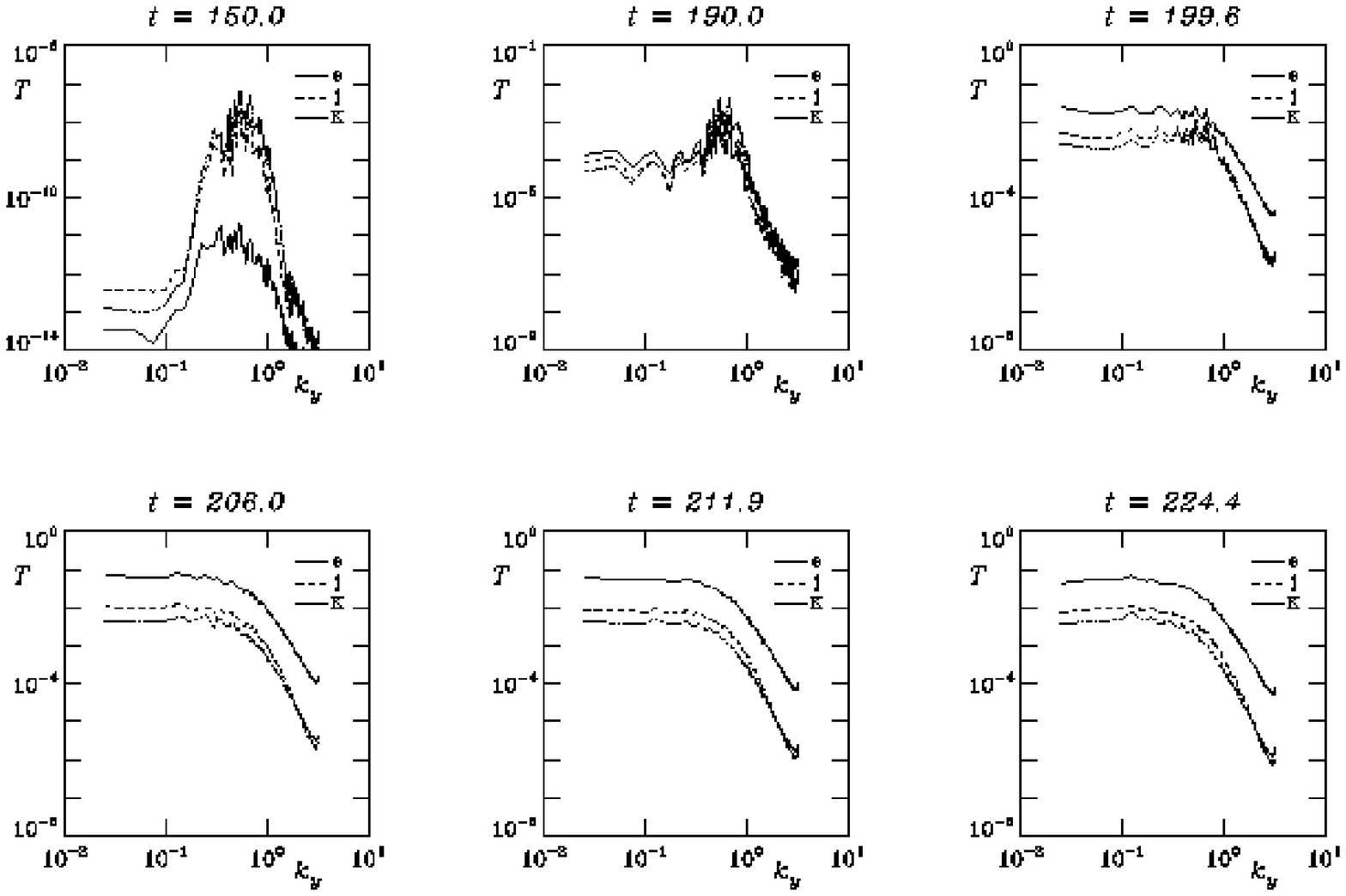}$$
}
{Figure \figwtransl.}{  Saturation and transition to turbulence,
as seen in the dynamical transfer spectra for the ExB vorticity, for the
linear stage to about $t=190$, the saturation stage around $t=200$, and
the nonlinear structure adjustment stage after about $t=210$.
Compare with Fig.~\figwtrans.
Initially the eigenmode is controlled by interchange forcing, with the
linear polarisation drift negligible (cf.\ Eq.~\eqresponse).  But as
the amplitude becomes finite the vorticity nonlinearity, the same one
which causes the drift wave nonlinear instability, emerges to become the
principal agent supporting nonadiabatic electron dynamics.  The
transition is extremely rapid, taking place within about one correlation
time for the fully developed turbulence.
(The exact moment of saturation is $t=201$, and the growth rate becomes
positive again at $t=212$, and negative again at $t=345$.)
}

\dofigure{
$$\psboxto(15 true cm;0cm){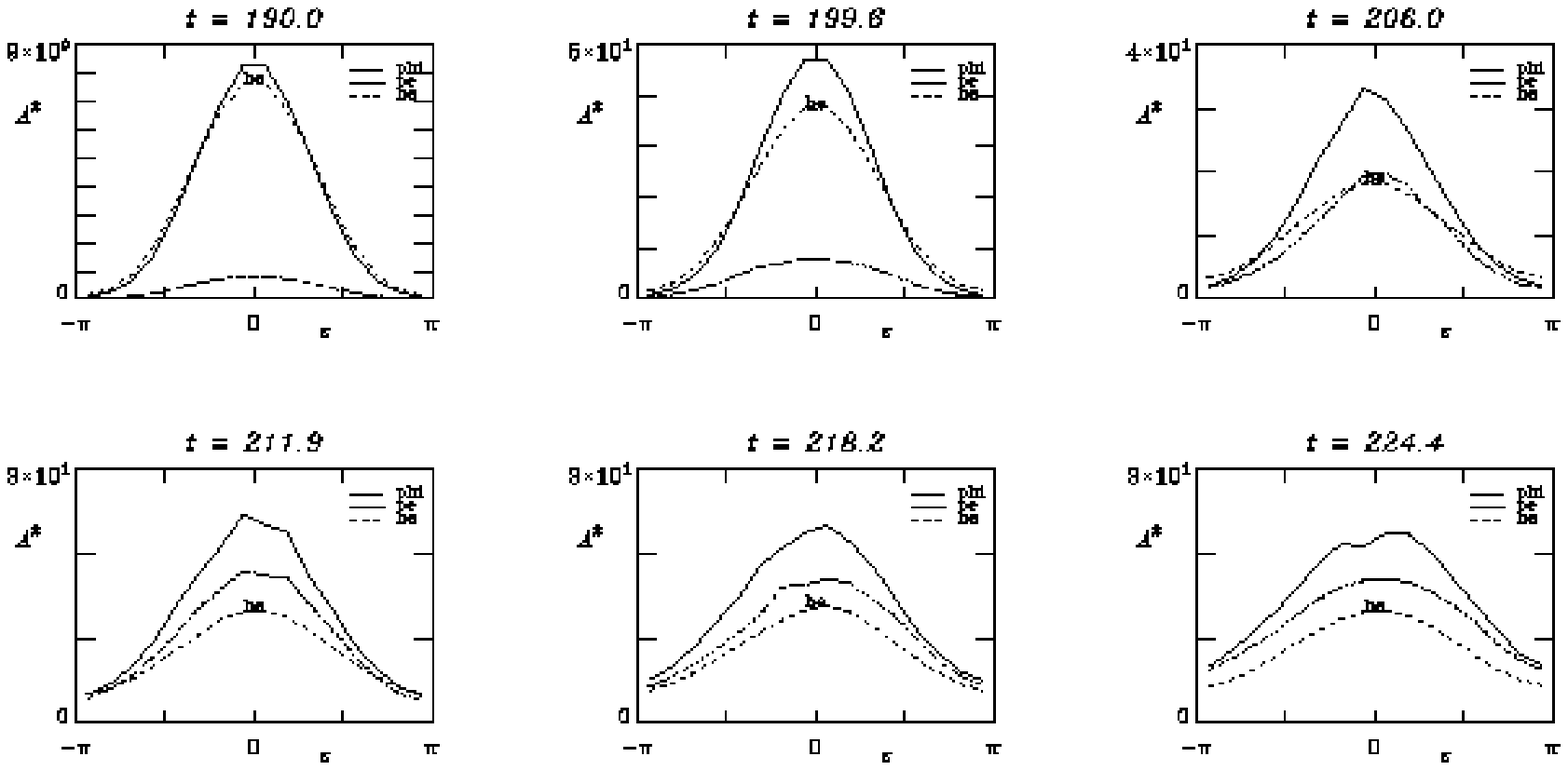}$$
}
{Figure \figplflsl.}{  Saturation and transition to turbulence,
as seen in the parallel envelope structure ($k_y\ne 0$ only), in the 
linear stage to about $t=190$, the saturation stage around $t=200$, and
the nonlinear structure adjustment stage after about $t=210$.
Compare with Fig.~\figplfls.
The principal signature of the transition is the emergence of $\phifl$
supported by the vorticity nonlinearity.  The interchange flows of the
linear stage are eliminated by the turbulent vorticity, and the
adiabatic response causes $\phifl$ to track $\pefl$.
The degree of asymmetry is also reduced, especially in $\hefl$.
This transition between linear ($\hefl\sim\pefl\gg\phifl$)
and nonlinear ($\phifl\sim\pefl\gg\hefl$)
mode structure is almost as rapid as in Fig.~\figwtransl.  
(The exact moment of saturation is $t=201$, and the growth rate becomes
positive again at $t=212$, and negative again at $t=345$.)
}

\dofigure{
$$\psboxto(15 true cm;0cm){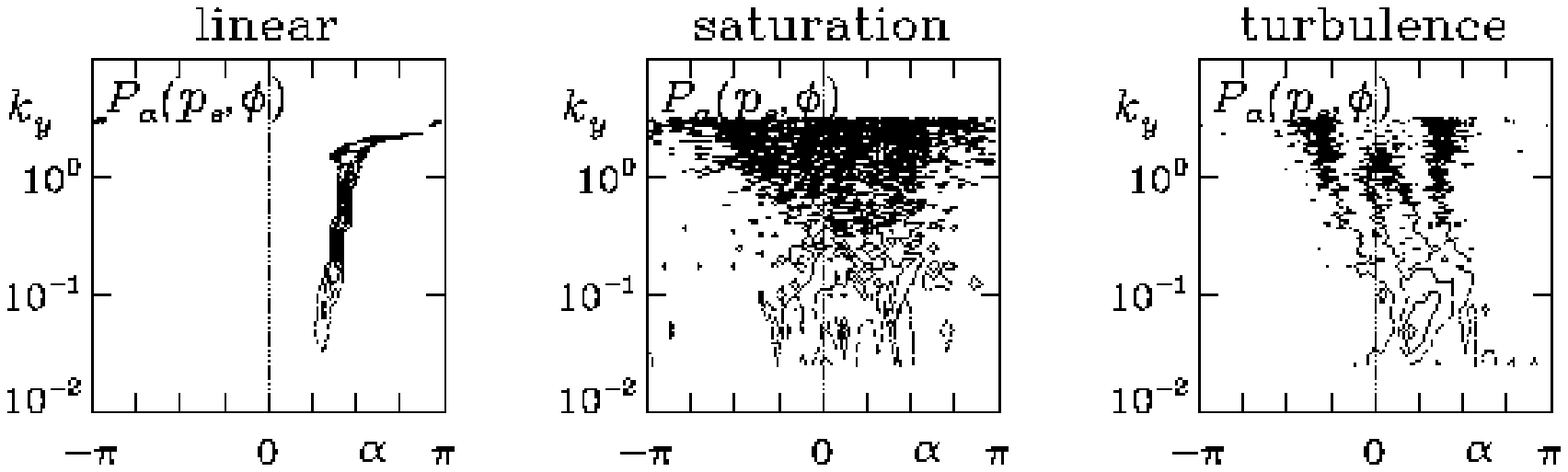}$$
}
{Figure \figphasel.}{  Phase shift distributions of $\pefl$
ahead of $\phifl$ at each $k_y$, 
for the linear stage, averaged over $50<t<150$, 
through saturation, averaged over $200<t<244$, 
and for the stage of fully developed turbulence, averaged over $502<t<611$.
Compare with Fig.~\figphasdist.
The turbulence emerges to supersede the linear structure with its own,
due to the fact that the rms vorticity level of the turbulence is larger
than the linear growth rate of the original instability.  The linear
instabilities in the range $0.3<k_y\rs<1.0$ have no role in the
turbulence.  Only in the fully developed stage does the interchange
dynamics for $k_y\rs<0.1$ emerge to make this case with $\nu_B=1.25$
the transitional one between turbulence of the drift wave and
interchange type.
}

\dofigure{
$$\psboxto(15 true cm;0cm){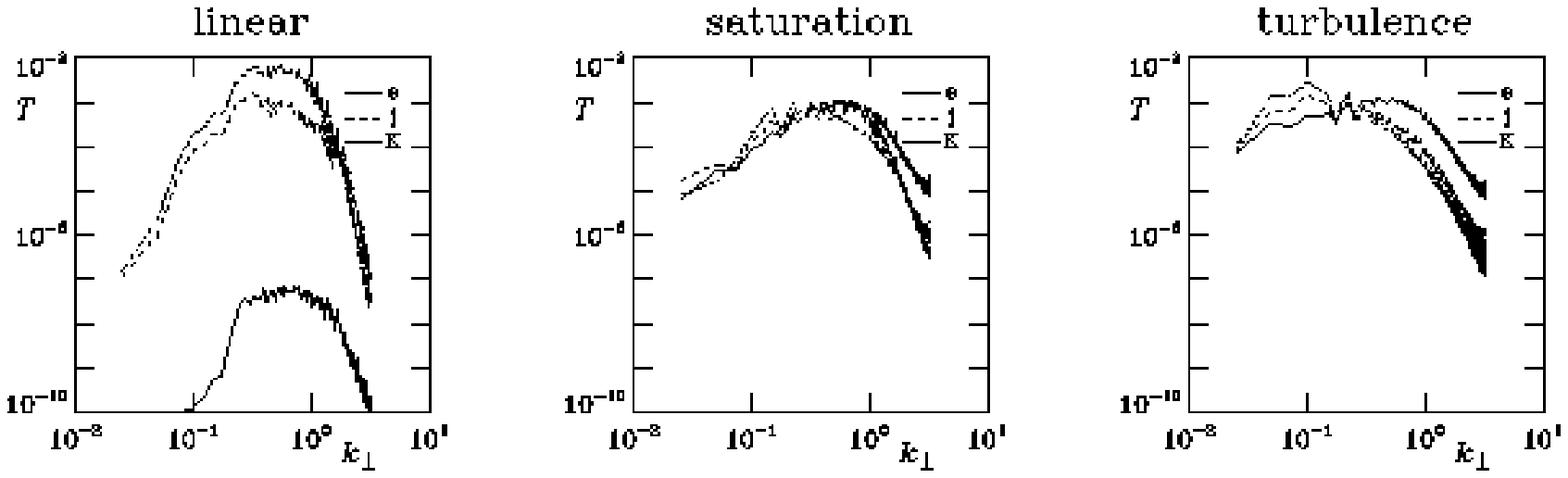}$$
}
{Figure \figwtransll.}{Dynamical transfer spectra for the ExB vorticity,
for the linear stage, averaged over $50<t<150$, 
through saturation, averaged over $200<t<244$, 
and for the stage of fully developed turbulence, averaged over $502<t<611$,
plotted against $\kpp$ rather than $k_y$, showing the scale of motion
rather than the wavelength in the drift direction.  The linear
instability is dominantly in the same range which is later dominated by
the turbulence, making the linear interchange dominated mode
irrelevant.  When interchange effects do enter, as in this transitional
case with $\nu_B=1.25$, they do so at larger scale where they more
easily overcome the native vorticity of the drift wave turbulence.
}


\dofigure{
$$\psboxto(15 true cm;0cm){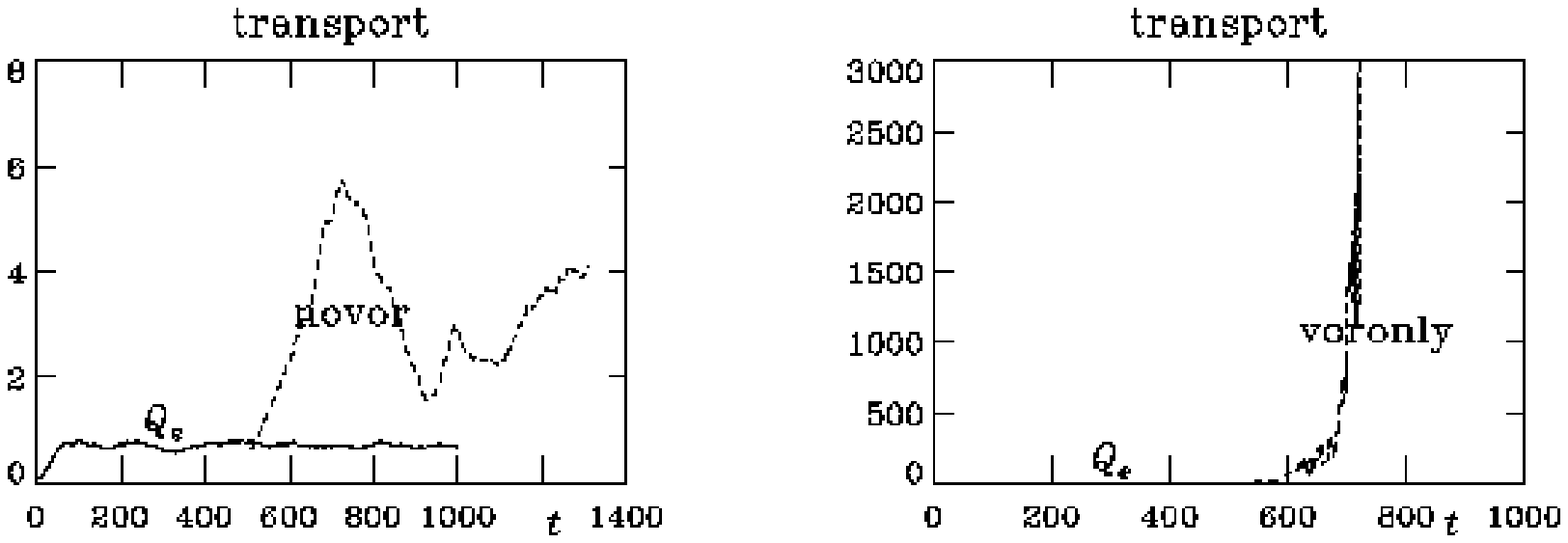}$$
}
{Figure \figtorsat.}{  Saturation mechanism in drift wave
turbulence in toroidal geometry.  The run with $\nu=2$
(hence $C=5.1$ and $\nu_B=0.25$) taken to $t=1000$ (solid curves)
is restarted from $t=500$ (dashed curves) with the
ExB vorticity nonlinearity either left out (`novor') or with all ExB
nonlinearities except the vorticity one left out (`voronly').
Without the vorticity nonlinearity, the linear drive 
is balanced by mixing via $\vedl\pefl$ and saturation occurs.  Without
the pressure nonlinearity, the vorticity is vigorously scattered via
$\vedl\vorfl$ with little dissipative effect, and this nonlinear
excitation continues indefinitely without saturation.
}


\dofigure{
$$\psboxto(15 true cm;0cm){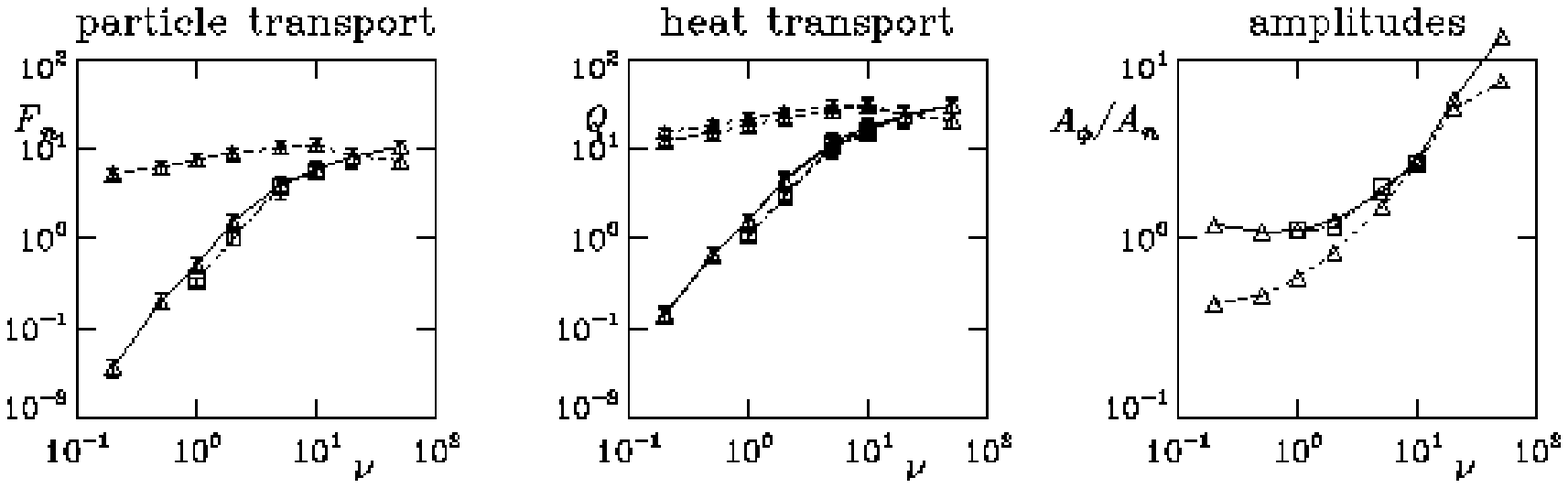}$$
}
{Figure \figbdmodeti.}{  Transport scaling (left two frames) and
amplitude ratios (right) of drift wave (solid lines) and
resistive MHD (dashed lines) turbulence in toroidal geometry under the
DALFTI model.  Both ion and electron heat transport are shown as a pair;
in each case the ion transport curve is the one lying slightly higher in
the pair.
The MHD model, which neglects the drift wave coupling
terms between $\pefl$ and $\Jfl$, is insensitive to collisionality
($C=2.55\nu$) because although $\phifl$ is too small, $\pifl$ is too
large, compared to the drift wave model.  The transition to the MHD
regime at a lower $\nu>3$ is assisted by the warm ion interchange 
physics, since $\tifl$ does not feel the adiabatic response.  The extra
points marked with squares for DW ($\nu=1,2,5,10$)
are with double resolution in the
drift plane.  Compare with Fig.~\figbdmode.
}

\dofigure{
$$\psboxto(15 true cm;0cm){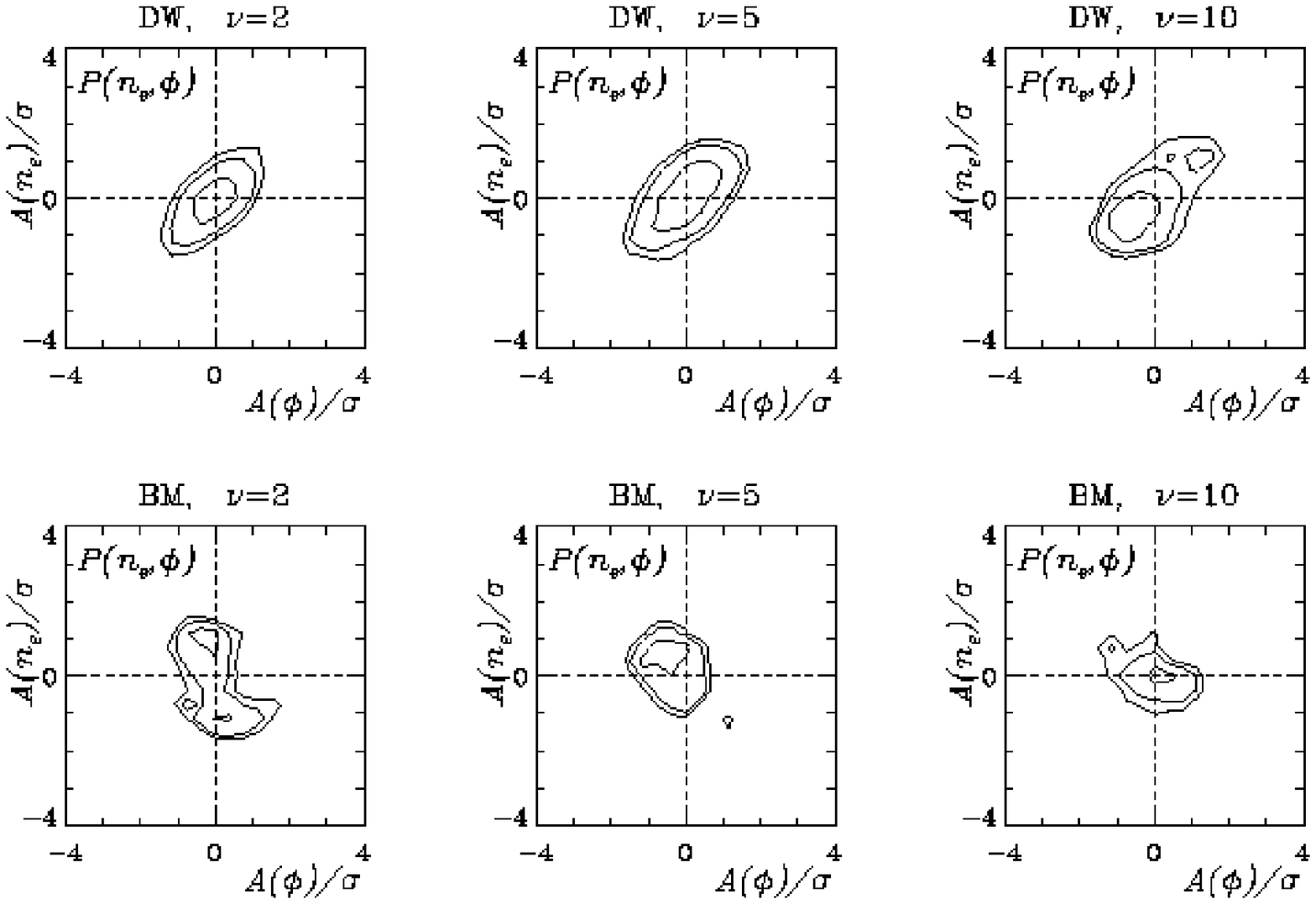}$$
}
{Figure \figplcoherti.}{  Cross coherence between $\nefl$ and
$\phifl$, for drift wave (top row) and interchange (bottom row)
turbulence, for 
$\nu=2$, $5$, and $10$ (left to right), where $C=2.55\nu$ and
$\nu_B=C\wcv$ as defined in Eq.~(\eqresbal).  
The turbulence makes the transition
from drift wave to resistive ballooning mode structure for $\nu_B$ of
about $0.5$ (here, $\nu=5$).
Compare to Fig.~\figplcoher.  
Interchange turbulence with
warm ions is very violent at small scales due to the effects of
gyroviscosity, reducing the timestep and shortening the runs.
}

\dofigure{
$$\psboxto(15 true cm;0cm){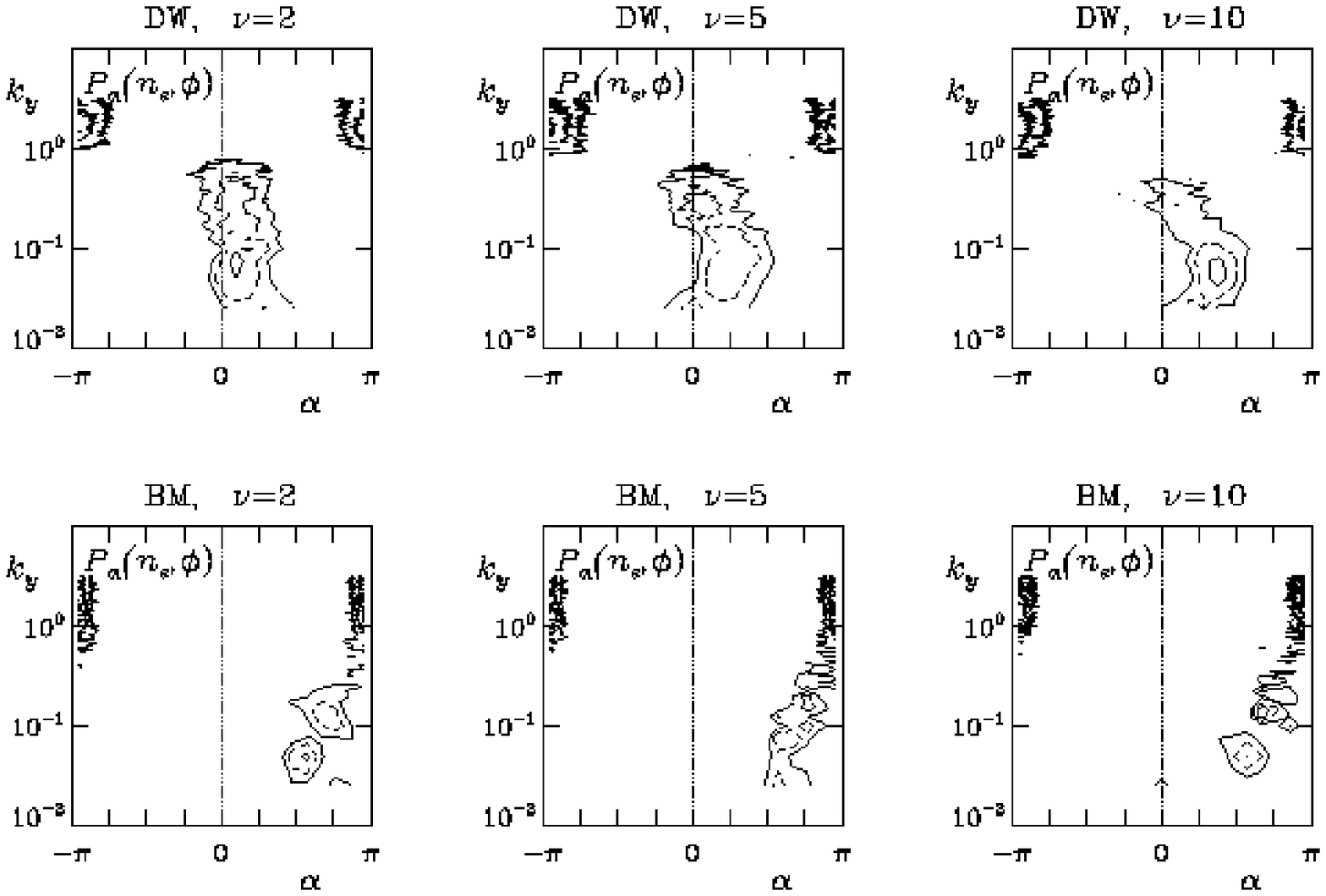}$$
}
{Figure \figphasdistti.}{  Phase shift distributions of $\nefl$
ahead of $\phifl$ at each $k_y$, 
for drift wave (top row) and interchange (bottom row)
turbulence, for 
$\nu=2$, $5$, and $10$ (left to right), where $C=2.55\nu$ and
$\nu_B=C\wcv$ as defined in Eq.~(\eqresbal).  
The turbulence makes the transition
from drift wave to resistive ballooning mode structure for $\nu_B$ of
about $0.5$.  Compare to Fig.~\figphasdist.
The tendency of the phase shift to go to $-\pi$ at small scales is the
signature of the nonlinearity in the gyroviscosity which conserves
energy but not vorticity for warm ions, producing the effects seen in
Fig.~\figplcoherti. 
As in the DALF3 model, the transition to resistive ballooning turbulence
in the DALF3 model occurs in the longest wavelengths, $k_y\rs<0.1$.  
}

\dofigure{
$$\psboxto(15 true cm;0cm){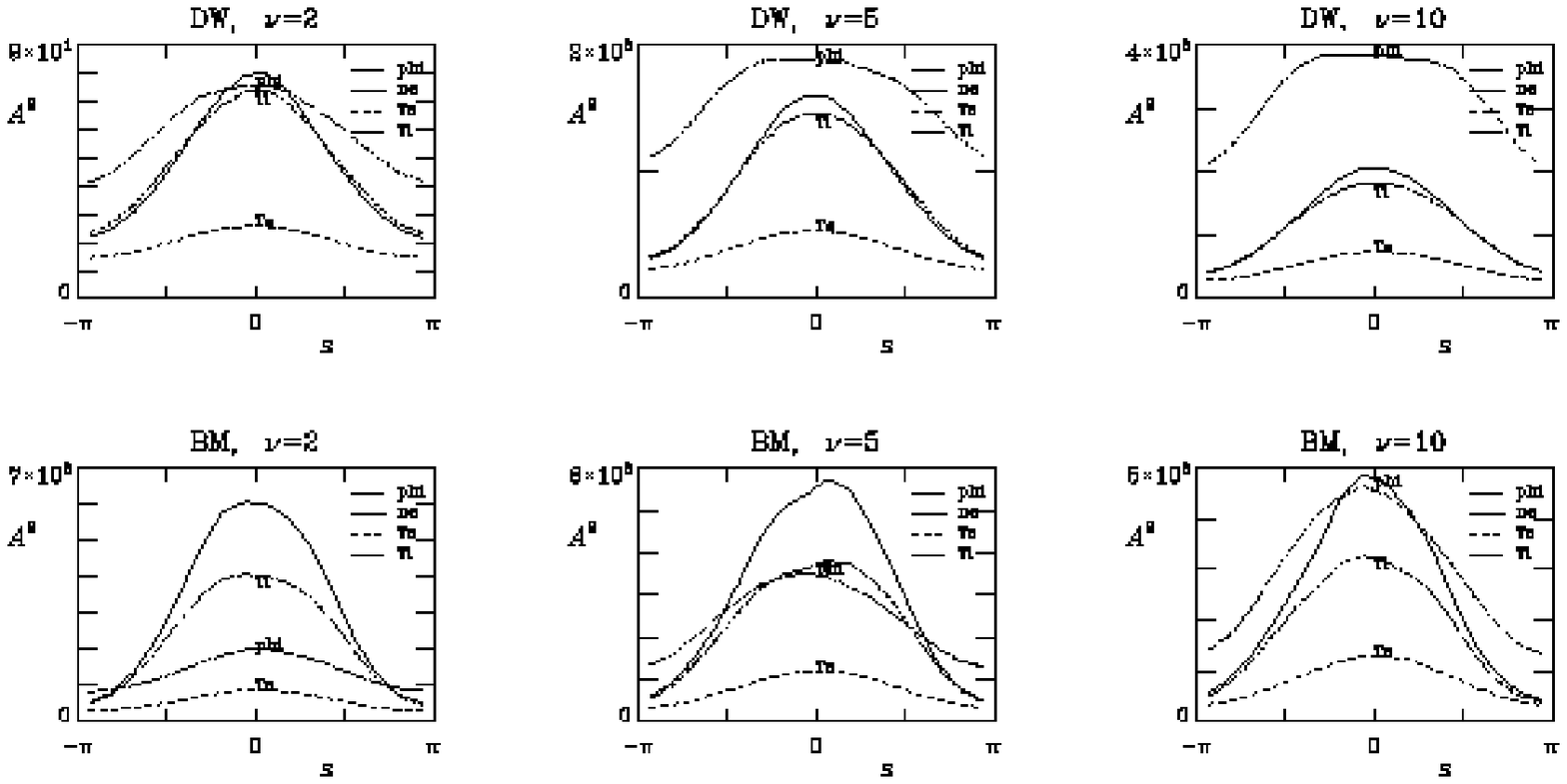}$$
}
{Figure \figplflsti.}{  Mean squared amplitude envelopes 
($k_y\ne 0$ only)
showing parallel structure of $\phifl$, $\nefl$, $\tefl$, and $\tifl$,
respectively labelled by `phi', `ne', `Te', and  `Ti',
for drift wave (top row) and interchange (bottom row) turbulence, 
for $\nu=2$, $5$, and $10$ (left to right), where $C=2.55\nu$ and
$\nu_B=C\wcv$ as defined in Eq.~(\eqresbal).
The turbulence makes the transition
from drift wave to resistive ballooning mode structure for $\nu_B$ of
about $0.5$, with additional effects due to the contribution of $\pifl$
to the vorticity, as described in the text.  
Note $\pefl=\nefl+\tefl$
and $\pifl=\tau_i\nefl+\tifl$ in this model.
}

\dofigure{
$$\psboxto(15 true cm;0cm){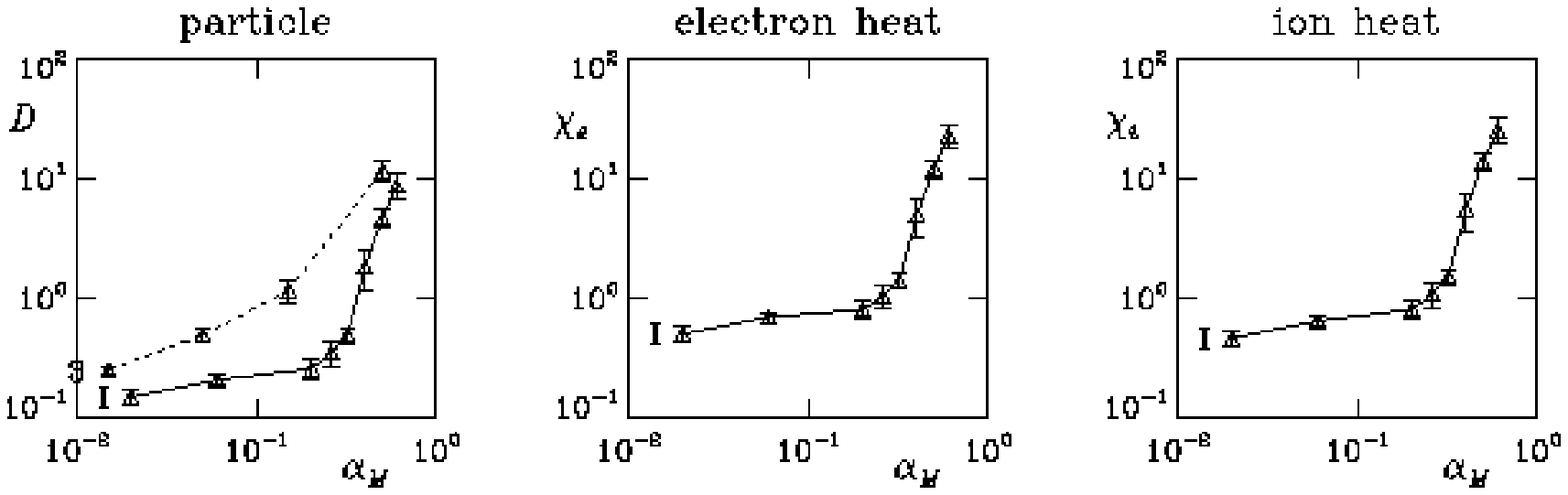}$$
}
{Figure \figtransbetas.}{  Transport scaling as a function of the
ideal ballooning parameter $\alpha_M$, for the DALF3 (`3') and
DALFTI (`I') models, expressed as diffusivities, with $\chi_{e,i}$
including the convective contributions.
The collisionality was $C=2.55$, in the drift wave
regime.  The transport is given in physical units ($\msqsec$,
assuming deuterium
ions, and $B=2.5\tesla$, and $\Lpp=4.2\cm$), compensating
for the effect of varying $\bhat$ on the normalisation scale
$\rs^2c_s/\Lpp$.  The ion temperature assists the transition to
ideal MHD in the measure and for the same reason as for the resistive
ballooning cases: more total pressure gradient, and $\tifl>\tefl$.
But the ideal ballooning threshold, now between $0.2$ and $0.6$,
is lowered by a factor of at least two relative to the linear analysis.
The experimentally interesting range is actually $\alpha_M>0.2$,
corresponding to $\bhat>1$ for the DALFTI model.
}

\dofigure{
$$\psboxto(15 true cm;0cm){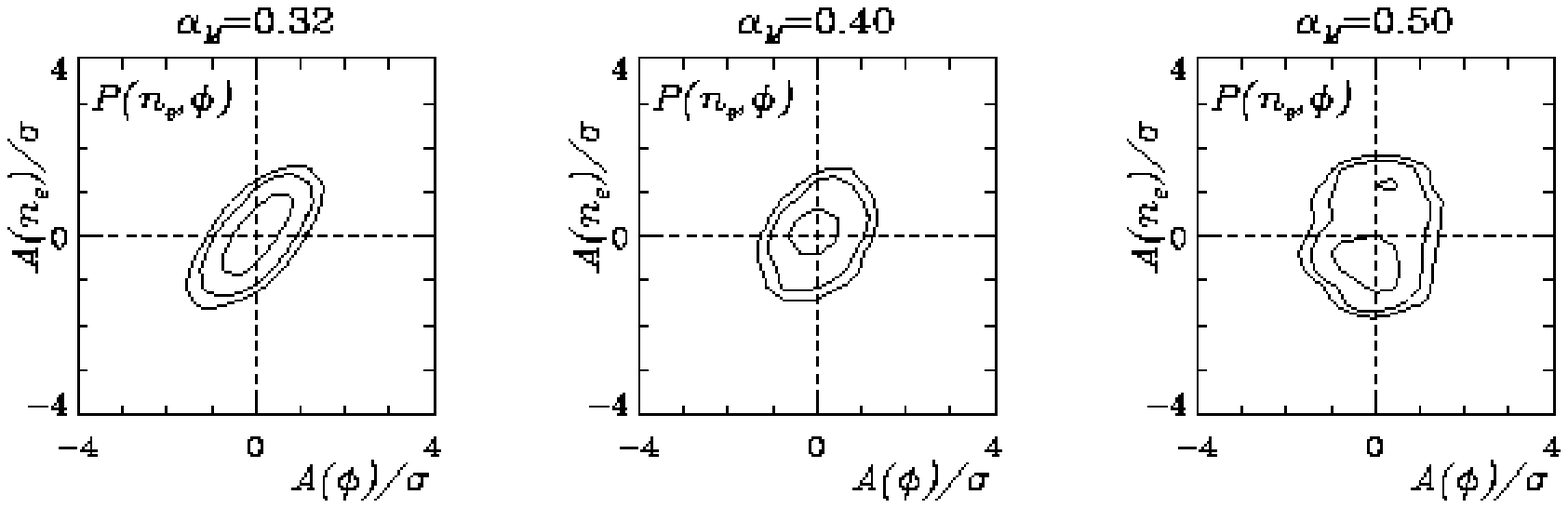}$$
}
{Figure \figplcoherbetas.}{  Cross coherence between $\nefl$
and $\phifl$, 
for the DALFTI model at $C=2.55$ hence $\nu_B=0.25$, for various
$\alpha_M$ in the range of the sharp transport rise, noting
$\alpha_M=0.2\bhat$.  The turbulence makes the transition 
from drift wave to ideal ballooning mode structure in this parameter
range.  Compare to Fig.~\figplcoherti.
}

\dofigure{
$$\psboxto(15 true cm;0cm){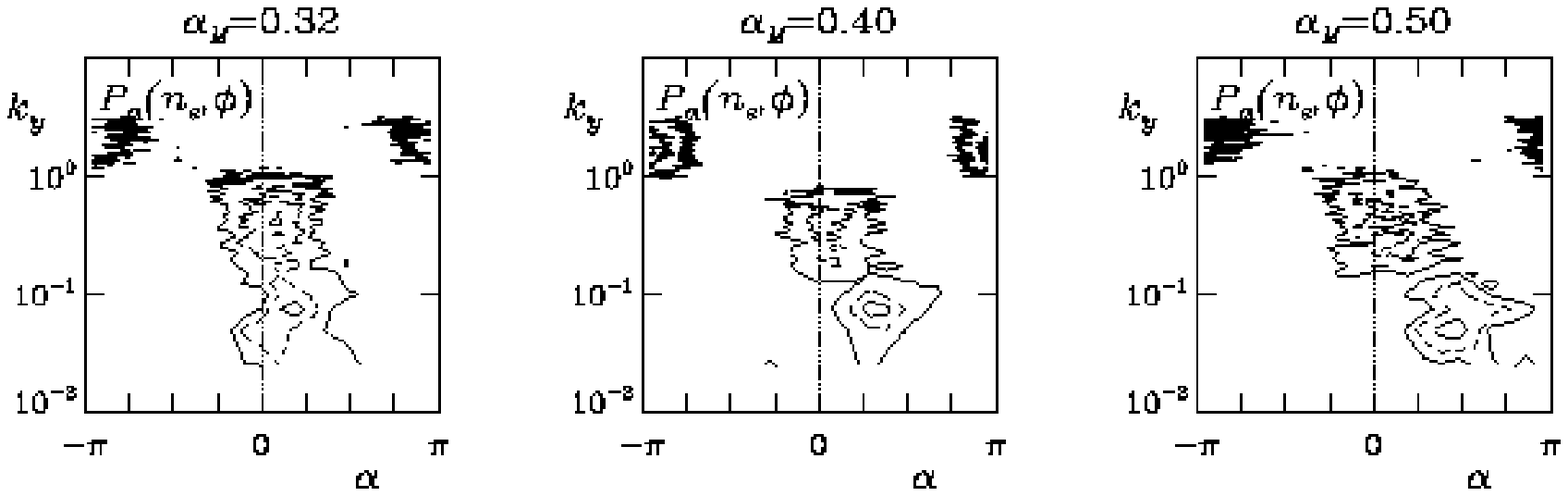}$$
}
{Figure \figphasdistbetas.}{  Phase shift distributions of $\nefl$
ahead of $\phifl$ at each $k_y$, 
for the DALFTI model at $C=2.55$ hence $\nu_B=0.25$, for various
$\alpha_M$ in the range of the sharp transport rise, noting
$\alpha_M=0.2\bhat$.  The turbulence makes the transition 
from drift wave to ideal ballooning mode structure in this parameter
range.  Compare to Fig.~\figphasdistti.
Here as well, the transition to resistive ballooning turbulence in the
DALF3 model occurs in the longest wavelengths, $k_y\rs<0.1$.  
}

\dofigure{
$$\psboxto(15 true cm;0cm){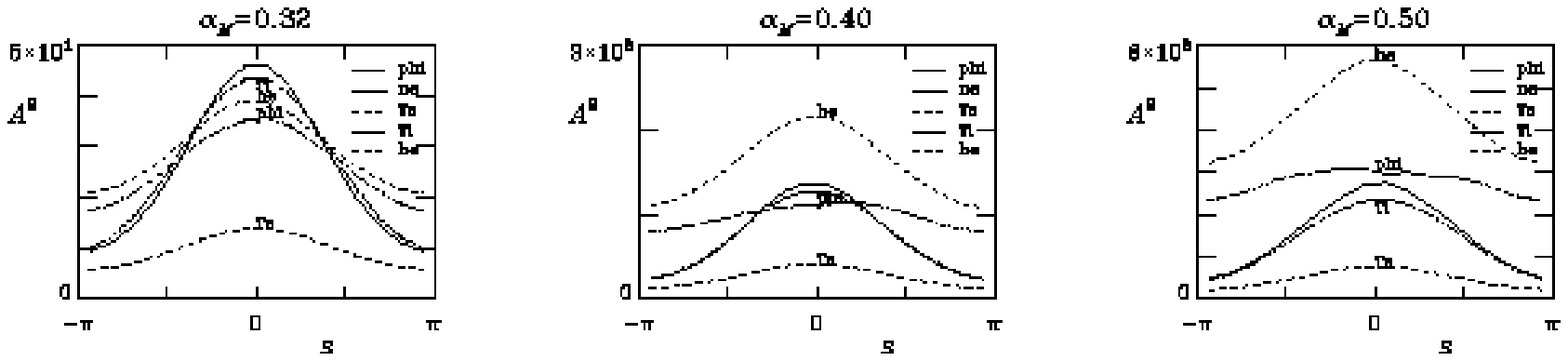}$$
}
{Figure \figplflsbetas.}{  Mean squared amplitude envelopes 
($k_y\ne 0$ only)
showing parallel structure of $\phifl$, $\nefl$, $\tefl$, and $\tifl$,
respectively labelled by `phi', `ne', `Te', and  `Ti',
for the DALFTI model at $C=2.55$ hence $\nu_B=0.25$, for various
$\alpha_M$ in the range of the sharp transport rise, noting
$\alpha_M=0.2\bhat$.  The turbulence makes the transition 
from drift wave to ideal ballooning mode structure in this parameter
range.  Compare to Fig.~\figplflsti.
Note $\pefl=\nefl+\tefl$ and $\pifl=\tau_i\nefl+\tifl$ in this model.
}

}

\def\emskip{\hskip 1 em}
\def\hfb{\hfil\break}
\def\etc{{\it etc.}}
\def\visavis{{\it vis-a-vis}\ }
\def\ie{{\it i.e.}}
\def\eg{{\it e.g.}}
\def\etal{{\it et al}}
\def\ua{u.a.\ }
\def\dh{d.h.\ }
\def\zb{z.B.\ }
\def\bzw{bzw.\ }
\def\usw{usw.\ }

\def\idelta{$i$-delta}


\def\half{ {1\over 2} }
\def\third{ {1\over 3} }
\def\fourth{ {1\over 4} }
\def\tth{ {2\over 3} }
\def\twothirds{ {2\over 3} }
\def\threehalves{ {3\over 2} }
\def\fivehalves{ {5\over 2} }
\def\fivethirds{ {5\over 3} }
\def\sevenhalves{ {7\over 2} }
\def\threeh{\threehalves}
\def\eps{\epsilon}
 
\def\grapprox{\mathop{\lower.5ex \hbox{$\buildrel{\fivesy >}\over{\fivesy\sim}$}} \nolimits}
\def\lsapprox{\mathop{\lower.5ex \hbox{$\buildrel{\fivesy <}\over{\fivesy\sim}$}} \nolimits}
\def\grls{\mathop{\lower.5ex \hbox{$\buildrel{\fivesy >}\over{\fivesy <}$}} \nolimits}

\def\vec#1{{\bf #1}}
\def\tsr#1{{\secfnt #1}}
\def\avg#1{\left\langle #1 \right\rangle}
\def\abs#1{\left\vert #1 \right\vert}
\def\prf#1{\overline{#1}}

\def\max{{}_{{\rm max}}}
\def\min{{}_{{\rm min}}}

\def\minus{\mathop{\hbox{--}}\nolimits}

\def\re{\mathop{\rm Re}\nolimits}
\def\im{\mathop{\rm Im}\nolimits}
\def\sech{\mathop{\rm sech}\nolimits}
\def\diag{\mathop{\rm diag}\nolimits}
\def\Max{\mathop{\rm Max}\nolimits}
\def\Min{\mathop{\rm Min}\nolimits}
\def\nint{\mathop{\rm NINT}\nolimits}
\def\mod{\mathop{\rm mod}\nolimits}
\def\det{\mathop{\rm det}\nolimits}
\def\Tr{\mathop{\rm Tr}\nolimits}
\def\sign{\mathop{\rm sign}\nolimits}

\def\LBR{\left\lbrace}
\def\RBR{\right\rbrace}
\def\LB{\left\lbrack}
\def\RB{\right\rbrack}
\def\LP{\left (}
\def\RP{\right )}
\def\qq{\qquad\qquad}
\def\qqq{\qquad\qquad\qquad}
\def\Det#1{\left\vert\matrix{#1}\right\vert}

\def\pt{\partial}

\def\pzz#1{{\partial #1\over\partial z}}
\def\pxx#1{{\partial #1\over\partial x}}
\def\pyy#1{{\partial #1\over\partial y}}
\def\pww#1{{\partial #1\over\partial w}}
\def\pss#1{{\partial #1\over\partial s}}
\def\prr#1{{\partial #1\over\partial r}}
\def\prhrh#1{{\partial #1\over\partial \rho}}
\def\pthth#1{{\partial #1\over\partial \theta}}
\def\pchch#1{{\partial #1\over\partial \chi}}
\def\ppsps#1{{\partial #1\over\partial \psi}}
\def\pzeze#1{{\partial #1\over\partial \zeta}}
\def\pphph#1{{\partial #1\over\partial \phi}}
\def\ptt#1{{\partial #1\over\partial t}}
\def\pVV#1{{\partial #1\over\partial V}}
\def\phh#1{{\partial #1\over\partial \theta}}
\def\pvhvh#1{{\partial #1\over\partial \vartheta}}
\def\pxixi#1{{\partial #1\over\partial \xi}}
\def\dtt#1{{d #1\over dt}}
\def\dss#1{{d #1\over ds}}
\def\drr#1{{d #1\over dr}}
\def\pprr#1{{\partial^2 #1\over\partial r^2}}
\def\pprhrh#1{{\partial^2 #1\over\partial \rho^2}}
\def\ppss#1{{\partial^2 #1\over\partial s^2}}
\def\ppxx#1{{\partial^2 #1\over\partial x^2}}
\def\ppxy#1{{\partial^2 #1\over\partial x\partial y}}
\def\ppyy#1{{\partial^2 #1\over\partial y^2}}
\def\ppzz#1{{\partial^2 #1\over\partial z^2}}
\def\pptt#1{{\partial^2 #1\over\partial t^2}}
\def\ppVV#1{{\partial^2 #1\over\partial V^2}}
\def\ppphph#1{{\partial^2 #1\over\partial \phi^2}}
\def\ppthth#1{{\partial^2 #1\over\partial \theta^2}}
\def\pphh#1{{\partial^2 #1\over\partial \theta^2}}
\def\ppvhvh#1{{\partial^2 #1\over\partial \vartheta^2}}
\def\ppxixi#1{{\partial^2 #1\over\partial \xi^2}}
\def\ppz#1{\partial #1/\partial z}
\def\ppx#1{\partial #1/\partial x}
\def\ppy#1{\partial #1/\partial y}
\def\ppw#1{\partial #1/\partial w}
\def\ppr#1{\partial #1/\partial r}
\def\pprh#1{\partial #1/\partial \rho}
\def\pps#1{\partial #1/\partial s}
\def\ppt#1{\partial #1/\partial t}
\def\ppV#1{\partial #1/\partial V}
\def\pph#1{\partial #1/\partial \theta}
\def\ppvh#1{\partial #1/\partial \vartheta}
\def\ppxi#1{\partial #1/\partial \xi}

\def\ddt#1{d #1/dt}
\def\pppz#1{\partial^2 #1/\partial z^2}
\def\pppx#1{\partial^2 #1/\partial x^2}
\def\pppy#1{\partial^2 #1/\partial y^2}
\def\pppr#1{\partial^2 #1/\partial r^2}
\def\ppprh#1{\partial^2 #1/\partial \rho^2}
\def\ppps#1{\partial^2 #1/\partial s^2}
\def\pppt#1{\partial^2 #1/\partial t^2}
\def\pppV#1{\partial^2 #1/\partial V^2}
\def\ppph#1{\partial^2 #1/\partial \theta^2}
\def\pppvh#1{\partial^2 #1/\partial \vartheta^2}
\def\pppxi#1{\partial^2 #1/\partial \xi^2}
\def\dddt#1{d^2 #1/dt^2}

\def\grad{\nabla}
\def\cross{{\bf \times}}
\def\div{\grad\cdot}
\def\divp{\grad_\perp\cdot}
\def\divpl{\grad_\parallel\cdot}
\def\curl{\grad\cross}
\def\dpl{\grad_\parallel}
\def\ddpl{\grad_\parallel^2}
\def\dpp{\grad_\perp}
\def\ddpp{\grad_\perp^2}
\def\delsq{\grad^2}
\def\delamb{ \mathchar"0274\hskip -.665em\mathchar"0275 }
\let\delam=\delamb
\def\lapl{\grad^2}
\def\lapldef{\ddpp=(\pt^2/\pt x^2)+K^2(\pt^2/\pt y^2)}

\def\pwww#1{{\partial #1\over\partial \vec w}}
\def\pwwpl#1{{\partial #1\over\partial w_\parallel}}
 
\def\pvv#1#2{{\partial #2\over\partial v_{#1}}}
\def\ppv#1#2{{\partial #2/\partial v_{#1}}}
\def\pvvv#1{{\partial #1\over\partial \vec v}}
\def\pvvp#1#2{{\partial #2\over\partial v'_{#1}}}
\def\ppvp#1#2{{\partial #2/\partial v'_{#1}}}
\def\pvvvp#1{{\partial #1\over\partial \vec v'}}
\def\pvvpl#1{{\partial #1\over\partial v_\parallel}}
 
\def\xunit{\vec{\hat x}}
\def\yunit{\vec{\hat y}}
\def\zunit{\vec{\hat z}}
\def\sunit{\vec{\hat s}}
\def\bunit{\vec{b}}
\def\eunit{\vec{\hat e}}
\def\nunit{\vec{\hat n}}
\def\dt{\Delta t}
\def\becomes{\leftarrow}
\def\from{\leftarrow}
\def\to{\rightarrow}
\def\fromto{\leftrightarrow}
\def\implies{\,\,\,\Longrightarrow\,\,\,}
\def\dotdot{\!:\!}

\def\meters{\,{\rm m}}
\def\invm{\,{\rm m}^{-3}}
\def\invsec{\,{\rm sec}^{-1}}
\def\cm{\,{\rm cm}}
\def\km{\,{\rm km}}
\def\invcc{\,{\rm cm}^{-3}}
\def\invcm{\,{\rm cm}^{-1}}
\def\mm{\,{\rm mm}}
\def\Vcm{\,{\rm V/cm}}
\def\Acm{\,{\rm A/cm^2}}
\def\kA{\,{\rm kA}}
\def\MA{\,{\rm MA}}
\def\degk{\,{\rm K}}
\def\ergs{\,{\rm erg}}
\def\eV{\,{\rm eV}}
\def\keV{\,{\rm keV}}
\def\MeV{\,{\rm MeV}}
\def\GeV{\,{\rm GeV}}
\def\kG{\,{\rm kG}}
\def\tesla{\,{\rm T}}
\def\kW{\,{\rm kW}}
\def\MW{\,{\rm MW}}
\def\radsec{\,{\rm rad/sec}}
\def\Hz{\,{\rm Hz}}
\def\kHz{\,{\rm kHz}}
\def\MHz{\,{\rm MHz}}
\def\mpersec{\,{\rm m}/{\rm sec}}
\def\msqsec{\,{\rm m^2}/{\rm sec}}
\def\cmsec{\,{\rm cm}/{\rm sec}}
\def\kmsec{\,{\rm km}/{\rm sec}}
\def\ccpersec{\,{\rm cm}^3/{\rm sec}}
\def\minutes{\,{\rm min}}
\def\yr{\,{\rm yr}}
\def\hr{\,{\rm hr}}
\def\Bar{\,{\rm bar}}
\def\sec{\,{\rm sec}}
\def\msec{\,{\rm msec}}
\def\usec{\,\mu{\rm sec}}
 
\def\bdel{\vec b\cdot\grad}
\def\Bdel{\vec B\cdot\grad}
\def\Jdel{\vec J\cdot\grad}
\def\bdot{\vec B\cdot}
\def\Bdot{\vec B\cdot}
\def\exb{\vec E\cross\vec B}
\def\jxb{\vec J\cross\vec B}
\def\uxb{\vec u\cross\vec B}
\def\vxb{\vec v\cross\vec B}
\def\wxb{\vec w\cross\vec B}
\def\ucxb{{\vec u\over c}\cross\vec B}
\def\vcxb{{\vec v\over c}\cross\vec B}
\def\wcxb{{\vec w\over c}\cross\vec B}
\def\jcxb{{\vec J\cross\vec B\over c}}

\def\vexb{\vec v_E}
\def\vpol{\vec v_p}
\def\upol{\vec u_p}
\def\vstar{\vec v_*}
\def\ustar{\vec u_*}
\def\Jstar{\vec J_*}
\def\Jpol{\vec J_p}
\def\vgradb{\vec v_{\grad B}}
\def\qstar{\vec q_\wedge}
\def\qestar{\vec q_e{}_\wedge}
\def\qistar{\vec q_i{}_\wedge}
\def\pistar{\vec\Pi_*}
\def\vR{\vec v_R}
\def\vdl{\vec v\cdot\grad}
\def\vdel{\vec v\cdot\grad}
\def\vedl{\vexb\cdot\grad}
\def\udl{\vec u\cdot\grad}
\def\udel{\vec u\cdot\grad}
\def\uidl{\vec u_i\cdot\grad}
\def\uidel{\vec u_i\cdot\grad}
\def\wdel{\vec w\cdot\grad}
\def\dedt#1{d_E #1/dt}
\def\dett#1{{d_E #1\over dt}}
\def\jpp{J_\perp}
\def\jperp{\vec\jpp}
\def\qpp{q_\perp}
\def\qperp{\vec\qpp}
\def\upp{u_\perp}
\def\uperp{\vec\upp}
\def\wpl{w_\parallel}
\def\wpp{w_\perp}
\def\wperp{\vec\wpp}
\def\vpp{v_\perp}
\def\vperp{\vec\vpp}
\def\lnb{\log B}
 
\def\rms{_{rms}}
 
\def\Jpl{J_\parallel}
\def\jpl{J_\parallel}
\def\Jpp{J_\perp}
\def\jpp{J_\perp}
\def\Jperp{\vec\Jpp}
\def\Bperp{\vec B_\perp}
\def\Apl{A_\parallel}
\def\apl{A_\parallel}
\def\App{A_\perp}
\def\app{A_\perp}
\def\Aperp{\vec\App}
\def\Epl{E_\parallel}
\def\epl{E_\parallel}
\def\Epp{E_\perp}
\def\epp{E_\perp}
\def\Eperp{\vec\Epp}
\def\upl{u_\parallel}
\def\vpl{v_\parallel}
\def\Upl{U_\parallel}
\def\vor{\grad_\perp^2\phi}
\def\kpl{k_\parallel}
\def\kkpl{k_\parallel^2}
\def\kpp{k_\perp}
\def\kperp{\vec\kpp}
\def\kkpp{k_\perp^2}
\def\xpl{{x_\parallel}}
\def\xpp{x_\perp}
\def\DD{\Delta_D}
\def\Dpl{D_\parallel}
\def\Dpp{\Delta_\perp}
\def\Depl{D_e{}_\parallel}
\def\Dipl{D_i{}_\parallel}
\def\Rpl{R_\parallel}
\def\qpl{q_\parallel}
\def\qepl{q_e{}_\parallel}
\def\qipl{q_i{}_\parallel}
\def\mupl{\mu_\parallel}
\def\mupp{\mu_\perp}
\def\nuei{\nu_{ei}}
\def\nuee{\nu_{ee}}
\def\nuii{\nu_{ii}}
\def\wpe{\omega_{pe}}
\def\wpi{\omega_{pi}}
\def\nudamp{\nu_d}
\def\zeff{Z_{\!e\!f\!f}}
\def\lmfp{\lambda_{\!m\!f\!p}}
\def\ws{{\omega_*}}
\def\wsi{{\omega_{*i}}}
\def\wn{\omega_n}
\def\wt{\omega_t}
\def\wi{\omega_i}
\def\wT{\omega_T}
\def\wp{\omega_p}
\def\wc{{\omega_c}}
\def\kappacv{{\cal K}}
\def\wcv{{\omega_B}}
\def\etai{\eta_i}
\def\taui{\tau_i}
\def\rs{\rho_s}
\def\ld{\lambda_D}
\def\Lpl{L_\parallel}
\def\Lpp{L_\perp}
\def\lcorpl{\lambda_\parallel}
\def\lcorpp{\lambda_\perp}
\def\rch{\rho_{ch}}
\def\npl{\eta_\parallel}
\def\etapl{\eta_\parallel}
\def\ald{a_L}
\def\alde{a_{Le}}
\def\aldi{a_{Li}}
\def\npp{\eta_\perp}
\def\etapp{\eta_\perp}
\def\kappapl{\kappa_\parallel}
\def\dprime{\Delta'}
\def\sk{{}_{\vec k}}
\def\sky{_{k_y}}
\def\gk{\gamma_k}
\def\vk{\vfl_k}
\def\nk{\nfl_k}
\def\tk{\tfl_k}
\def\dk{\Delta k}
\def\gd{\gamma_0}
\def\mwn{\Delta_n}
\def\mwh{\Delta_h}
\def\gamT{\Gamma_T}
\def\gamn{\Gamma_n}
\def\gamt{\Gamma_t}
\def\gami{\Gamma_i}
\def\gamc{\Gamma_c}
\def\gamk{\Gamma_k}
\def\gams{\Gamma_s}
\def\gaml{\Gamma_l}
\def\gamr{\Gamma_r}
 
\def\ptb{\widetilde}
\def\psifl{\widetilde\psi}
\def\phifl{\widetilde\phi}
\def\ffl{\widetilde f}
\def\fe{f_e}
\def\fefl{\widetilde f_e}
\def\fifl{\widetilde f_i}
\def\nfl{\widetilde n}
\def\hfl{\widetilde h}
\def\tfl{\widetilde T}
\def\nefl{\widetilde n_e}
\def\nifl{\widetilde n_i}
\def\tefl{\widetilde T_e}
\def\tifl{\widetilde T_i}
\def\pfl{\widetilde p}
\def\pefl{\widetilde p_e}
\def\pifl{\widetilde p_i}
\def\hefl{\widetilde h_e}
\def\vx{\widetilde v_x}
\def\vfl{\widetilde v}
\def\vefl{\widetilde \vexb}
\def\vxfl{\widetilde v_x}
\def\vyfl{\widetilde v_y}
\def\vrfl{\widetilde v_r}
\def\vppfl{\widetilde v_\perp}
\def\vplfl{\widetilde \vpl}
\def\Bfl{\widetilde \vec B}
\def\Bflpp{\widetilde B_\perp}
\def\Aplfl{\widetilde A_\parallel}
\def\Appfl{\widetilde A_\perp}
\def\Aperpfl{\widetilde {\vec A}_\perp}
\def\ufl{\widetilde u_\parallel}
\def\vorfl{\grad_\perp^2\phifl}
\def\jfl{\widetilde J_\parallel}
\def\qfl{\widetilde q_\parallel}
\def\qefl{\widetilde q_e{}_\parallel}
\def\qifl{\widetilde q_i{}_\parallel}
\def\jppfl{\widetilde J_\perp}
\def\jperpfl{\widetilde {\vec J}_\perp}
\def\Afl{\ptb A_\parallel}
\def\Jfl{\ptb J_\parallel}
\def\efl{\widetilde E_\parallel}
\def\Efl{\widetilde E_\parallel}
\def\Eppfl{\widetilde E_\perp}
\def\Eperpfl{\widetilde {\vec E}_\perp}
\def\etafl{\widetilde\eta}
\def\isatfl{\widetilde I_{{\rm sat}}}
\def\phiflfl{\widetilde\phi_{{\rm fl}}}
 
\def\teplfl{\widetilde T_e{}_\parallel}
\def\teppfl{\widetilde T_e{}_\perp}
\def\qeplfl{\widetilde q_e{}_\parallel}
\def\qeppfl{\widetilde q_e{}_\perp}
\def\tiplfl{\widetilde T_i{}_\parallel}
\def\tippfl{\widetilde T_i{}_\perp}
\def\qiplfl{\widetilde q_i{}_\parallel}
\def\qippfl{\widetilde q_i{}_\perp}

\def\tepl{ T_e{}_\parallel}
\def\tepp{ T_e{}_\perp}
\def\qepl{ q_e{}_\parallel}
\def\qepp{ q_e{}_\perp}
\def\tipl{ T_i{}_\parallel}
\def\tipp{ T_i{}_\perp}
\def\qipl{ q_i{}_\parallel}
\def\qipp{ q_i{}_\perp}

\def\peplfl{\widetilde p_e{}_\parallel}
\def\peppfl{\widetilde p_e{}_\perp}
\def\piplfl{\widetilde p_i{}_\parallel}
\def\pippfl{\widetilde p_i{}_\perp}

\def\pepl{ p_e{}_\parallel}
\def\pepp{ p_e{}_\perp}
\def\pipl{ p_i{}_\parallel}
\def\pipp{ p_i{}_\perp}


\def\phinn{ {e\phifl\over T} }
\def\nnn{ {\nfl\over n} }
\def\tnn{ {\tfl\over T} }
\def\unn{ {\ufl\over c_s} }
\def\vornn{ \rho_s^2\ddpp\phinn }
\def\jnn{ {\jfl\over ne} }
\def\qnn{ {\qfl\over nT} }
\def\psinn{ {\psifl\over B\rho_s} }

\def\ahat{\hat\alpha}
\def\ehat{\hat\eta}
\def\khat{\hat\kappa}
\def\shat{\hat s}
\def\bhat{\hat\beta}
\def\muhat{\hat\mu}
\def\epss{\hat\epsilon}
\def\bigpoint#1{
    \par\bigskip
    {\baselineskip=\normalbaselineskip
    \parindent=0 pt
    {\hfill\vbox{ #1  }\hfill}}
    \par\bigskip
    }
 
\def\jfm#1{{\it J. Fluid. Mech.} {\secfnt #1}}
\def\prl#1{{\it Phys. Rev. Lett.} {\secfnt #1}}
\def\physletta#1{{\it Phys. Lett. A} {\secfnt #1}}
\def\physlettb#1{{\it Phys. Lett. B} {\secfnt #1}}
\def\pf#1{{\it Phys. Fluids} {\secfnt #1}}
\def\pfa#1{{\it Phys. Fluids A} {\secfnt #1}}
\def\pfb#1{{\it Phys. Fluids B} {\secfnt #1}}
\def\physp#1{{\it Phys. Plasmas} {\secfnt #1}}
\def\nf#1{{\it Nucl. Fusion} {\secfnt #1}}
\def\njp#1{{\it New J. Phys.} {\secfnt #1}}
\def\cpp#1{{\it Contrib. Plasma Phys.} {\secfnt #1}}
\def\ppcf#1{{\it Plasma Phys. Contr. Fusion} {\secfnt #1}}
\def\plasphys#1{{\it Plasma Phys.} {\secfnt #1}}
\def\revpp#1{{\it Rev. Plasma Phys.} {\secfnt #1}}
\def\iaea#1#2{in {\it Plasma Physics and Controlled Nuclear Fusion
    Research #1}, Proceedings of the #2th International Conference}
\def\EPS#1#2#3{in {\it Proceedings of the
{#1}th European Conference on Controlled Fusion and Plasma Physics,
{#2}, {#3}} (European Physical Society, {#2}, {#3})}
\def\jcp#1{{\it J. Comput. Phys.} {\secfnt #1}}
\def\jetp#1{{\it Sov. Phys. JETP} {\secfnt #1}}
\def\sovjpp#1{{\it Sov. J. Plasma Phys.} {\secfnt #1}}
\def\jnm#1{{\it J. Nucl. Mat.} {\secfnt #1}}
\def\rsi#1{{\it Rev. Sci. Inst.} {\secfnt #1}}
\def\adv#1{{\it Adv. Phys.} {\secfnt #1}}
\def\apjl#1{{\it Astrophys. J. Lett.} {\secfnt #1}}
\def\apj#1{{\it Astrophys. J.} {\secfnt #1}}
\def\aa#1{{\it Astron. Astrophys.} {\secfnt #1}}
\def\vol#1{\ {\secfnt #1}}

\parskip 6 pt



\def\Dpp{\Delta_\perp}
\def\Dpl{\Delta_\parallel}
\def\Kpl{K_\parallel}

\def\vv{\vec v}
\def\uu{\vec u}
\def\xx{\vec x}
\def\ww{\vec w}
\def\ee{\vec E}
\def\bb{\vec B}
\def\ff{\vec F}
\def\jj{\vec J}
\def\qq{\vec q}
\def\aa{\vec A}
\def\kk{\vec k}

\def\nefl{\widetilde n_e}
\def\nifl{\widetilde n_i}
\def\Wfl{\widetilde W}

\def\kkpp{k_\perp^2}

\def\bbfl{\ptb\bb}
\def\bbpp{\ptb\bb_\perp}
\def\Bperp{\ptb\vec B_\perp}

\def\lseq{\le}
\def\chiv{\hat\chi}
\def\chip{\hat\chi}
\def\psip{\hat\psi}
\def\gjac{g}
\def\lp{{l'}}
\def\Vp{V'}
\def\qp{q'}
\def\shat{\hat s}
\def\ky{{k_y}}
\def\psibar{\bar\Psi}
\def\dalpha{\Delta\alpha}

\def\drift{{c\over B^2}\bb\cross}

\def\lcorx{\lambda_x}
\def\lcory{\lambda_y}
\def\tcor{\tau_c}

\def\vemu{v_E^\mu}
\def\bflmu{b^\mu}
\def\pxxmu#1{{\pt #1\over\pt x^\mu}}
\def\pyyk#1{{\pt #1\over\pt y_k}}
\def\ppyyk#1{{\pt^2 #1\over\pt y_k^2}}
\def\pww#1{{\pt #1\over\pt \wpl}}

\def\vex{v_E^x{}}
\def\vey{v_E^y{}}
\def\ves{v_E^s{}}

\def\bflx{b^x{}}
\def\bfly{b^y{}}
\def\bfls{b^s{}}

\def\dofigure#1#2#3{\vskip 10 pt#1
	{\hfill\vbox{\hsize=12cm \baselineskip 13 pt \noindent
	{\secfnt #2}#3
	}\hfill}
	\par\vfill\eject}

\def\secintro{I}
\def\secmodel{II}
\def\secenergetics{III}
\def\seccontrol{IV}
\def\secbdmode{V}
\def\secforcing{VI}
\def\seclinear{VII}
\def\secsat{VIII}
\def\secdalfti{IX}
\def\secideal{X}
\def\secsummary{XI}

\def\figctrldgdw{1}
\def\figcontrol{2}
\def\figbdmode{3}
\def\figplcoher{4}
\def\figphasdist{5}
\def\figwtrans{6}
\def\figpldwa{7}
\def\figpldwb{8}
\def\figplfls{9}
\def\figplbal{10}

\def\figdgdwl{11}
\def\figplctrl{12}
\def\figplctrlb{13}
\def\figwtransl{14}
\def\figplflsl{15}
\def\figphasel{16}
\def\figwtransll{17}

\def\figtorsat{18}

\def\figbdmodeti{19}
\def\figplcoherti{20}
\def\figphasdistti{21}
\def\figplflsti{22}

\def\figtransbetas{23}
\def\figplcoherbetas{24}
\def\figphasdistbetas{25}
\def\figplflsbetas{26}

\title{Drift Wave versus Interchange Turbulence in Tokamak Geometry:}
\title{Linear versus Nonlinear Mode Structure}
\author{Bruce D. Scott}
\address{Max-Planck-Institut f\"ur Plasmaphysik}
\address{EURATOM Association}
\address{D-85748 Garching, Germany}
\date{Feb 2001}
 
\abstract{
The competition between drift wave and interchange physics in general
E-cross-B drift turbulence is studied with computations in three
dimensional tokamak flux tube geometry.  For a given set of background
scales, the parameter space can be covered by the plasma beta and drift
wave collisionality.  At large enough plasma beta the turbulence breaks
out into ideal ballooning modes and saturates only by depleting the free
energy in the background pressure gradient.  At high collisionality it
finds a more gradual transition to resistive ballooning.  At moderate
beta and collisionality it retains drift wave character, qualitatively
identical to simple two dimensional slab models.  The underlying cause
is the nonlinear vorticity advection through which the self sustained
drift wave turbulence supersedes the linear instabilities, scattering
them apart before they can grow, imposing its own physical character on
the dynamics.  This vorticity advection catalyses the gradient drive,
while saturation occurs solely through turbulent mixing of
pressure disturbances.  This situation persists in the whole of tokamak
edge parameter space.  Both simplified isothermal models and complete
warm ion models are treated.  }

\vfill

\noindent PACS numbers: 

52.55.Fa---Tokamaks

52.35.Ra---Plasma turbulence, 

52.65---Plasma simulation.

\eject

\section{\secintro. Introduction --- More Than One Eigenmode in Turbulence}

Edge turbulence in tokamak flux tube geometry has been treated by models
working from a drift wave [\dwtor,\dalfloc]
or resistive ballooning [\zeiler,\rogers,\xxu]
paradigm.  Beyond computing edge transport, the main purpose of these
computations is to understand the underlying physical character of the
turbulence: drive and saturation mechanisms and free energy transfer
channels.  

The pure situation of drift wave turbulence is slab geometry in a
sheared magnetic field, and all of its basic processes can be found in
the simplified two dimensional models in which it was first studied
[\hasmim,\wakhas,\waltz,\biskamp,\ssdw,\sorgdw].  The idealised
state is one of adiabatic electrons, towards which the electron
dynamics parallel to the magnetic field $\bb$ tends:
$$n_e e\Bdel\phi\to \Bdel p_e  
	\eqno\eqname\eqadiabatic
$$
(where equality means the electrons are adiabatic, and
where $\phi$ is the electrostatic potential and $n_e$ and $p_e$ are the
electron density and pressure, and $e$ is the electronic charge).
This is the adiabatic response, and if it carries all the way to
adiabatic electrons there can be no free
energy release, because the average transport is zero if there is no
phase shift in the direction of the electron drift given by
$-\bb\cross\grad p_e$.  The pressure force
$\Bdel p_e$ in the Ohm's law conserves energy against the compression of
parallel currents 
along field lines, $\div(\Jpl\bb/B)$, which provides the back reaction in
the electron pressure.  Together, these two terms constitute the
adiabatic coupling mechanism, which allows the free energy liberated
from the background gradient to enter the parallel current.
The more familiar Alfv\'enic coupling with $\Bdel\phi$ then allows the
free energy to enter the ExB flow eddies.  This is the free energy pathway
between the background gradient and the turbulence which gives rise to
drift waves, and collectively the adiabatic and Alfv\'enic couplings
constitute the drift wave coupling mechanism
$\pefl\fromto\Jfl\fromto\phifl$ among the respective disturbances.  It
is important to note in this context that the parallel gradient cannot
vanish for finite sized disturbances on closed magnetic flux surfaces
(the flux tube geometry is globally consistent [\fluxtube]), 
so this adiabatic response is always excited by electron pressure
disturbances. 

The pure situation of resistive ballooning mode turbulence or ideal
ballooning instabilities is an MHD (magnetohydrodynamics) model
[\strauss].  In this case, the adiabatic coupling mechanism is
specifically neglected so that dissipative Alfv\'enic activity merely
damps $e\phifl/T_e$ towards zero instead of towards $\pefl/p_e$
(here, $T_e$ is the electron temperature in energy units, and we assume
$p_e=n_eT_e$).
The
only way that free energy can be transferred to ExB motion from the
background pressure gradient in an MHD model is through the magnetic
curvature and gradient, which 
give rise to a finite compressibility of the diamagnetic current and
hence the well known interchange effect.
In toroidal geometry, this
leads to the ballooning mode since the interchange effect is
destabilising only on the outer part of the torus, where the radius of
curvature and gradient-B vectors become parallel to the pressure gradient.
The ideal form of this dynamics as an instability in toroidal geometry
was discovered in attempts to explain curious observations of magnetic
fluctuations localised to the outboard side of the PLT tokamak
[\origballooning].  These fluctuations were explained as ideal
ballooning modes, which were treated conventionally [\straussbal] 
as well as with what became known as the
ballooning transformation [\coppi,\connor,\glasser].  Resistive
ballooning [\straussresbal]
was invoked to explain the existence of tokamak edge
turbulence at pressure gradient values below the ideal stability limit
and at amplitudes seemingly unattainable by other mechanisms
[\carreras,\resbal], and has also been the focus of many linear MHD
calculations [\rbmlinear].  More recently [\guzdar], the ballooning
paradigm was extended to situations including the two fluid Ohm's
law (\ie, the adiabatic response), leading to the ballooning mode
approach to edge turbulence [\zeiler,\rogers,\xxu], coinciding with the
development of the earliest treatments of flux tube geometry, which were
originally constructed with ballooning modes specifically in mind
[\cowley,\beer]. 

The new phenomenon which arose with the three dimensional
flux tube models is the simultaneous presence of these two quite different
types of eigenmode in the same dynamical system.  Obviously, there is
only one electrostatic potential, so it is not germane to treat both
processes separately and simply add up the resulting growth rates and
hence mixing length transport estimates (which one could do for linear
eigenmodes if the wavenumbers and frequencies were significantly different).
We have a situation of ExB turbulence in the presence of a background
pressure gradient, which is the basic free energy source for both
eigenmode types. 
What differs is the energy transfer channel, that is, parallel dynamics
or interchange forcing.  The MHD model contains the interchange
mechanism but neglects the drift wave mechanism by omitting the
adiabatic coupling.  The slab drift wave model neglects the interchange
mechanism by omitting the gradient-B and magnetic curvature effects.  
One must obviously
retain these interchange effects in a toroidal model, so the principal
consideration becomes the adiabatic response:
the neglect of that limits the validity of the MHD model to
situations in which the disturbances satisfy $\pefl/p_e\ll e\phifl/T_e$
(cf.\ Eq.~\eqadiabatic).  For
purely parallel (shear Alfv\'en) dynamics this is satisfied if
$\kpp\rs\ll 1$ [\kalf], where $\rs$
is the drift scale at which simple linear drift waves principally occur.
However, for ExB motion driven by a pressure gradient, the typical
dynamical frequencies are comparable to $\ws$, the diamagnetic
frequency, and hence ExB advection of the pressure will render
$\pefl/p_e\sim e\phifl/T_e$, which is the situation of drift waves (note
that this does {\it not} mean adiabatic electrons, as the phase remains
arbitrary and the equality need not be very close).  The MHD paradigm is
therefore not useful in addressing this situation; moreover, it is
impossible even to discuss drift waves within MHD due to the missing
adiabatic coupling mechanism.

The competition between the two types of eigenmode is strongly
complicated by nonlinear physics, when one addresses turbulence.  For
linear drift waves in the presence of interchange forcing, it can be
shown that the interchange forcing can be the principal mechanism
leading to a 
finite parallel current divergence and therefore departures from the
adiabatic state, hence phase shift, over a wide range of parameters.
This phenomenon is what gave rise to the ``drift resistive
ballooning'' paradigm, in which it was originally attempted to order
$\rs$ small in the dynamics [\guzdar].  Indeed, the scales of motion in
that paradigm are derived from the assumption that interchange forcing
balances the parallel current divergence.  On the other hand, it was
already known that while the linear polarisation drift by itself could
only cause a weak nonadiabatic response, the nonlinear dynamics greatly
transforms the mode structure of the turbulence [\wakhas], leading to
robust turbulence even in models with no linear instability at all
[\ssdw].  Early attempts to treat simple interchange forcing within such
models showed that under certain circumstances the interchange forcing
did not affect the turbulence even if it could cause significant linear
instabilities [\slabcurv], a result which was confirmed by three
dimensional studies [\dwtor].  
If the nonlinear polarisation drift is strong enough to overcome the
interchange forcing in the turbulence, it will provide the balance with
the parallel current thereby maintaining electron nonadiabaticity, and
the principal scale of motion
will vary mostly with $\rs$ even though it is larger than $\rs$.  This
is even true in two dimensional slab turbulence, where there is also
dependence on dimensionless parameters such as the collisionality
[\ssdw].  In three dimensional electromagnetic computations, 
a finite plasma beta was found to enhance the nonlinear drift wave as
well as the interchange (ideal ballooning) effects [\dalfloc].
It therefore becomes crucial to resolve $\rs$ in any
computation of tokamak edge turbulence, and crucial further to use a
formulation of the magnetic geometry which represents both slab and
toroidal mode structure equally well.  For the question of what type of
mode structure obtains is closely related to which of the polarisation
or diamagnetic currents provide the balance with the parallel current
under the total divergence.

In this Article we treat the competition between drift wave and
interchange physics for control of the turbulence in detail.  Firstly
working within the simplest model treating both eigenmode types
(isothermal electrons, cold ions, where the difference to MHD is in a
single pair of terms) we choose parallel and perpendicular
scales typical of edge profiles in modern tokamaks and sweep the plasma
beta and collisionality parameters, which is the same thing as varying
temperature and density at fixed gradient scale length.  We compare the
details of the mode structure --- amplitude ratios, parallel structure,
phase shifts, cross coherence, and drive spectrum --- to control cases
reflecting the pure situation of both eigenmode types.  We also directly
measure the details of the vorticity dynamics --- mean squared
divergences of the pieces of the total current and mean squared transfer
effects --- to find which two mechanisms are in statistical balance.
This gives a rather more complete picture than one would have merely
with the usual time traces of transport levels and contour plots showing
spatial morphology.

We find that the transitions to both ideal and resistive ballooning
turbulence are well defined, and that the ``working area'' of tokamak edge
density/temperature parameter space is occupied by drift wave
turbulence.  By studying situations where an interchange driven linear
eigenmode makes the transition to fully developed turbulence, we also show
that the reason for the drift wave character is the nonlinear
polarisation drift, which replaces the diamagnetic current as the main
balance for parallel currents and nonadiabatic electron activity.  It is
always dangerous to form one's insight for turbulence from the
properties of linear eigenmodes, and this error is essentially what happened
to the drift resistive ballooning paradigm.  The conclusions of linear
theory are not valid for turbulence, because drift wave turbulence is
very much more robust than linear drift waves are: in a variant of the
well known ExB shear suppression scenario [\exbshear], the rms vorticity
level of the turbulence is larger than the ideal interchange growth rate
in this steep gradient regime, and so the small scale linear
instabilities arising from ballooning effects are superseded by the
nonlinear drift wave instability which imposes its own physical
character on the dynamics.  
Finally, though the
subject deserves its own treatment in a separate paper, we address the
question of the saturation mechanism, with the result that the vorticity
nonlinearity always drives while the pressure one provides saturation,
as in the pure drift wave case.

\section{\secmodel. The DALF3 Model}

The most basic model of ExB drift turbulence containing both drift wave
and interchange physics is a four field model in tokamak flux tube
geometry which we call DALF3, for three dimensional drift Alfv\'en
turbulence [\dalfloc].  
The magnetic geometry is assumed to be a set of nested,
toroidal, axisymmetric flux surfaces, on which one can define a globally
consistent flux tube [\fluxtube]; this is a matter of defining Hamada
coordinates aligned to the equilibrium magnetic field $\bb$ for which
there is only one nonvanishing contravariant component, which is a flux
function, and applying the parallel boundary condition after one
connection length $\Lpl$, ensuring that all degrees of freedom in the
flux tube actually exist on the flux surface for which it is a model.
A technique called ``shifted metric'' then re-adapts the geometry so as
to represent slab and toroidal mode structure equally well, by providing
for a perpendicular metric which is locally diagonal within each drift
plane perpendicular to $\bb$ [\shifted].  Following the local drift
scale ordering, the dependence of the metric on all coordinates except
the one parallel to $\bb$ is neglected (cf.\ Ref.\ [\beer]).

The dynamics consists of the fluctuating
electrostatic potential ($\phifl$) and electron pressure ($\pefl$) as
state variables, those acting as force potentials, and the 
parallel current ($\Jfl$) and parallel ion velocity ($\ufl$) as their
corresponding flux
variables.  Their set of equations is formed by the two
conservation laws for charge ($\div\jj=0$) and electron thermal energy,
and then the parallel component of the
equations of motion for the electrons and ions.  The
parallel magnetic potential ($\Afl$) is given by $\Jfl$ through
Ampere's law, and then $\phifl$ and $\Afl$ serve as stream functions for
the ExB velocity ($\vexb$) and the magnetic field disturbances ($\Bperp$),
respectively.  Traditional flute mode ordering is assumed, with parallel
and perpendicular components satisfying
$\Lpl^{-1}\sim\dpl\ll\Lpp^{-1}\lsapprox\dpp$, where 
$\Lpp$ is the perpendicular scale of the background.  The background is
a set of constant parameters except for where $\vedl$ or
$\Bperp\cdot\grad$ act on 
the background pressure gradient.  The only nonlinearities are $\vedl$
and $\Bperp\cdot\grad$, the latter which is part of $\dpl$.  The
derivation is standard; one can consult Ref.~[\dalfloc] for an example
(this model is derived from that one by setting $\tefl$ to zero).  With
the neglect of the temperatures the model is not different from the one
by Braginskii [\brag] under the local drift approximation [\hinton],
retaining electromagnetic parallel dynamics [\drifttm].

A more general model will be briefly invoked in the Section \secdalfti\
to show that the results are not dependent on the neglect of the
temperatures, but the usefulness of DALF3 is its closeness to resistive,
reduced MHD: the two models differ only in the presence or absence of
the adiabatic coupling mechanism and of the (nearly negligible)
diamagnetic heat flux in the pressure equation.  This allows
investigation of the special effects of the adiabatic $\pefl\fromto\Jfl$
coupling in direct fashion.

The equations are normalised in terms of drift scales $\rs$ and
$c_s/\Lpp$, defined by
$$\rs^2={c^2M_iT_e\over e^2B^2} \qquad\qquad
	c_s^2 = {T_e\over M_i} \qquad\qquad
	\eqno\eqname\eqrsdef
$$
This is the standard gyro-Bohm normalisation, leading to a diffusion
level $D_{GB}=\rs^2 c_s/\Lpp$ and a flux velocity level $\Gamma_{GB}=c_s
(\rs/\Lpp)^2$. 
The dependent variables are scaled in terms of 
$e\phifl/T_e$,
$\pefl/p_e$, 
$\Jfl/n_e e c_s$, 
and
$\ufl/c_s$, 
and the auxiliary one in terms of
$\Afl/\bhat B\rs$, 
where
$\bhat$ is one of the parameters discussed below.
The dependent variables are further scaled with an additional factor of
$\rs/\Lpp$, such that a normalised $\phifl=1$ refers to an unnormalised
$e\phifl/T_e = \rs/\Lpp$.
The parallel gradient and divergence are normalised against $\Lpl/2\pi$,
while the perpendicular gradient is normalised against $\Lpp$.
The normalised equations appear as
$${1\over B^2}\dtt{}\vorfl = B\dpl{\Jfl\over B} - \kappacv(\pefl)
	\eqno\eqname\eqvor$$
$$\bhat\ptt{\Afl}+\muhat\dtt{\Jfl} = \dpl(p_e+\pefl-\phifl) - C\Jfl
	\eqno\eqname\eqohm$$
$$\dtt{}(p_e+\pefl) = B\dpl{\Jfl-\ufl\over B} - \kappacv(\pefl-\phifl)
	\eqno\eqname\eqpe$$
$$\epss\dtt{\ufl} = -\dpl(p_e+\pefl) + \mupl\ddpl\ufl
	\eqno\eqname\equi$$ 
with the Ampere's law
$$\Jfl = -\ddpp\Afl
	\eqno\eqname\eqamp$$
serving as a constitutive relation (since the eigenvalues of $-\ddpp$
are positive, $\Jfl$ and $\Afl$ are in essence proportional and can be
used interchangeably as dependent variables).
Note that in the DALF3 model $\pefl=\nefl$ (we thereby avoid the
misconception that drift wave dynamics is driven solely by the density
gradient --- cf.\ Ref.\ [\ssdw]).

For the present purpose we use the simplified form of the sheared flux
tube 
geometry, neglecting all finite aspect ratio effects except for the
existence of the interchange operator $\kappacv$.  The factors of the
magnetic field strength in Eqs.~(\eqvor,\eqpe) are set to unity.  The
connection length $\Lpl$ becomes $2\pi qR$, with $q$ the standard field
line pitch parameter and $R$ the major radius.  We use the shifted
metric coordinate system [\shifted], which uses a different
$y$-coordinate at each 
location in $s$, labelled $y_k$ for the particular field aligned
coordinate system which has a locally orthogonal perpendicular Laplacian
operator at $s=s_k$.  When taking derivatives in $s$ we then have a
shift in the $y$-coordinate, since the various $y_k$ are
different.  The perpendicular differential operators for the $k$-th
coordinate system at $s=s_k$ are
$$\ddpp = \ppxx{} + \ppyyk{} \qquad\qquad
	\kappacv = \wcv\LP\sin s\pxx{}+\cos s\pyyk{}\RP
	\eqno\eqn$$
where $\wcv$ is a parameter,
and the centered partial $s$-derivative at $s=s_k$ is given by
$$\pss{f}(x,y_k-\alpha_k,s_k) 
	= {f(x,y_{k+1}-\alpha_{k+1},s_{k+1})
		- f(x,y_{k-1}-\alpha_{k-1},s_{k-1})
		\over s_{k+1}-s_{k-1}}
	\eqno\eqn
$$
with shifts given by
$$\alpha_k = \shat s_k x
	\eqno\eqn$$
and $\shat$ parameterising the magnetic shear.  
We note that in each coordinate system
$\Bdel y_k=0$ for the equilibrium $\bb$, because $\Bdel x=0$.  We note
as well that the shifts do not affect the partial $y$-derivative, which
is hereafter written without the subscript $k$.  Similar centered
differences are defined at the mid-node positions $s_{k+1/2}$, which
shift $y_k$ and $y_{k+1}$ onto $y_{k+1/2}$, as required by the second
order upwind numerical scheme (cf.\ Ref.\ [\gyrofluid], which adapts the
second order MUSCL scheme [\vanleer,\colella] to this problem).  
Accordingly, the entire set of geometric quantities (metric, magnetic
field strength and unit vector, interchange operator) are defined at
both nodes and mid-nodes $k$ and $k+1/2$.  It is important to note that
the shifting does not merely reverse the field aligning, as with each
$s_k$ a global constant, we still have $B^y=0$ since each $\alpha_k$ is
strictly a function only of $x$.

The advective and parallel derivatives are given by
$$\dtt{}=\ptt{}+\vedl \qquad\qquad \dpl = \bdel
	\eqno\eqn$$
where $\vexb$ and $\bunit$ have contravariant components given by
$$\vex = -\pyy{\phifl} \qquad \vey = \pxx{\phifl} \qquad
	b^x = \bhat\pyy{\Afl} \qquad b^y = -\bhat\pxx{\Afl} \qquad
	b^s = 1
	\eqno\eqn$$
The appearance of the background $p_e$ under derivatives represents the
gradient drive terms, through
$$p_e = -\wp x
	\eqno\eqn$$
where $\wp$ is a parameter (note choosing $\wp=1$ is equivalent to
choosing $\Lpp=L_p$).

The dependent variables are defined on a domain
$$-{\pi\over AK}<x<{\pi\over AK} 
	\qquad\qquad -{\pi\over K}<y<{\pi\over K} 
	\qquad\qquad -\pi<s<\pi
	\eqno\eqn$$
where $K$ and $A$ are parameters.  The boundary conditions
are Dirichlet for $x$ (dependent variables vanish at the boundaries),
periodic for $y$, and in $s$ they satisfy
$$f(x,y_{k+N},s_{k+N})=f(x,y_k,s_k)
	\eqno\eqn$$
where $N$ is the number of nodes, noting that the two points are on
differing coordinate systems, as in the parallel derivatives.
The profiles are maintained by boundary source terms, damping the
$k_y=0$ component of $\pefl$ by a Gaussian of amplitude $0.1$ and a
$1/e$ width of $0.1$ times the domain width, centered at each boundary.

The physical parameters are
$$\bhat = {4\pi p_e\over B^2}\LP{qR\over\Lpp}\RP^2 \qquad
  \muhat = {m_e\over M_i}\LP{qR\over\Lpp}\RP^2 \qquad
  C = 0.51{\Lpp\over c_s\tau_{\nu e}}{m_e\over M_i}\LP{qR\over\Lpp}\RP^2
	\eqno\eqn$$
for the dissipative kinetic shear Alfv\'en dynamics, and
$$\wp = {\Lpp\over L_p} \qquad 
  \wcv = {2\Lpp\over R} \qquad 
  \epss = \LP{qR\over\Lpp}\RP^2
	\eqno\eqn$$
for the gradient drive, interchange forcing, 
and the sound waves, with $\mupl$ an
arbitrary parallel diffusion coefficient.  Here, $\tau_{\nu e}$ is the
standard 
Braginskii collision time for the electrons [\brag].
Nominal parameters corresponding to a typical tokamak plasma edge are
$$\bhat = 1 \qquad \muhat = 5 \qquad C = 7.65
  \qquad \wcv = 0.05 \qquad \epss = 18350 \qquad \shat=1
	\eqno\eqname\eqparms
$$
with $\wp=1$, roughly reflecting physical parameters:
$$ \matrix{ n_e = 3\times 10^{13}\invcc \qquad
   T_e = 70\eV \qquad B = 2.5\tesla \cr
		\null\hfill\cr
   R = 165\cm \qquad \Lpp = 4.25\cm \qquad q = 3.5 \cr}
	\eqno\eqname\eqphysparms
$$
The grid for all cases consists of $64\times256\times16$
equidistant nodes in $\{x,y,s\}$, with a domain size of
$20\pi\times80\pi\times2\pi$, respectively ($K=0.025$ and $A=4$).  The
time step is $0.05$, and the initial state is a random phase
distribution for $\pefl$, with amplitude spectrum
$[1+(\kpp^2/\sqrt{128}K^2)^4]^{-1}$ normalised to an rms amplitude of
$3.0$, and with $\phifl=\Afl=\ufl=0$. 

\section{\secenergetics. DALF3 Energetics}

The physics of turbulence cannot be assessed from transport scalings or
pictorial morphology; quantitative diagnostics of the amplitude, energy,
and energetics spectra, as well as several cross-variable statistical
measurements relating the transported quantities to the ExB flows that
transport them, are needed
[\dwtor,\dalfloc,\ssdw,\sorgdw,\slabcurv,\gyrofluid,\sfdw,\warm].  In
the present case, the diagnosis of the energy transfer channels between
the equations for $\pefl$ and $\phifl$ are among the more important.

Since the energetics of the DALF3 model will form a central part of the
analysis, we review it once more, within the treatment of magnetic
coordinates just outlined.  To form the energy theorem we multiply
Eqs.~(\eqvor--\equi) by $-\phifl$, $\Jfl$, $\pefl$, and $\ufl$,
respectively, and integrate over the spatial domain, assuming that total
divergences vanish.  We find the following, in
which the integration operation is denoted by the angle brackets,
$$\ptt{}\avg{\half{1\over B^2}\abs{\dpp\phifl}^2} 
	= \avg{\Jfl\dpl\phifl} - \avg{\pefl\kappacv\phifl}
	\eqno\eqname\eqdriftenergy$$
$$\ptt{}\avg{\half\Jfl\LP\bhat\Afl+\muhat\Jfl\RP}
	= \avg{\Jfl\dpl\pefl} - \avg{\Jfl\dpl\phifl} 
	- C\avg{\abs{\Jfl}^2}
	\eqno\eqn$$
$$\ptt{}\avg{\half\abs{\pefl}^2} 
	= - \avg{\Jfl\dpl\pefl} + \avg{\ufl\dpl\pefl} 
	+ \avg{\pefl\kappacv\phifl}
	+ \wp\avg{-\pefl\pyy{\phifl}}
	\eqno\eqn$$
$$\ptt{}\avg{{\epss\over 2}\abs{\ufl}^2}
	= - \avg{\ufl\dpl\pefl} + \wp\bhat\avg{\ufl\pyy{\Afl}}
	- \mupl\avg{\abs{\dpl\ufl}^2}
	\eqno\eqn$$
The quantities under the partial time derivatives are the ion drift
energy, magnetic and electron kinetic energy, thermal free energy, and
sound wave energy.
Terms appearing in the right sides of
two equations with opposite sign represent the
transfer channels, while isolated terms reflect the sources and sinks.
The only important source is the one in the pressure equation, ExB
advection down the gradient.  Currents flowing down the gradient cancel
out of the energetics, and sound wave motion down the gradient is
always negligible.  The sinks are resistive friction ($C$), sound
wave viscosity ($\mupl$), and the subgrid dissipation resulting from the
nonlinearities, especially $\vedl\pefl$, transferring free energy to
arbitrarily small scales --- out of the spectrum on the high-$\kpp$ 
side.\footnote{${}^1$}{Conventional treatments treat this through
artificial viscosity terms; herein, the subgrid dissipation is an
integral part of the numerical scheme; in any case, it is a defining
feature of high Reynolds number turbulence, with the total dissipation
independent of the size or form of the dissipation operator.}

There are two ways for free energy to be channelled from the source in
the pressure equation into the ExB turbulence: the drift wave mechanism,
which couples $\pefl\fromto\Jfl\fromto\phifl$ through the parallel dynamics
($\dpl$), and the MHD compression mechanism, which couples
$\pefl\fromto\phifl$ directly
through the interchange forcing ($\kappacv$).  One of
our tasks will be to determine which of these is the more prominent.  If
we remove the $\kappacv$ terms only the drift wave mechanism is
present; if we remove the $\dpl(p_e+\pefl)$ term in the Ohm's law
(Eq.~\eqohm) and the $\dpl\Jfl$ term in the pressure equation
(Eq.~\eqpe) only the interchange mechanism is present.  These form the
two relevant control cases.  The linear transfer effects ($\dpl$ and
$\kappacv$) would proceed unidirectionally for linear or coherent
nonlinear systems, but with the $\vedl$ terms providing a quasirandom
character the transfer effects should be understood as exchange
mechanisms --- as we will see, their rms levels are often much greater
than their mean values.

Half squared amplitude, free energy, and source and sink profiles and
spectra can be defined, for example,
$$A_\phi(k_y) = {1\over N_k}\sum_k {1\over N_i}\sum_i
		\half\abs{\phifl(x_i,k_y,s_k)}^2
	\eqno\eqname\eqspecdef
$$
for the half squared amplitude of $\phifl$,
with $k_y=lK$ for the $l$-th Fourier component, and the $N$'s giving the
number of grid nodes in each dimension.  Half squared means simply $1/2$
of mean squared; the factor of $1/2$ is used
because some of the half squared amplitudes (\eg, $A_p$ for $\pefl$)
also appear as free energy components.  Further details of how these
quantities are defined and measured can be found in
Refs.~[\dwtor,\ssdw].  

\section{\seccontrol.  Drift Wave and Interchange Control Cases}

Here we re-examine two dimensional drift wave and interchange turbulence
models, in order to provide a control for the analysis below.  From the
DALF3 model we neglect the $s$-direction, and set $\bhat=\muhat=0$,
substituting the resulting collisional Ohm's law into
Eqs.~(\eqvor,\eqpe).  We replace $-C^{-1}\ddpl$ with the positive
constant parameter $D$ for the dynamics
involving $\Jfl$, and neglect sound wave motion.
This reverts to the well known Hasegawa-Wakatani model
[\wakhas] with interchange forcing.  We also add a damping term 
$\gamma_d\phifl$ to allow the interchange case to saturate.
The resulting two dimensional
control model is
$$\dtt{}\vorfl = D(\phifl-\pefl) - \wcv\pyy{\pefl} - \gamma_d\phifl
	\eqno\eqname\eqvorhw$$
$$\dtt{}(p_e+\pefl) = D(\phifl-\pefl) - \wp\pyy{\phifl} 
	- \wcv\pyy{}(\pefl-\phifl)
	\eqno\eqname\eqpehw$$
The drift wave version uses $D=0.05$ and $\wcv=\gamma_d=0$, and the
interchange version uses $D=0$, $\wcv=0.05$, and $\gamma_d=0.01$.  For
the drift wave model we eliminate the variations of the $k_y=0$
components at each time step.
Within the single drift plane, we use the same domain size and initial
state as in the three dimensional model.  We start both models with the
given initial state and run to $t=1000$.  The time evolution of the
half squared amplitudes of $\phifl$ and $\pefl$ are shown for both
cases in Fig.~\figctrldgdw, demonstrating saturation.

These two models differ fundamentally in the degree of coupling between
$\pefl$ and $\phifl$, and we will now see that this produces an
unequivocal distinction in their mode structure.  For the saturated
stage $500<t<1000$ we show the average cross coherence and the phase
shift probability distribution spectrum between $\pefl$ and $\phifl$ for
both models.  For the cross coherence we sample $\pefl$ and $\phifl$ at
each node in the domain and at 50 equidistant time intervals, normalise
each to its standard deviation, and plot the contours of the resulting
histogram.  For the phase shifts, we sample
$\alpha_{p\phi}(l)=\im\log(\pefl{}_l^*\phifl_l)$ for each Fourier
component $k_y=lK$, at each node in $x$ and the same 50 equidistant
time intervals, and plot the contours of the spectrum of resulting
histograms.  The signature of the drift wave coupling mechanism is very
clear: For the drift wave case the cross coherence is significant and
the phase shift distributions are narrow and close to zero (essentially
all activity below about $\alpha_{p\phi}=\pi/4$).  It is important to
note that this affects every mode in the spectrum, not just the dominant
drive range of $k_y\rs\sim 0.2$.  For the interchange case we find the
opposite situation of nearly no cross coherence and phase shifts close
to $\pi/2$ since every mode is strongly driven.  

This is the mode structure information against which we will compare the
results of the three dimensional computations.
On the other hand, the drive spectra are very similar and it is
important to note that these and other information relying principally
or solely on the properties of $\nefl$ or $\pefl$ alone cannot
distinguish these two eigenmode types.

\section{\secbdmode. Direct Comparison of DALF3 to Resistive MHD}

We now run computations within the full DALF3 model and compare them to
the resistive MHD model at the same parameters.  The MHD model is a
reduced MHD case, in which under the drift approximation the ExB flow is
dynamically incompressible, and so the perpendicular electric field is
electrostatic, just as in the drift wave model.  The pressure is the
total pressure in MHD, but within a cold ion model the total pressure is
simply $p_e$.  The cold ion model without temperature effects is
therefore the closest to the corresponding MHD case.  The only
difference in the physics is the adiabatic coupling mechanism
$\pefl\fromto\Jfl$, represented by the terms $\dpl p_e$ in the Ohm's law
and $\dpl\Jpl$ in the equation for $p_e$.  In the MHD model this
coupling is absent, and the diamagnetic flux divergence is neglected in
the pressure equation; in all other respects the DALF3 and MHD models
are identical (in MHD the distinction between warm or cold ions is
lost through the neglect of the diamagnetic flow vorticity and
gyroviscosity, and the neglect of the adiabatic coupling mechanism in
the electrons).  The MHD model appears as
$${1\over B^2}\dtt{}\vorfl = B\dpl{\Jfl\over B} - \kappacv(\pefl)
	\eqno\eqname\eqvorm$$
$$\bhat\ptt{\Afl}+\muhat\dtt{\Jfl} = -\dpl\phifl - C\Jfl
	\eqno\eqname\eqohmm$$
$$\dtt{}(p_e+\pefl) = - B\dpl{\ufl\over B} + \kappacv(\phifl)
	\eqno\eqname\eqpem$$
$$\epss\dtt{\ufl} = -\dpl(p_e+\pefl) + \mupl\ddpl\ufl
	\eqno\eqname\equim$$
with the Ampere's law,
$$\Jfl = -\ddpp\Afl
	\eqno\eqname\eqampm$$
All other considerations are as in Section III.  Simple comparison to
Eqs.~(\eqvor--\equi) shows that the resistive MHD model is an entirely
included subset of the DALF3 model.

We run both models with the fixed parameter set shown above, and vary
$C$ from very low to very high values.  The cases shown are for $\nu =
\{1,2,5,10,20,50,100,200\}$, with $C=2.55\times\nu$,
and $\mupl=0$ (allowing the numerical scheme to provide dissipation at
the very shortest parallel as well as perpendicular wavelengths).  The
normalised $\nu$ is $\Lpp/c_s\tau_{\nu e}$, and $C=0.51\muhat\nu$.

The first quantities to show are the practical ones, the transport and
the relative amplitudes between $\phifl$ and $\pefl$, which appear in
Fig.~\figbdmode, with the four main drift wave cases with $\nu=1,2,5,10$
checked with doubled resolution in the drift plane.  If the MHD
model predicted these correctly the DALF3 model would represent an
unnecessary level of complication.  We find that the transport scaling
does in 
fact disagree, strongly.  We also find that the rms amplitude ratios
$\phifl/\pefl$ disagree.  In drift wave dynamics, $\pefl$ and $\phifl$
are held together by the dissipative adiabatic response, but in MHD this
Alfv\'en dynamics dissipates $\phifl$ only.  Since MHD neglects the
electron pressure in the Ohm's law, it implicitly assumes that
$\phifl\gg\pefl$ in any situation in which the geometry sets the
parallel scale as it does here.  But with only $\phifl$ damped by the
Alfv\'en effects, the MHD model actually delivers a result contrary to
its validity, that $\pefl>\phifl$.  The DALF3 model finds the result of
the adiabatic response, namely, $\phifl\sim\pefl$.  Both models give a
power law for this amplitude ratio at low $C$.  The position at which
these two lines cross defines the regime boundary between the two
models.  For $C$ larger than this critical value, both models find
$\phifl>\pefl$.  It follows that the resistive MHD model is valid only
for $C$ beyond this value.  Because the interchange forcing enters
through the parameter $\wcv$, we find the boundary for the transition to
resistive ballooning turbulence is at
$$\nu_B=C\wcv=1.02 {c_s\over R\tau_{\nu e}}\LP {qR\over V_e}\RP^2
	\grapprox 1
	\eqno\eqname\eqresbal
$$
The transport curves both reach maxima and would eventually converge for
very large $C$; this is due to the fact that both approach two
dimensional interchange dynamics moderated only by sound waves and
nonlinear $\vedl\pefl$ cascades for values of $C$ so large that $\Jfl$
is essentially negligible.  Such values are however well out of the
parameter range of interest for tokamak edge turbulence.

Beyond the simple scalings, we turn to the mode structure in the
transition range given by $\nu=\{5,10,20\}$.  The most important
diagnostics are the ones which tell us about the relationship between
$\phifl$ and $\pefl$, since this is where the models are different.  In
Fig.~\figplcoher\ we examine the cross coherence between $\pefl$ and
$\phifl$ for both models.  The control cases appear in
Section~\seccontrol\ for drift wave and interchange turbulence.  We find
that the MHD model shows uncorrelated behaviour for all three values of
$\nu$, while the DALF3 model shows moderate correlation even for the
largest $\nu$.  The adiabatic response coupling $\phifl$ to $\pefl$
through $\Jfl$ is always significant in this parameter range.

Turning to the phase shift distributions for each $k_y$ component in the
spectrum, we find the MHD model predicting the ideally unstable values
of $\alpha_{p\phi}=\pi/2$ for all $k_y$, while the DALF3 model shows the
transition between $0$ and $\pi/2$ at low $k_y$ as $\nu$ is increased.
We see why we need these two diagnostics together: the cross coherence
reflects the turbulence as a whole, while the phase shift spectrum shows
each component.  We can see that the transition is in the longer
wavelengths, while the cross coherence is influenced also by the smaller
scale activity which remains in a drift wave mode structure.  In the
drift wave regime, however, the entire spectrum functions as a unit, and
to the extent the longer wavelengths cannot separate and form the
transition to interchange dominated turbulence, the longer wavelengths
are controlled by the shorter ones.

For the fundamental reasons for this state of affairs we turn to the
energetics.  Recall that the only transfer mechanism between $\pefl$ and
$\phifl$ in the MHD model is the interchange forcing.  For drift wave
turbulence in 
slab geometry this coupling occurs through the adiabatic response,
mediated by $\Jfl$, and so the free energy liberated from the background
gradient can only reach the ExB eddies through the parallel dynamics.
In the DALF3 model both coupling mechanisms are available, and so the
question arises as to which is most important energetically as well as
in the correlations.  In Fig.~\figwtrans, we measure the energy transfer
terms affecting the ion drift energy Eq.~(\eqdriftenergy), including the
nonlinearity which 
operates between various wavelengths but within the ion drift energy.
The relevant quantities are the rms levels of these transfer channels,
since their importance to the turbulence is in their maintenance of 
each part of the energetics at finite levels, whatever the instantaneous
sign.  The results may be compared to those for drift wave and for
interchange turbulence.  

For the drift wave cases the nonlinear polarisation drift
($\vedl\vorfl$) is largest, followed by the
nonadiabaticity ($\dpl\Jfl$), while the interchange
forcing ($\kappacv\pefl$) is subdominant.  It is especially important to
note that the interchange forcing cannot account for the rms level of
the nonadiabaticity, while the nonlinear polarisation drift is more than
large enough.  The nonlinear ($\vedl\vorfl$) and linear
($\ddpp[\ppt{\phifl}]$) parts of the polarisation drift largely balance
each other, with the excess maintaining the nonadiabaticity.

For pure (MHD) interchange turbulence, especially at low collisionality
where the 
resistive dissipation keeps $\phifl$ small, the interchange forcing and
parallel current divergence are in rough balance, and when $\phifl$ and
hence the polarisation drift is larger, the interchange forcing becomes
larger than the nonadiabaticity.  In the DALF3 model containing both
eigenmode types, the transition from drift wave to interchange transfer
dynamics occurs in the $\nu>10$ range, at which all three effects are of
comparable importance.

We note now as in the end of Section \seccontrol\ the important fact
that if we look only at the morphology of one dependent variable,
$\pefl$ for example, that we cannot easily see the difference between
the two models.  We can see this in the spectra and parallel structure,
although if we have this data for both $\pefl$ and $\phifl$ the
conclusion is obvious.  We show the amplitude spectra for the same six
cases, in Fig.~\figpldwa.  The spectra of $\pefl$ look quite similar for
the three cases $\nu=\{5,10,20\}$ in both models, but the relationship
to $\phifl$ is different.  In Fig.~\figpldwb\ the drive spectra are
shown, and these too are very similar.  Most importantly, the scale of
motion is not a distinguishing factor between the two models.
{\it This means that experiments that
measure the statistics or morphology only of density fluctuations, which
are the most accessible in a hot plasma, can do little to ascertain what
the dynamics of the turbulence actually is.}

We find a similar state of affairs in the parallel structure.  The
amplitude envelopes of the state variables are shown in Fig.~\figplfls,
and the variation of the transport flux with $s$ is shown in
Fig.~\figplbal, each for the same six cases (for these
parallel structure figures the $k_y=0$ mode is stripped so as to
concentrate on the part of the disturbances which lead to net 
transport down the gradient, noting that $v_E^x=-\ppy{\phifl}$).
Here we find that the outboard to inboard asymmetry of the density
disturbance activity is also not a distinguishing factor, as the
mere presence of interchange forcing, even at the low levels implied by
$\wcv=0.05$, is sufficient to make a difference in the overall
dynamics.  But this difference is a quantitative difference.  The
important part of the parallel structure is the much lower
degree of asymmetry in
$\hefl$ compared to either $\phifl$ or $\pefl$ in the DALF3 model.
Although $\pefl$ is strongly ballooned, the quantity which actually
liberates free energy is actually $\hefl$, whose relative lack of
ballooning in DALF3 for $\nu_B\lsapprox 1$
reflects the importance of field line connection and drift
wave coupling.  The fact that $\pefl$ (or $\phifl$) is ballooned is
actually not very relevant to the question of what type of turbulence
this is.

Taking all these diagnostics together, the qualitative difference of the
DALF3 model with $\nu_B<1$ to the two dimensional slab drift wave model
is negligible.  Although the presence of the interchange effects is
noticeable, the omission of the adiabatic $\pefl\fromto\Jfl$ coupling
leads to catastrophic changes.
The basic mode structure features which are able to tell
the difference between drift wave and interchange turbulence decide the
contrast unequivocally for drift wave mode structure, and therefore
dynamics, for the regime described by $\nu_B<1$.

\section{\secforcing. The Role of Interchange Forcing in Drift Wave
Turbulence} 

It is instructive to examine the effect of the interchange terms, which do
the interchange forcing, on simple linear drift waves with prescribed,
constant $\kpp$ and $\kpl$.  This version of the linearised equations takes
a single Fourier component by setting $\ddpp\to -\kkpp$, and $\dpl\to
i\kpl$, and $\ppy{}\to ik_y$, and for the interchange forcing
$\kappacv\to i\wcv
k_y$.  To find the fate of the instabilities, all we
need to know is the response of $\hefl$ to $\phifl$ in order to find
the dispersion relation, because the resulting phase shift is closely
related to the linear growth rate.  The vorticity equation and Ohm's
law, Eqs.~(\eqvor) and (\eqohm), are all we need.  Using the relations
$\ws=\wp k_y$ and $\wc=\wcv k_y$, their linearised form is
$$\omega\kkpp\phifl\sk=\kpl\Jfl\sk - \wc\pefl\sk
	\eqno\eqname\eqvork$$
$$\bhat\LP\omega-\ws\RP\Afl\sk+\LP\muhat\omega+iC\RP\Jfl\sk 
	= -\kpl\LP\pefl\sk-\phifl\sk\RP
	\eqno\eqname\eqjek$$
Eliminating $\Jfl\sk$, we find the equation for the
response.  It is simplest to do this in the collisional regime, where
$C$ is the dominant effect in Eq.~(\eqjek).  We then find
$$\pefl\sk = {\kkpl - iC\kkpp\omega\over\kkpl + iC\wc}\,\phifl\sk
	\eqno\eqname\eqresponse$$
As with drift waves proper, the consideration of the phase shift is only
important when $\kpl$ is finite, enough so that $\kkpl$ is the largest
term in both numerator and denominator of Eq.~(\eqresponse).  If this is
so, then the phase shift is small.  It then follows that the eigenmode
has the same properties as a drift wave, merely with a different
excitation mechanism for the phase shift.  But the role of the
interchange forcing in this case is simply to excite a phase shift and
thereby cause the turbulence to be driven by the background gradient.
This is the same role the polarisation drift has for drift waves, and
the same role its nonlinear version, the vorticity nonlinearity, has for
drift wave turbulence.  Even in this regime dominated by parallel
dynamics, the interchange effects can assume control and change the mode
structure if they dominantly cause the phase shift.  We note here that
the finiteness of $\kpl$ for finite $k_y$ is guaranteed by the
combination of magnetic shear and toroidal topology (closed flux
surfaces), that is, field line connection [\fluxtube].

\section{\seclinear.  Linear versus Nonlinear Mode Structure}

The reason that the result of this investigation for the turbulence is
so different than what one might expect by looking at linear
instabilities is that the nonadiabatic character of drift waves,
especially, is greatly changed by their associated nonlinearities.  
The one with the
strongest effect is the nonlinear polarisation drift --- equivalently,
the self advection of unsteady ExB vortices, or eddies --- which is the
mechanism through which the turbulence has such strengthened parallel
current dynamics, and hence more net nonadiabaticity of the electrons
than those linear eigenmodes.  This is the reason for and importance of
the distinction between drift waves and drift wave turbulence.

We can investigate this directly by starting the dynamics with an
initial state in the linear stage with a linear instability strongly
driven by the interchange forcing, and then tracking the evolution of the
mode structure through the saturation stage, which is the transition
from the linear stage towards the turbulence, and then into the fully
developed turbulence in which there is no detailed memory of whatever
transpires several correlation times before.

The initial state is the same as before, just with an amplitude of
$a_0=3\times 10^{-10}$ instead of the usual $3.0$.  We take a single
case, that given by the nominal parameters in Eq.~(\eqparms), but 
with $\nu=10$ (hence
$C=25.5$ and $\nu_B=1.25$) and $\mupl=0$.  The resolution in the
$x$-direction is increased fourfold to diminish the effects
of linear grid modes (the same domain size as before
is used in all three dimensions, but the node count is now
$256\times256\times16$; grid modes result from the failure to resolve
the rational surface density in $x$ for the largest $k_y$ component on
the grid).
The time traces of the
amplitudes, transport, and energetics are shown in Fig.~\figdgdwl.  The
linear instability develops, and reaches a constant growth rate which is
quite large, about $0.15$ in the units of $c_s/\Lpp$ (in which the ideal
interchange growth rate is about $\wcv^{1/2}=0.22$).
When the disturbances reach finite amplitude, the growth rate 
rapidly drops to zero.  This is the onset of saturation (the exact
moment for this case is $t=201$).  Then, the
growth rate begins to fluctuate, rises away from zero for a short while,
and then approaches zero again up to some statistical fluctuation on the
scale of a correlation time.  
This is the mode structure adjustment
stage, beginning for this case with the next rise of $\Gamma_T$ above
zero at $t=212$ and ending at $t=345$ with the next drop of $\Gamma_T$
back to zero. 
The transport overshoots
before settling, peaking near $t=200$, quickly dropping by half to about
$t=210$, and then slowly strengthening throughout the adjustment stage
until statistical saturation is reached after about $t=400$ for the
turbulence and after $t=600$ for the larger scale ExB flows.
For lower $C$ the linear instability is
confined to rather smaller scales, with more immediately efficient
nonlinear decorrelation, and the overshoot is barely noticeable.  
In either case, the transport and
overall growth rate reach statistical saturation only some several
$\dt=100$ after 
initial saturation, by which time the drive and amplitude spectra have
their final form.  
(Preliminary studies of these effects with only $32\times 128$ grid
nodes in the drift plane showed much stronger overshoot, with a longer
period of statistical fluctuation before saturation --- by comparison to
the result shown here with full resolution one can conclude such
behaviour is a manifestation of too few degrees of freedom in the
turbulence.)  
This is the
stage of fully developed turbulence, which we can see has a very
distinct separation in time from the prior two stages.  Full saturation
also occurs only after the zonal flows (the $k_y=0$ and $\kpl=0$
component of $\phifl$), to which the turbulence is
energetically coupled [\zfhahm,\zfdiamond], reach statistical
equilibrium, in this case after about $t=600$.  

These time traces show evidence for complete supersession of the linear
instability by the drift wave turbulence.
The last frame (lower
right) of Fig.~\figdgdwl\ compares the rms vorticity (mostly
at $0.3<\kpp\rs<1.0$; see below) to the linear growth and nonlinear
drive rates. 
In the conventional mixing picture of turbulence driven by a linear
instability, the amplitude grows until the nonlinear rms vorticity
$\Omega_{{\rm rms}}$ (where $\Omega=\abs{\curl\vexb}=\ddpp\phifl$)
is comparable to the linear growth rate $\gamma_L$.  The
eddies comprising the turbulence are then said to be driven by the
instability and decorrelated nonlinearly by their vorticity, leading to
the estimates of linear mixing length models:
$$\gamma_L\sim \kpp\vec v_{\kpp} \qquad\qquad
	D_{{\rm mix}} \sim {v_{\kpp}\over \kpp} \sim {\gamma_L\over\kkpp}
	\eqno\eqn$$
where $D_{{\rm mix}}$ is the turbulent diffusivity.  In this situation,
however, we 
have found not only that $\Omega_{{\rm rms}}$ is much larger than $\gamma_L$,
but also that the overall drive rate $\Gamma_+$ is significantly {\it
lower} for fully saturated turbulence than either the drive or overall
growth rates in the linear stage.  This means that the turbulence is
producing its own vorticity through the process of nonlinear self
sustainment [\ssdw], and the reason that the linear instability at small
scale has so little effect on the turbulence mode structure, as we have
seen, is that small scale structures are scattered apart by the
vorticity before they can grow.  The dynamics therefore does not feel
the linear instability in these scales.  This is a
generalisation of the ideas of linear mixing length models as well as of
the linear ExB shear-suppression scenario [\exbshear], simply instead of
the instability being suppressed by background ExB vorticity
($\Omega_E=\ppx{V_E}$), it is suppressed by turbulent ExB vorticity:
$$\Omega_{{\rm rms}}\gg\gamma_L \implies 
	D_{{\rm scatter}}\gg {\gamma_L\over\kkpp}
	\eqno\eqname\eqturbsuppress$$
where $D_{{\rm scatter}}$ is the scattering diffusivity of the
turbulence (nonlinear dissipation effect goes mostly through
$\vedl\pefl$).  
This is the ultimate reason for the mode structure changes from the
linear to the turbulent stages, which we now document.

Fig.~\figplctrl\ shows the spatial morphology in the linear, saturation,
and fully developed turbulent stages (only $1/4$ of the $y$-domain is
shown).  Up until initial saturation, the 
properties of radial ($\grad x$) 
interchange flows are pronounced, with the spatial
morphology looking very much like that of buoyancy driven flow plumes
(cf.~$t=191$). 
Initial saturation occurs when the nonlinearity is strong enough to
radially displace the density structures and to break up the plumes
($t\grapprox 200$). 
The turbulence begins as both
these structures and the ExB flows isotropise as a 
result of both main ($\vedl$ on the state variables) nonlinearities
($215\lsapprox t\lsapprox 230$). 
It
then becomes established as the structures and eddies (coupled by the
adiabatic response) expand in scale until the various linear and
nonlinear transfer processes are in balance ($400\sim t\sim 600$).
The isotropisation process is the same as that observed in the earliest
collisional drift wave turbulence computations [\wakhas].
On top of the turbulence, the zonal flows develop as eddies
serendipitously line up in the $y$-direction and then reinforce
themselves by tilting the eddies into an energy-losing relationship
vis-a-vis the flow [\sfdw];
these are visible as vertical stripes in the pattern of $\phi(x,y)$.
These striking visual changes provide motivation for the
statement that the 
properties of linear instabilities have no {\it a priori} relevance to
the turbulence.  But to really find out whether the linear stage is
relevant, we have to examine the mode structure in both stages.  

The diagnostics we have been using for the turbulence show important
differences between the linear, saturation, and fully developed
turbulent stages.  The dynamical transfer spectra for the ExB vorticity 
are shown for the transition stage in Fig.~\figwtransl, noting that the
state quickly approaches what we see in Fig.~\figwtrans.
In the linear stage,
the nonlinear polarisation drift is
obviously negligible and there is a tight balance
between interchange forcing
and nonadiabaticity (which also shows that the linear polarisation
drift, which involves the partial time derivative, is subdominant).
As in the interchange turbulence cases,
the linear stage of the DALF3 model shows a close balance between
interchange forcing and the parallel current divergence.
We therefore have an interchange-dominated linear instability.  Through
the saturation stage, the 
polarisation drift enters, and in the end state the polarisation drift
has become the dominant cause of the finite $\dpl\Jfl$ and hence
electron nonadiabaticity $\hefl=\pefl-\phifl$ and hence the phase shifts
necessary for finite free energy access and, ultimately, net transport,
both of which are proportional to $-\oint dy\,\pefl(\ppy{\phifl})$.
The parallel structure, displayed in Fig.~\figplflsl, also shows
interesting changes between the linear and turbulent stages.  The degree
of ballooning, especially in $\hefl$ is markedly reduced in turbulence.
Development of the nonlinear mode structure takes longer for this
particularly three dimensional signal, following the slowness of the
transport of the adiabatic part of the system along field lines, through
sound wave dynamics.

Finally, the phase shift distributions are examined in the stages of
linear growth, nonlinear structural adjustment, and fully developed
turbulence in Fig.~\figphasel.  For this choice of parameters the linear
instability is very strong ($\gamma_L\approx 0.15$), a significant
fraction of the nominal ideal interchange growth rate
($\gamma_I=\wcv^{1/2}=0.22$).  Its dominant range is about
$0.2<k_y\rs<1.0$.  When this range becomes nonlinear, the first
consequence for the phases is that the large phase shifts in the range
$k_y\rs>0.3$ are eliminated, replaced by a wide distribution which also
includes negative values.  At late times the long wavelength component
finds its structure --- the frame on the right of Fig.~\figphasel\ shows
the same character as the corresponding case in Fig.~\figphasdist.
Indeed, the saturated state is independent of the previous history
beyond a few correlation times; here, the correlation time is about
$6$, and long term, large scale memory persists no longer than about
ten of these.
With the short correlation 
time of the turbulence, the initial state with which the turbulence is
reached is not relevant to consideration of its mode structure or
underlying dynamics.  It follows that this applies also to linear
instabilities.  Whether these are relevant depends on the particular
case and can be demonstrated only by diagnostics taken on the fully
developed turbulence in its own context.  

In the present case, the
interchange forcing is relevant to the longer wavelengths $k_y\rs<0.1$,
not to the range of largest linear growth rate.  The reason is that the
rms vorticity of the turbulence is strongly scale dependent, able to
supersede the linear instability in the first decade longward of the
$\rs$ scale, but quickly reduced in robustness at larger scale due to
the factors of $\kpp$ in the vorticity nonlinearity (note the factors of
$\kpp$ in Eq.~\eqturbsuppress).  This is made clear
if we plot the dynamical transfer spectra for the ExB vorticity 
against $\kpp$ rather than
$k_y$, as in Fig.~\figwtransll.  
The spectra are normalised to the total ExB energy.
The linear growth stage ($50<t<150$) shows
the linear instability peaking at $\kpp\rs\grapprox 0.3$.  In the
transitional stage ($200<t<250$) the vorticity nonlinearity imposes its
own character in that same spectral range.  In the stage of fully
developed turbulence, the larger scales find their saturated state, and
in this case the interchange effects clearly control the range
$\kpp\rs<0.1$.  Whether the turbulence is ultimately of the drift wave
or interchange type depends on whether this larger scale range can
decouple from the nonlinear vorticity dynamics which is mostly at
$0.3<\kpp\rs<1.0$.  If the linear effects are important {\it only} in
this small scale range, then the linear instabilities are completely
superseded by the turbulence and do not play a role in its dynamics.

The linear analysis shows that $\wcv$ is often the dominant effect in
creating a nonzero phase shift $\alpha_{p\phi}$ and hence the linear
instability, giving it a ballooning physical (not just morphological)
character.  But as the linear eigenmode reaches finite amplitude and the
nonlinearities emerge, the turbulence not only saturates but changes its
physical character.  The mechanism maintaining the finite level of the
parallel current changes from interchange forcing to the polarisation
drift.  We saw in the
vorticity dynamical spectra in Fig.~\figwtrans, that in the fully
developed turbulence the interchange forcing does not assume control
until the collisionality reaches what we found as the
resistive ballooning threshold.  The threshold is at about
$$(\nu_B)_{{\rm crit}} = (C\wcv)_{{\rm crit}} = 1
	\eqno\eqn$$
which is the boundary beyond which the interchange effects overcome the
adiabatic response.  At that threshold the transport does not sharply
change, but the mode structure does so.  So we can properly think of
this transition as an eigenmode regime boundary, and of the turbulence
to the low collisionality side of it as drift wave turbulence.

The supersession of the
linear instability spectrum found by the preceding analysis is the same
as was recently shown in a detailed study of the drift 
wave nonlinear instability [\focusdw].  Not only does this form of self
sustained turbulence cause turbulence in a sheared slab geometry when
initialised at finite amplitude [\ssdw], it can also
impose itself in favour of the linear instability mechanisms provided
these are active in the same spectral range.  The vorticity dynamics is
active principally in the spectral range 
$0.3<\kpp\rs<1.0$.  Linear instabilities in this range are only relevant
if they are strong enough to overcome the dynamical rate indicated by
the rms vorticity of the turbulence --- a reasonable rule of thumb would
be a diamagnetic level of vorticity given by 
$$\Omega_D = v_D/L_p
	\eqno\eqn$$
(unity in normalised units), where $v_D=c_s\rs/L_p$ is the diamagnetic
velocity.  ExB flow shear must be stronger than this to suppress the
turbulence [\gyrofluid].  Short wavelength linear instabilities will be
relevant only if
$$\gamma_L > \Omega_D
	\eqno\eqn$$
\ie, their growth rate overcomes this.  Long
wavelength linear instabilities are always relevant because they can
easily take over the spectrum against what is in this range a relatively
weak vorticity nonlinearity.  The result is, though resistive ballooning
is not important to tokamak edge turbulence, ideal ballooning and other
such large scale MHD instabilities are always relevant.

The result that the turbulence changes character from the initial linear
eigenmode due to supersession by vorticity scattering ($\Omega_{{\rm
rms}}>\gamma_L$ but with little direct dissipation) is also interesting
because the resulting nonlinear drive rate $\Gamma_+$
is quite smaller than the original $\gamma_L$, even for this borderline
drift wave/interchange case.  It shows that contrary to usual practice
(cf.\ Ref.\ [\guzdar]), the judgement as to whether
a linear or nonlinear instability has principal relevance should not
rest on which one in its own native model produces the larger anomalous
transport coefficient (\ie, net turbulent flux).  If a simple comparison
is to be made, the one between 
$\Omega_{{\rm rms}}$ and $\gamma_L$ is more useful.  But one should be
careful about the $\kpp$-dependence of both, and ultimately a
comprehensive set of numerical solutions to the nonlinear equations
containing both models is required to make an informed judgement.  The
same is true of the nonlinear saturation mechanism, to which we now
turn. 

\section{\secsat.  Drive and Saturation Mechanisms}

The roles of the principal nonlinearities, $\vedl\vorfl$ and
$\vedl\pefl$, are the same in toroidal geometry as in pure slab
drift wave turbulence.  We can show this
by comparing the DALF3 model to the two dimensional drift wave model in
a simple test in which we remove either of the two principal
nonlinearities from the model.  As a baseline case we take the DALF3
model run to $t=1000$, well into saturation, for the nominal parameters
given in Eq.~(\eqparms), but 
with $\nu=2$ (hence $C=5.1$ and $\nu_B=0.25$) and $\mupl=0$.  
Then, we go back to $t=500$ and restart with
either $\vedl\vorfl$ removed (``novor'') or all $\vedl$ nonlinearities
except $\vedl\vorfl$ removed (``voronly'')
or $\vedl\pefl$ removed, and run again to
$t=1000$. The results are shown in Fig.~\figtorsat.

The vorticity nonlinearity very strongly excites the
turbulence; it does not saturate even for Dirichlet boundary conditions,
as without $\vedl\pefl$ there is no nonlinear mixing of the pressure.
The pressure gradient continues to feed free energy into the
disturbances, with the vorticity nonlinearity maintaining sufficient
$\dpl\Jfl$ for the turbulence to access regions of statistical phase
space with substantial phase shifts.  The pressure nonlinearity on the
other hand does not lead to arbitrarily strong turbulence, but saturates
the overall amplitude at a somewhat higher level.  The linear
instability is back at work in the absence of the nonlinear polarisation
drift, balancing $\dpl\Jfl$ with $\kappacv(\pefl)$.  
This result obtained for each of the five cases for which it was checked
($\nu=1,2,5$ for the DW cases and $\nu=2,5$ for the BM cases); in no
case does the turbulence saturate through the $\vedl\vor$ nonlinearity.

There are subtle differences to this result in the two dimensional (2D)
systems.  Though most of the subgrid dissipation in all the models goes
through $\vedl\pefl$, it remains true for the 2D Hasegawa-Wakatani model
(where $-C^{-1}\ddpl$ is replaced by the constant $D$) that for many
parameter combinations both nonlinearities are required in order to
achieve saturation.  Nevertheless, the strongest growth is obtained when
$\vedl\vorfl$ is present but $\vedl\pefl$ is absent.  The main
restriction of the 2D Hasegawa-Wakatani model is that $\ddpl$ cannot
change its characteristic magnitude in response to changes in amplitude
of the turbulence.

The 2D model in which the nonlinear instability was first demonstrated is
the the 2D sheared-slab model, a
simple generalisation of the Hasegawa-Wakatani model to incorporate
magnetic shear: the collisional, electrostatic limit is obtained by
taking $\bhat=\muhat=0$, neglecting any dependence in the $s$-direction,
and replacing $\dpl$ with $x(\ppy{})$, that is, the coordinate system is
not field aligned [\ssdw].  In that model the details of the dynamics
are complicated by the way $\kpl$ is tied to $x$; among the
nonlinearities, retaining only $\vedl\vorfl$ causes the entire
$x$-domain of the turbulence to be narrowed (as is the case with
adiabatic electron turbulence [\biskamp]), while retaining only
$\vedl\pefl$ causes the $x$-domain to spread.  These results were found
by taking the 2D slab restrictions within the present DALF3 code and
running with a resolution of $128\times 512$ grid nodes in $x$ and $y$,
within the same $xy$-domain size of $20\times80$ in units of $\pi\rs$,
respectively.  The cases run were $C=1$, $10$, and $100$.  The effects
on the morphology due to the absence of one of the nonlinearities was
the same as in the 3D model: $\vedl\vorfl$ produces isotropic, monopolar
vortices in $\phifl$ which excite $\pefl$ unevenly but to large
amplitude.  In this case however, $\vedl\pefl$ by itself can produce the
nonlinear instability at much the same level as with both
nonlinearities, by acting indirectly through the partial time
derivatives on both state variables.  These details of nonlinear and
drive and saturation deserve their own study (to be produced elsewhere);
it suffices here to have presented the broad outlines.

In three dimensions the degrees of freedom represented by $x$ and $\kpl$
are independent, and
the situation is clearer: the slab model which one
obtains by setting $\wcv=0$ produces the same results [\gyrofluid] both
quantitatively and qualitatively as in the toroidal case shown in this
Section, except for the effect of the toroidal case's
linear instability in the absence of $\vedl\vorfl$.
Removal of the vorticity nonlinearity leads to linear drive (in the
toroidal case) and
saturation.  Removal of the pressure nonlinearity leads to ever growing
turbulence and no saturation.  The result is that in three dimensions
the vorticity
nonlinearity always drives and the pressure nonlinearity always
saturates.  One important consequence is that all scenarios which rely
on Kelvin-Helmholtz effects on turbulent flows for saturation of
tokamak edge turbulence are ruled out.

\section{\secdalfti.  The DALFTI Model and Results for Warm Ion
Turbulence} 

Here we briefly confirm that the foregoing does not depend on the fact
that we restricted to a model with isothermal electrons and cold ions.
That was done to make the physics of the system as transparent as
possible.  To put the temperature dynamics back in, we refer to the
methods of Ref.~[\dalfloc] to treat the parallel dynamics; the heat
fluxes are themselves dynamical variables so that the two pairs of
additional dependent variables $\{\tefl,\qefl\}$ and $\{\tifl,\qifl\}$
are treated on the same self consistent footing as $\{\nefl,\Jfl\}$ in
the DALF3 model (in which $\pefl=\nefl$).  This treats time dependent
Landau damping, extending the model to the weakly collisional regime in
which the Braginskii equations lose validity (mainly by overestimating
damping of $\tefl$ and especially $\tifl$, since the turbulence is about
two orders of magnitude faster than the ion collision frequency).
We now have $\pefl=\nefl+\tefl$ and $\pifl=\tau_i\nefl+\tifl$ as
normalised pressure disturbances, where $\tau_i=T_i/T_e$ gives the
warmness of the ions.  The only remaining subtlety is gyroviscosity.
The form of the diamagnetic cancellation used here is
$$n_iM_i\ustar\cdot\grad\uu + \div\pistar = \grad\chi
	\eqno\eqname\eqgyroviscosity
$$
in physical units,
where $\ustar$ is the ion diamagnetic velocity, $\pistar$ is the
gyroviscosity tensor (the diamagnetic momentum flux), and $\chi$ is a
scalar involving finite gyroradius effects and divergences of both the
heat flux and the ion velocity [\smolyakov].  The sole effect of the
gyroviscosity is to cancel $\ustar$ in the advection.  The resulting
normalised equations are [\warm]
$${1\over B^2}\LB\dtt{}\ddpp(\phifl+\pifl) 
	+ (\grad\vexb)\dotdot(\grad\grad\pifl)\RB
	= B\dpl{\Jfl\over B} 
	- \kappacv(\pefl+\pifl)
	\eqno\eqname\eqvorti$$
$$\dtt{}(n_e+\nefl)
	= B\dpl{\Jfl-\ufl\over B}
	- \kappacv(\pefl-\phifl)
	\eqno\eqn$$
$$\threehalves\dtt{}(T_e+\tefl)
	= B\dpl{\Jfl-\ufl-\qefl\over B}
	- \kappacv(\pefl-\phifl) - \fivehalves\kappacv(\tefl)
	\eqno\eqn$$
$$\threehalves\dtt{}(T_i+\tifl)
	= B\dpl{\tau_i(\Jfl-\ufl)-\qifl\over B}
	- \tau_i\kappacv(\pefl-\phifl) + \fivehalves\tau_i\kappacv(\tifl)
	\eqno\eqn$$
$$\bhat\ptt{\Afl}+\muhat\dtt{\Jfl} 
	= \dpl(p_e+\pefl-\phifl)
	- \muhat\nu_e\LB\eta\Jfl
		+{\alpha_e\over\kappa_e}(\qefl+\alpha_e\Jfl)\RB
	\eqno\eqn$$
$$\epss\dtt{\ufl} 
	=  - \dpl(p_e+p_i+\pefl+\pifl) + \mupl\ddpl\ufl
	\eqno\eqn$$
$$\muhat\dtt{\qefl} + a_L{}_e\LP\qefl\RP
	= -\fivehalves\dpl(T_e+\tefl)
	- {5/2\over \kappa_e}\muhat\nu_e(\qefl+\alpha_e\Jfl)
	\eqno\eqn$$
$$\epss\dtt{\qifl} + a_L{}_i\LP\qifl\RP
	= -\fivehalves\tau_i\dpl(T_i+\tifl)
	- {5/2\over\kappa_i}\epss\nu_i\qifl
	\eqno\eqn$$
with current, pressures, and temperature ratio
$$\Jfl = -\ddpp\Afl \qquad
	\pefl=\nefl+\tefl \qquad \pifl=\tau_i\nefl+\tifl
	\qquad \tau_i={T_i/T_e}
	\eqno\eqn$$
and Landau damping operators
$$a_L{}_e = \muhat^{1/2}\LP 1-0.2\dpl^2\RP \qquad
a_L{}_i = (\tau_i\epss)^{1/2}\LP 1-0.2\dpl^2\RP
	\eqno\eqn
$$
including linear gradient terms through
$$n_e=-\wn x \qquad T_e=-\wt x \qquad T_i=-\tau_i\wi x
	\qquad 
	p_e=-(\wn+\wt) x
	\qquad
	p_i=-\tau_i (\wn+\wi) x
	\eqno\eqn
$$
The geometry is the same as described in Section II.  This is called the
DALFTI model, extending drift Alfv\'en turbulence to the warm ion regime
with Landau electrons and ions.

The extra parameters describing collisional dissipation are
the normalised collisional frequencies,
$$\nu_e = {L_p\over c_s\tau_{\nu e}} \qquad\qquad
  \nu_i = {L_p\over c_s\tau_{\nu i}}
	\eqno\eqn
$$
where the $\tau_{\nu e,i}$ are the standard Braginskii collision times
[\brag], 
in which terms the drift wave collisionality is $C=\eta\muhat\nu$.
The numerical constants are
$$\eta=0.51 \qquad \alpha_e=0.71 \qquad \kappa_e=3.2
	\qquad \kappa_i=3.9
	\eqno\eqname\eqbragcoeff
$$
for pure hydrogen, treating resistivity, the thermoelectric effect, and
electron and ion thermal conduction, respectively.  With these numbers
$C$ has the same meaning as in the DALF3 model.  If the $\nu_{e,i}$
become arbitrarily large we recover the Braginskii equations, as
discussed in Ref.~[\dalfloc], but we are not in that regime for typical
tokamak edge parameters, especially for ions.

We briefly report the reconsideration of the cases in Section V for the
DALFTI model, with the same nominal parameters as in Eqs.~(\eqparms),
for which $\nu=3$, and with $\wn=\wt=\wi=\tau_i=1$ placing the model in
the drift wave gradient regime (ion temperature gradient turbulence,
with larger values of $\wi/\wn$, will be treated elsewhere; cf.\ also
Refs.\ [\gyrofluid] and [\warm]).
The warm ion MHD model is formed analogously to DALF3: the combinations
$\phifl+\pifl$ and $\phi-\pefl$ are all replaced simply by $\phifl$
(including the background gradient pieces where they appear under
$\vedl$ and $\Bperp\cdot\grad$).
In Fig.~\figbdmodeti\ we find the
transport scaling and the relative disturbance amplitudes for cases with
various $\nu$, with the four main drift wave cases with $\nu=1,2,5,10$
checked with doubled resolution in the drift plane.  
These results may be compared to Fig.~\figbdmode.  Two
things should be noted: The results are somewhat clouded by the presence
of the ion temperature disturbance, $\tifl$, which always exhibits
interchange dynamics due to the fact that the nominal sound wave transit
frequency $c_s/qR$ is typically smaller than the dynamical frequencies
in the range of $0.2 c_s/\Lpp$, so while the interchange effects are
relatively weak compared to the turbulence they are still much stronger
than the parallel ion dynamics.  We also have the fact that $\tifl$
plays no role in the parallel electron dynamics, and is the only state
variable for which this is true.  Secondly, $\pefl$ includes both
$\nefl$ and $\tefl$, so that the ratio $\phifl/\nefl$ is also influenced
by changes in the relative role of the temperature and density
gradients.  This follows from the fact that more nonadiabatic electron
dynamics in the stronger turbulence for higher collisionality leads to
comparatively stronger effect due to $\grad T_e$ [\dalfloc,\ssdw].
Nevertheless,
although the transition regime is wider than in cases with pure electron
dynamics (cold ions), it still occurs for $\nu_B$ only slightly less
than unity.

The transition to resistive ballooning for these warm ion cases is shown
in Figs.~\figplcoherti--\figplflsti.  
As $\nu$ is swept through the values $2$, $5$, and
$10$, we find the gradual increase in the amplitude of $\phifl$ relative
to the other state variables.  In the drift wave regime $\tefl$ roughly
has the same structure as $\hefl$ in the DALF3 model (compare with
Fig.~\figplfls); indeed, just as in slab drift wave turbulence these two
quantities control the release of free energy from the electron thermal
gradient and hence play the same role in the turbulence [\ssdw].  The
fact that they are less ballooned than the density or ion temperature
(or the transport fluxes) is the reason that drift wave mode structure
persists in turbulence in toroidal geometry, just as for $\hefl$ in the
DALF3 model.  The arrival of the pure MHD regime for $\nu=10$ is also
facilitated by the contribution of $\pifl$ to the total ion flow stream
function, $\Wfl=\phifl+\pifl$, and hence the vorticity, $\ddpp\Wfl$.
This adds to the power of the total nonlinear polarisation drift and
removes the linear forcing terms from dominance at lower collisionality
than in a pure MHD model for which the ion flow is just the ExB flow
(cf.\ also [\dwtor]).
Tests have shown that the presence of the nonlinear gyroviscosity has a
large effect in the practical drift wave regime $1<\nu<5$, but we leave
the details of this to a subsequent publication
(without the gyroviscosity, the vorticity in Eq.~\eqvorti\ is advected
by the total flow expressed with $\Wfl$, and there is no correction; in
contrast to the actual situation, both energy and enstrophy are
conserved in the case that gyroviscosity is neglected).  Due mostly to
the new effects brought in by the finite ion temperature, the onset
of the transition to the resistive ballooning regime is moved to about
$$(\nu_B)_{{\rm crit}} = 0.5
	\eqno\eqn$$
that is, downward by about a factor of two, generally reflecting the
equal and additive effects of the electron and ion pressures.

\section{\secideal.  The Ideal Ballooning Boundary}

Up to now we have focussed on the comparison between drift wave and
resistive ballooning turbulence, where only the collisionality is
varied.  The other boundary is better known: the ideal MHD boundary,
which gives the onset of dissipation free ballooning modes called ideal
modes [\straussbal].  
Experimentally this is thought to give the beta limit, which is
known both globally and locally [\opdiagram].  In Fig.~\figtransbetas\
we show this for both the
DALF3 and DALFTI models, setting $\nu=1$ to get a sharper transition.
The transport is computed as a flux as before,
but this time it is given in terms of the traditional transport
coefficients in physical units ---
ambiguity with regard to the trend
is minimised by choosing all gradient scale
lengths equal to $\Lpp$ and by avoiding the effect of hidden
normalisation, as the diffusivities scale with $\rs^2 c_s/\Lpp$ and
hence with $(\bhat/C)^{1/2}$ if $B$ and the scale lengths are all held
fixed.  We find a clear boundary in both models 
though it is accentuated by the presence of $\tifl$.
The jump in
the transport appears at different values of $\bhat$ because the total
pressure differs in the models.  The standard ideal MHD stability
parameter is
$$\alpha_M = -q^2R\grad\beta = \bhat\wcv[(\wn+\wt)+\tau_i(\wn+\wi)]
	\eqno\eqname\eqideal
$$
in both physical and normalised units, where the two combinations in
parentheses give the normalised $\grad p_e$ and $\grad p_i$,
respectively, giving a factor of $4$ for these parameters for DALFTI and
a factor of $1$ for DALF3.
The transport amplitude is plotted against $\alpha_M$, showing that the
effective ideal ballooning boundary for the turbulence is given by
$$0.2 < (\alpha_M)_{{\rm crit}} < 0.6$$
for $\shat=1$.  The regime change starts closer to $0.2$ but the mode
structure changes are not complete until the longer wavelengths separate
from the rest of the dynamics, for $\alpha_M$ about $0.6$.

The mode structure changes in the ideal ballooning transition are shown
in Figs.~\figplcoherbetas--\figplflsbetas, for the three cases
$\bhat=\{0.3,1.0,3.0\}$ for the DALFTI model, measured in the same way
as in Section \secbdmode.  The signature of the change in character of
the turbulence is very clear in these figures.  When the ideal
ballooning takes over, it does so in the longest wavelengths as their
phase shifts separate from the rest of the spectrum.  At the same
parameters, the cross coherence between $\nefl$ and $\phifl$ is lost.
The signatures of the change in character of the turbulence in this
transition is sharper and clearer than in the one for resistive
ballooning.  All vestige of drift wave character in the cross coherence
and parallel structure diagnostics is lost when the ideal ballooning
boundary is crossed.  The phase shift diagnostic shows that the long
wavelength piece splits off and is no longer influenced by the vorticity
dynamics coming from the shorter wavelengths, with the result that the
entire spectrum no longer acts as a self consistently causal unit.  The
same behaviour is exhibited by both the DALF3 and DALFTI models.

The ideal ballooning cases saturate only by depleting the pressure
profile:  With periodic boundary conditions in $x$ these cases do not
saturate at all, instead forming a wide jet in the $x$-direction with
$k_x=0$ and $k_y L_x\approx\pi$ where $L_x=2\pi/AK$ is the $x$-domain width.
With Dirichlet boundaries the jet becomes a more or less isotropic cell
with $k_x\approx k_y\approx\pi/L_x$, saturating only as the resulting
flux balances the source terms maintaining the profile in the edge
damping zones.  This means that saturation proceeds only directly on the
profile, with the combined action of the nonlinearities unable to
suffice.  In no case was the vorticity nonlinearity observed to provide
saturation through Kelvin-Helmholtz effects.  This has been found to be
a feature of the ``thin atmosphere'' situation with aspect ratio $A\gg
1$, contrary to other efforts [\rogers] which use $A=1$, artificially
constricting the formation of wide down-gradient flows.

\section{\secsummary.  Summary --- Turbulence in Context}

What these results show is that it is important to consider the
turbulence in context when making judgments about its character,
especially when in a situation like this there is more than one possible
mode structure into which it can arrange itself.  
Moreover, it is essential to set up computations with equal regard for
the properties of both or all the possible eigenmode types, so that the
results are not prematurely anticipated.
We find that this physical character undergoes strong changes
no only in spatial structure but also in energetics
as the linear
instability makes the transition into turbulence.  Since the turbulence
has no detailed memory beyond several (rather short)
correlation times, the character the dynamics has in the linear 
stage is not only not relevant, it delivers an incorrect paradigm as to
understanding of the basic nature of the turbulence.  This serves to
underscore the danger of relying on linear instability theory in the
formative stages of a body of work whose aim is to understand turbulence
and transport.  The linear instability is relevant whenever it can act
on the longer wavelengths (low $\kpp$), where the vorticity nonlinearity
is weak, but at high $\kpp\rs\sim 1$ (\ie, ``high-$n$ ballooning'') they
lose relevance because these short wavelength instabilities are
superseded by the self sustained drift wave turbulence.  Since this
turbulence has an rms vorticity greater than the ideal interchange
growth rate, it scatters the linear eigenmodes apart before they can
grow (in analogous fashion to the way a background ExB vorticity
suppresses instabilities).

The important thing to note about these results is that they all speak
together; there are no contradictions.  In all cases, the mode structure
for low collisionality and low beta exhibits
clear drift wave character,
and the transition to resistive MHD is found to start for $\nu>3$, which
for this choice of parameters corresponds to $\nu_*=41$,
where $\nu_*=(qR/\tau_{\nu e} V_e)\eps^{-3/2}$ and in normalised units
$\nu_*=\nu\muhat^{1/2}\eps^{-3/2}$ or about $13.6\nu_e$ and $5.3C$ for
the nominal scale ratios in Eq.~(\eqparms),
having assumed an inverse aspect ratio of $\eps=0.3$.
Therefore, having examined the properties of the turbulence
in its native context, we can
determine where the MHD model begins to be valid, and where we need the
drift wave model not only for computation but for fundamental
understanding.  Where the MHD model loses validity due to the adiabatic
response becoming important, it also loses usefulness as a paradigm,
because the basic physics of how the disturbances in pressure and ExB
flow communicate undergoes fundamental changes.
Generally, MHD character 
is found when the interchange forcing on the pressure
overcomes the adiabatic response.  For ideal MHD the regime boundary is
the ballooning limit at the critical value of $\alpha_M=q^2
R\abs{\grad\beta}$ for the particular geometry; in the circular tokamak
model it is near $(\alpha_M{})_{{\rm crit}}=\shat$.  
For resistive ballooning the ideal
Alfv\'en response is replaced by its dissipative MHD limit, and
the boundary is near $\nu_B=C\wcv=1$.  The factors of order
unity will vary with
magnetic geometry and some of the complications of the temperatures, but
generally for the low beta and moderate collisionality regime in which
fusion plasmas are found, one either finds drift wave turbulence if the
equilibrium is steady, or ideal ballooning phenomena when there are
disruptive events which typically collapse the pressure gradient.

Concerning the situation of linear instabilities vis-a-vis self
sustained drift wave turbulence, we find a general indication that short
wavelength ($k_y\rs\grapprox 0.3$) linear instabilities are relevant to
the turbulence if and only if their linear growth rates are stronger
than the general diamagnetic vorticity level:
$\gamma_L>\Omega_D$.  Long wavelength  ($k_y\rs\lsapprox 0.1$) linear
instabilities, on the other hand, are always relevant because the
vorticity nonlinearity is relatively weak in that spectral range.
Borderline cases can only be decided with well resolved computations
assisted by detailed diagnosis.

It is important to note that the question of drift wave or ballooning
character is one which is decided by the physical processes responsible
for free energy generation, transfer, and saturation, not by the general
look and feel of the turbulence which would be a very arbitrary issue.
The formative literature for both eigenmode types made very clear
statements of what these processes should be, and in order to decide
which is most relevant one must diagnose those processes directly and
within the context of the turbulence itself.  It is precisely that which
we have done herein.

{
\parindent 0 pt
\frenchspacing
\parskip=10pt plus 1pt minus 1pt
\def\ref##1.##2\par{\par\hangindent 15pt [##1]##2}
\par\section{References}

\ref\dwtor.
    B. Scott, \ppcf{39} (1997) 471.

\ref\dalfloc.
    B. Scott, \ppcf{39} (1997) 1635.

\ref\zeiler.
    A. Zeiler, D. Biskamp, J. F. Drake, P. N. Guzdar,
    \physp{3} (1996) 2951.

\ref\rogers.
    B. Rogers and J. F. Drake, \prl{79} (1997) 229.

\ref\xxu.
    X. Q. Xu, R. H. Cohen, T. D. Rognlien, and J. R. Myra,
    \physp{7} (2000) 1951.

\ref\hasmim.
    A. Hasegawa and K. Mima, \prl{39} (1977) 205; \pf{21} (1978) 87.

\ref\wakhas.
    M. Wakatani and A. Hasegawa, \pf{27} (1984) 611.

\ref\waltz.
    R. E. Waltz, \pf{28} (1985) 577.

\ref\biskamp.
    D. Biskamp and M. Walter, {\it Phys. Lett. A}\vol{109} (1985) 34.

\ref\ssdw.
    B. Scott, \prl{65} (1990) 3289; \pfb{4} (1992) 2468.

\ref\sorgdw.
    B. Scott, H. Biglari, P. W. Terry, and P. H. Diamond,
	\pfb{3} (1991) 51.

\ref\fluxtube.
    B. Scott, \physp{5} (1998) 2334.

\ref\strauss.
    H. Strauss, \pf{19} (1976) 134.

\ref\origballooning.
    A. M. M. Todd, M. S. Chance, J. M. Greene, R. C. Grimm,
    J. L. Johnson, and J. Manickam, \prl{38} (1977) 826.

\ref\straussbal.
    H. Strauss, \pf{20} (1977) 1354.

\ref\coppi.
    B. Coppi, \prl{39} (1977) 939.

\ref\connor.
    J. W. Connor, R. J. Hastie, and J. B. Taylor, \prl{40} (1978) 396.

\ref\glasser.
    R. L. Dewar and A. H. Glasser, \pf{26} (1983) 3038.

\ref\straussresbal.
    H. Strauss, \pf{24} (1981) 2004.

\ref\carreras.
    B. A. Carreras, L. Garcia, and P. H. Diamond, \pf{30} (1987) 1388.

\ref\resbal.
    B. A. Carreras and P. H. Diamond, \pfb{1} (1989) 1011.

\ref\rbmlinear.
    Two prominent ones are:
    B. A. Carreras, P. H. Diamond, M. Murakami, J. L. Dunlap,
	J. D. Bell, H. R. Hicks, J. A. Holmes, E. A. Lazarus, 
	V. K. Pare, P. Similon, C. E. Thomas, and R. M. Wieland,
	\prl{50} (1983) 503;
    T. C. Hender, B. A. Carreras, W. A. Cooper, J. A. Holmes,
	P. H. Diamond, and P. L. Similon, \pf{27} (1984) 1439.

\ref\guzdar.
    D. R. McCarthy, P. N. Guzdar, J. F. Drake, 
	T. M. Antonsen, Jr., and A. B. Hassam, \pfb{4} (1992) 1846;
    P. N. Guzdar, J. F. Drake, \pfb{5} (1993) 3712.

\ref\cowley. 
    S. C. Cowley, R. M. Kulsrud, and R. N. Sudan, \pfb{3}, 2767 (1991).

\ref\beer. 
    M. A. Beer, S. C. Cowley, and G. W. Hammett, \physp{2}, 2687 (1995).

\ref\kalf.
    A. Hasegawa and L. Chen, \pf{19} (1976) 1924.

\ref\slabcurv.
    B. Scott, in {\it Plasma Physics and
	Controlled Nuclear Fusion Research 1992} (IAEA, Vienna 1993),
	Vol. 2, p. 203.

\ref\exbshear.
    H. Biglari, P. H. Diamond, and P. W. Terry, \pfb{2} (1990) 1;
    Ch. P. Ritz, H. Lin, T. L. Rhodes, and A. J. Wootton, 
	\prl{65} (1990) 2543;
    T. S. Hahm, \physp{1} (1994) 2940;
    T. S. Hahm and K. H. Burrell, \physp{2} (1995) 1648.

\ref\shifted.
    B. Scott, \physp{8} (2001) 447.

\ref\brag. 
    S. I. Braginskii, \revpp{1} (1965) 205.

\ref\hinton.
    F. L. Hinton and C. W. Horton, Jr, \pf{14} (1971) 116.

\ref\drifttm.
    B. Scott, A. B. Hassam, and J. F. Drake, \pf{28} (1985) 275.

\ref\gyrofluid.
    B. Scott, \physp{7} (2000) 1845.

\ref\vanleer.
    B. Van Leer, \jcp{32} (1979) 101.

\ref\colella.
    P. Colella, \jcp{87} (1990) 171.

\ref\zfhahm.
    T. S. Hahm, M. A. Beer, Z. Lin, G. W. Hammett, W. W. Lee, and
	W. M. Tang, \physp{6} (1999) 922.

\ref\zfdiamond.
    M. A. Malkov, P. H. Diamond, and A. Smolyakov,
	\physp{8} (2001) 1553.

\ref\sfdw.
    B. Scott, \ppcf{34} (1992) 1977.

\ref\focusdw.
    B. Scott, \njp{4} (2002) 52.

\ref\smolyakov.
    A. Smolyakov, {\it Canadian J. Phys.}\vol{76} (1998) 321.

\ref\warm.
    B. Scott, \cpp{38} (1998) 171.

\ref\opdiagram.
    W. Suttrop, M. Kaufmann, H. J. de Blank, B. Br\"usehaber, K. Lackner, 
	V. Mertens, H. Murmann, J. Neuhauser, F. Ryter, H. Salzmann, 
	J. Schweinzer, J. Stober, H. Zohm and the ASDEX Upgrade Team, 
	\ppcf{39} (1997) 2051.

\par

}

\par\vfill\eject
\def\fig##1.##2\par{\item{{\secfnt Fig.\ ##1.}}##2}
\frenchspacing
\parskip 6pt plus 1pt minus 1pt
\parindent 0 pt
\par\section{Figures}
\def\temp{1.34}%
\let\tempp=\relax
\expandafter\ifx\csname psboxversion\endcsname\relax
  \message{PSBOX(\temp) loading}%
\else
    \ifdim\temp cm>\psboxversion cm
      \message{PSBOX(\temp) loading}%
    \else
      \message{PSBOX(\psboxversion) is already loaded: I won't load
        PSBOX(\temp)!}%
      \let\temp=\psboxversion
      \let\tempp= 
    \fi
\fi
\tempp
\let\psboxversion=\temp
\catcode`\@=11
%
%
\def\psfortextures{
\def\PSspeci@l##1##2{%
\special{illustration ##1\space scaled ##2}%
}}%
\def\psfordvitops{
\def\PSspeci@l##1##2{%
\special{dvitops: import ##1\space \the\drawingwd \the\drawinght}%
}}%
\def\psfordvips{
\def\PSspeci@l##1##2{%
\d@my=0.1bp \d@mx=\drawingwd \divide\d@mx by\d@my
\includegraphics{##1\space}}}%
\def\psforoztex{
\def\PSspeci@l##1##2{%
\special{##1 \space
      ##2 1000 div dup scale
      \number-\psllx\space \number-\pslly\space translate
}}}%
\def\psfordvitps{
\def\psdimt@n@sp##1{\d@mx=##1\relax\edef\psn@sp{\number\d@mx}}
\def\PSspeci@l##1##2{%
\special{dvitps: Include0 "psfig.psr"}
\psdimt@n@sp{\drawingwd}
\special{dvitps: Literal "\psn@sp\space"}
\psdimt@n@sp{\drawinght}
\special{dvitps: Literal "\psn@sp\space"}
\psdimt@n@sp{\psllx bp}
\special{dvitps: Literal "\psn@sp\space"}
\psdimt@n@sp{\pslly bp}
\special{dvitps: Literal "\psn@sp\space"}
\psdimt@n@sp{\psurx bp}
\special{dvitps: Literal "\psn@sp\space"}
\psdimt@n@sp{\psury bp}
\special{dvitps: Literal "\psn@sp\space startTexFig\space"}
\special{dvitps: Include1 "##1"}
\special{dvitps: Literal "endTexFig\space"}
}}%
\def\psfordvialw{
\def\PSspeci@l##1##2{
\special{language "PostScript",
position = "bottom left",
literal "  \psllx\space \pslly\space translate
  ##2 1000 div dup scale
  -\psllx\space -\pslly\space translate",
include "##1"}
}}%
\def\psforptips{
\def\PSspeci@l##1##2{{
\d@mx=\psurx bp
\advance \d@mx by -\psllx bp
\divide \d@mx by 1000\multiply\d@mx by \xscale
\incm{\d@mx}
\let\tmpx\dimincm
\d@my=\psury bp
\advance \d@my by -\pslly bp
\divide \d@my by 1000\multiply\d@my by \xscale
\incm{\d@my}
\let\tmpy\dimincm
\d@mx=-\psllx bp
\divide \d@mx by 1000\multiply\d@mx by \xscale
\d@my=-\pslly bp
\divide \d@my by 1000\multiply\d@my by \xscale
\at(\d@mx;\d@my){\special{ps:##1 x=\tmpx, y=\tmpy}}
}}}%
\def\psonlyboxes{
\def\PSspeci@l##1##2{%
\at(0cm;0cm){\boxit{\vbox to\drawinght
  {\vss\hbox to\drawingwd{\at(0cm;0cm){\hbox{({\tt##1})}}\hss}}}}
}}%
\def\psloc@lerr#1{%
\let\savedPSspeci@l=\PSspeci@l%
\def\PSspeci@l##1##2{%
\at(0cm;0cm){\boxit{\vbox to\drawinght
  {\vss\hbox to\drawingwd{\at(0cm;0cm){\hbox{({\tt##1}) #1}}\hss}}}}
\let\PSspeci@l=\savedPSspeci@l
}}%
%
%
\newread\pst@mpin
\newdimen\drawinght\newdimen\drawingwd
\newdimen\psxoffset\newdimen\psyoffset
\newbox\drawingBox
\newcount\xscale \newcount\yscale \newdimen\pscm\pscm=1cm
\newdimen\d@mx \newdimen\d@my
\newdimen\pswdincr \newdimen\pshtincr
\let\ps@nnotation=\relax
{\catcode`\|=0 |catcode`|\=12 |catcode`|
|catcode`#=12 |catcode`*=14
|xdef|backslashother{\}*
|xdef|percentother{
|xdef|tildeother{~}*
|xdef|sharpother{#}*
}%
\def\R@moveMeaningHeader#1:->{}%
\def\uncatcode#1{%
\edef#1{\expandafter\R@moveMeaningHeader\meaning#1}}%
\def\execute#1{#1}
\def\psm@keother#1{\catcode`#112\relax}
\def\executeinspecs#1{%
\execute{\begingroup\let\do\psm@keother\dospecials\catcode`\^^M=9#1\endgroup}}%
\def\@mpty{}%
\def\matchexpin#1#2{
  \fi%
  \edef\tmpb{{#2}}%
  \expandafter\makem@tchtmp\tmpb%
  \edef\tmpa{#1}\edef\tmpb{#2}%
  \expandafter\expandafter\expandafter\m@tchtmp\expandafter\tmpa\tmpb\endm@tch%
  \if\match%
}%
\def\matchin#1#2{%
  \fi%
  \makem@tchtmp{#2}%
  \m@tchtmp#1#2\endm@tch%
  \if\match%
}%
\def\makem@tchtmp#1{\def\m@tchtmp##1#1##2\endm@tch{%
  \def\tmpa{##1}\def\tmpb{##2}\let\m@tchtmp=\relax%
  \ifx\tmpb\@mpty\def\match{YN}%
  \else\def\match{YY}\fi%
}}%
\def\incm#1{{\psxoffset=1cm\d@my=#1
 \d@mx=\d@my
  \divide\d@mx by \psxoffset
  \xdef\dimincm{\number\d@mx.}
  \advance\d@my by -\number\d@mx cm
  \multiply\d@my by 100
 \d@mx=\d@my
  \divide\d@mx by \psxoffset
  \edef\dimincm{\dimincm\number\d@mx}
  \advance\d@my by -\number\d@mx cm
  \multiply\d@my by 100
 \d@mx=\d@my
  \divide\d@mx by \psxoffset
  \xdef\dimincm{\dimincm\number\d@mx}
}}%
%
\newif\ifNotB@undingBox
\newhelp\PShelp{Proceed: you'll have a 5cm square blank box instead of
your graphics (Jean Orloff).}%
\def\s@tsize#1 #2 #3 #4\@ndsize{
  \def\psllx{#1}\def\pslly{#2}%
  \def\psurx{#3}\def\psury{#4}
  \ifx\psurx\@mpty\NotB@undingBoxtrue
  \else
    \drawinght=#4bp\advance\drawinght by-#2bp
    \drawingwd=#3bp\advance\drawingwd by-#1bp
  \fi
  }%
\def\sc@nBBline#1:#2\@ndBBline{\edef\p@rameter{#1}\edef\v@lue{#2}}%
\def\g@bblefirstblank#1#2:{\ifx#1 \else#1\fi#2}%
{\catcode`\%=12
\xdef\B@undingBox{
\def\ReadPSize#1{
 \readfilename#1\relax
 \let\PSfilename=\lastreadfilename
 \openin\pst@mpin=#1\relax
 \ifeof\pst@mpin \errhelp=\PShelp
   \errmessage{I haven't found your postscript file (\PSfilename)}%
   \psloc@lerr{was not found}%
   \s@tsize 0 0 142 142\@ndsize
   \closein\pst@mpin
 \else
   \if\matchexpin{\GlobalInputList}{, \lastreadfilename}%
   \else\xdef\GlobalInputList{\GlobalInputList, \lastreadfilename}%
     \immediate\write\psbj@inaux{\lastreadfilename,}%
   \fi%
   \loop
     \executeinspecs{\catcode`\ =10\global\read\pst@mpin to\n@xtline}%
     \ifeof\pst@mpin
       \errhelp=\PShelp
       \errmessage{(\PSfilename) is not an Encapsulated PostScript File:
           I could not find any \B@undingBox: line.}%
       \edef\v@lue{0 0 142 142:}%
       \psloc@lerr{is not an EPSFile}%
       \NotB@undingBoxfalse
     \else
       \expandafter\sc@nBBline\n@xtline:\@ndBBline
       \ifx\p@rameter\B@undingBox\NotB@undingBoxfalse
         \edef\t@mp{%
           \expandafter\g@bblefirstblank\v@lue\space\space\space}%
         \expandafter\s@tsize\t@mp\@ndsize
       \else\NotB@undingBoxtrue
       \fi
     \fi
   \ifNotB@undingBox\repeat
   \closein\pst@mpin
 \fi
\message{#1}%
}%
%
%
\def\psboxto(#1;#2)#3{\vbox{
   \ReadPSize{#3}%
   \divide\drawingwd by 1000
   \divide\drawinght by 1000
   \d@mx=#1
   \ifdim\d@mx=0pt\xscale=1000
         \else \xscale=\d@mx \divide \xscale by \drawingwd\fi
   \d@my=#2
   \ifdim\d@my=0pt\yscale=1000
         \else \yscale=\d@my \divide \yscale by \drawinght\fi
   \ifnum\yscale=1000
         \else\ifnum\xscale=1000\xscale=\yscale
                    \else\ifnum\yscale<\xscale\xscale=\yscale\fi
              \fi
   \fi
   \divide\pswdincr by 1000 \multiply\pswdincr by \xscale
   \divide\pshtincr by 1000 \multiply\pshtincr by \xscale
   \divide\psxoffset by1000 \multiply\psxoffset by\xscale
   \divide\psyoffset by1000 \multiply\psyoffset by\xscale
   \global\divide\pscm by 1000
   \global\multiply\pscm by\xscale
   \multiply\drawingwd by\xscale \multiply\drawinght by\xscale
   \ifdim\d@mx=0pt\d@mx=\drawingwd\fi
   \ifdim\d@my=0pt\d@my=\drawinght\fi
   \message{scaled \the\xscale}%
 \hbox to\d@mx{\hss\vbox to\d@my{\vss
   \global\setbox\drawingBox=\hbox to 0pt{\kern\psxoffset\vbox to 0pt{
      \kern-\psyoffset
      \PSspeci@l{\PSfilename}{\the\xscale}%
      \vss}\hss\ps@nnotation}%
   \advance\pswdincr by \drawingwd
   \advance\pshtincr by \drawinght
   \global\wd\drawingBox=\the\pswdincr
   \global\ht\drawingBox=\the\pshtincr
   \baselineskip=0pt
   \copy\drawingBox
 \vss}\hss}%
  \global\psxoffset=0pt
  \global\psyoffset=0pt
  \global\pswdincr=0pt
  \global\pshtincr=0pt 
  \global\pscm=1cm 
  \global\drawingwd=\drawingwd
  \global\drawinght=\drawinght
}}%
%
%
\def\psboxscaled#1#2{\vbox{
  \ReadPSize{#2}%
  \xscale=#1
  \message{scaled \the\xscale}%
  \advance\drawingwd by\pswdincr\advance\drawinght by\pshtincr
  \divide\pswdincr by 1000 \multiply\pswdincr by \xscale
  \divide\pshtincr by 1000 \multiply\pshtincr by \xscale
  \divide\psxoffset by1000 \multiply\psxoffset by\xscale
  \divide\psyoffset by1000 \multiply\psyoffset by\xscale
  \divide\drawingwd by1000 \multiply\drawingwd by\xscale
  \divide\drawinght by1000 \multiply\drawinght by\xscale
  \global\divide\pscm by 1000
  \global\multiply\pscm by\xscale
  \global\setbox\drawingBox=\hbox to 0pt{\kern\psxoffset\vbox to 0pt{
     \kern-\psyoffset
     \PSspeci@l{\PSfilename}{\the\xscale}%
     \vss}\hss\ps@nnotation}%
  \advance\pswdincr by \drawingwd
  \advance\pshtincr by \drawinght
  \global\wd\drawingBox=\the\pswdincr
  \global\ht\drawingBox=\the\pshtincr
  \baselineskip=0pt
  \copy\drawingBox
  \global\psxoffset=0pt
  \global\psyoffset=0pt
  \global\pswdincr=0pt
  \global\pshtincr=0pt 
  \global\pscm=1cm
  \global\drawingwd=\drawingwd
  \global\drawinght=\drawinght
}}%
%
\def\psbox#1{\psboxscaled{1000}{#1}}%
\newif\ifn@teof\n@teoftrue
\newif\ifc@ntrolline
\newif\ifmatch
\newread\j@insplitin
\newwrite\j@insplitout
\newwrite\psbj@inaux
\immediate\openout\psbj@inaux=psbjoin.aux
\immediate\write\psbj@inaux{\string\joinfiles}%
\immediate\write\psbj@inaux{\jobname,}%
%
%
\def\toother#1{\ifcat\relax#1\else\expandafter%
  \toother@ux\meaning#1\endtoother@ux\fi}%
\def\toother@ux#1 #2#3\endtoother@ux{\def\tmp{#3}%
  \ifx\tmp\@mpty\def\tmp{#2}\let\next=\relax%
  \else\def\next{\toother@ux#2#3\endtoother@ux}\fi%
\next}%
%
%
\let\readfilenamehook=\relax
\def\re@d{\expandafter\re@daux}
\def\re@daux{\futurelet\nextchar\stopre@dtest}%
\def\re@dnext{\xdef\lastreadfilename{\lastreadfilename\nextchar}%
  \afterassignment\re@d\let\nextchar}%
\def\stopre@d{\egroup\readfilenamehook}%
\def\stopre@dtest{%
  \ifcat\nextchar\relax\let\nextread\stopre@d
  \else
    \ifcat\nextchar\space\def\nextread{%
      \afterassignment\stopre@d\chardef\nextchar=`}%
    \else\let\nextread=\re@dnext
      \toother\nextchar
      \edef\nextchar{\tmp}%
    \fi
  \fi\nextread}%
\def\readfilename{\vbox\bgroup%
  \let\\=\backslashother \let\%=\percentother \let\~=\tildeother
  \let\#=\sharpother \xdef\lastreadfilename{}%
  \re@d}%
%
%
\xdef\GlobalInputList{\jobname}%
\def\psnewinput{%
  \def\readfilenamehook{
    \if\matchexpin{\GlobalInputList}{, \lastreadfilename}%
    \else\xdef\GlobalInputList{\GlobalInputList, \lastreadfilename}%
      \immediate\write\psbj@inaux{\lastreadfilename,}%
    \fi%
    \ps@ldinput\lastreadfilename\relax%
    \let\readfilenamehook=\relax%
  }\readfilename%
}%
\expandafter\ifx\csname @@input\endcsname\relax    
  \immediate\let\ps@ldinput=\input\def\input{\psnewinput}%
\else
  \immediate\let\ps@ldinput=\@@input
  \def\@@input{\psnewinput}%
\fi%
\def\nowarnopenout{%
 \def\warnopenout##1##2{%
   \readfilename##2\relax
   \message{\lastreadfilename}%
   \immediate\openout##1=\lastreadfilename\relax}}%
\def\warnopenout#1#2{%
 \readfilename#2\relax
 \def\t@mp{TrashMe,psbjoin.aux,psbjoint.tex,}\uncatcode\t@mp
 \if\matchexpin{\t@mp}{\lastreadfilename,}%
 \else
   \immediate\openin\pst@mpin=\lastreadfilename\relax
   \ifeof\pst@mpin
     \else
     \errhelp{If the content of this file is so precious to you, abort (ie
press x or e) and rename it before retrying.}%
     \errmessage{I'm just about to replace your file named \lastreadfilename}%
   \fi
   \immediate\closein\pst@mpin
 \fi
 \message{\lastreadfilename}%
 \immediate\openout#1=\lastreadfilename\relax}%
{\catcode`\%=12\catcode`\*=14
\gdef\splitfile#1{*
 \readfilename#1\relax
 \immediate\openin\j@insplitin=\lastreadfilename\relax
 \ifeof\j@insplitin
   \message{! I couldn't find and split \lastreadfilename!}*
 \else
   \immediate\openout\j@insplitout=TrashMe
   \message{< Splitting \lastreadfilename\space into}*
   \loop
     \ifeof\j@insplitin
       \immediate\closein\j@insplitin\n@teoffalse
     \else
       \n@teoftrue
       \executeinspecs{\global\read\j@insplitin to\spl@tinline\expandafter
         \ch@ckbeginnewfile\spl@tinline
       \ifc@ntrolline
       \else
         \toks0=\expandafter{\spl@tinline}*
         \immediate\write\j@insplitout{\the\toks0}*
       \fi
     \fi
   \ifn@teof\repeat
   \immediate\closeout\j@insplitout
 \fi\message{>}*
}*
\gdef\ch@ckbeginnewfile#1
 \def\t@mp{#1}*
 \ifx\@mpty\t@mp
   \def\t@mp{#3}*
   \ifx\@mpty\t@mp
     \global\c@ntrollinefalse
   \else
     \immediate\closeout\j@insplitout
     \warnopenout\j@insplitout{#2}*
     \global\c@ntrollinetrue
   \fi
 \else
   \global\c@ntrollinefalse
 \fi}*
\gdef\joinfiles#1\into#2{*
 \message{< Joining following files into}*
 \warnopenout\j@insplitout{#2}*
 \message{:}*
 {*
 \edef\w@##1{\immediate\write\j@insplitout{##1}}*
\w@{
\w@{
\w@{
\w@{
\w@{
\w@{
\w@{
\w@{
\w@{
\w@{
\w@{\string\input\space psbox.tex}*
\w@{\string\splitfile{\string\jobname}}*
\w@{\string\let\string\autojoin=\string\relax}*
}*
 \expandafter\tre@tfilelist#1, \endtre@t
 \immediate\closeout\j@insplitout
 \message{>}*
}*
\gdef\tre@tfilelist#1, #2\endtre@t{*
 \readfilename#1\relax
 \ifx\@mpty\lastreadfilename
 \else
   \immediate\openin\j@insplitin=\lastreadfilename\relax
   \ifeof\j@insplitin
     \errmessage{I couldn't find file \lastreadfilename}*
   \else
     \message{\lastreadfilename}*
     \immediate\write\j@insplitout{
     \executeinspecs{\global\read\j@insplitin to\oldj@ininline}*
     \loop
       \ifeof\j@insplitin\immediate\closein\j@insplitin\n@teoffalse
       \else\n@teoftrue
         \executeinspecs{\global\read\j@insplitin to\j@ininline}*
         \toks0=\expandafter{\oldj@ininline}*
         \let\oldj@ininline=\j@ininline
         \immediate\write\j@insplitout{\the\toks0}*
       \fi
     \ifn@teof
     \repeat
   \immediate\closein\j@insplitin
   \fi
   \tre@tfilelist#2, \endtre@t
 \fi}*
}%
\def\autojoin{%
 \immediate\write\psbj@inaux{\string\into{psbjoint.tex}}%
 \immediate\closeout\psbj@inaux
 \expandafter\joinfiles\GlobalInputList\into{psbjoint.tex}%
}%
%
%
%
\def\centinsert#1{\midinsert\line{\hss#1\hss}\endinsert}%
\def\psannotate#1#2{\vbox{%
  \def\ps@nnotation{#2\global\let\ps@nnotation=\relax}#1}}%
\def\pscaption#1#2{\vbox{%
   \setbox\drawingBox=#1
   \copy\drawingBox
   \vskip\baselineskip
   \vbox{\hsize=\wd\drawingBox\setbox0=\hbox{#2}%
     \ifdim\wd0>\hsize
       \noindent\unhbox0\tolerance=5000
    \else\centerline{\box0}%
    \fi
}}}%
%
\def\at(#1;#2)#3{\setbox0=\hbox{#3}\ht0=0pt\dp0=0pt
  \rlap{\kern#1\vbox to0pt{\kern-#2\box0\vss}}}%
%
\newdimen\gridht \newdimen\gridwd
\def\gridfill(#1;#2){%
  \setbox0=\hbox to 1\pscm
  {\vrule height1\pscm width.4pt\leaders\hrule\hfill}%
  \gridht=#1
  \divide\gridht by \ht0
  \multiply\gridht by \ht0
  \gridwd=#2
  \divide\gridwd by \wd0
  \multiply\gridwd by \wd0
  \advance \gridwd by \wd0
  \vbox to \gridht{\leaders\hbox to\gridwd{\leaders\box0\hfill}\vfill}}%
%
\def\fillinggrid{\at(0cm;0cm){\vbox{%
  \gridfill(\drawinght;\drawingwd)}}}%
%
%
\def\textleftof#1:{%
  \setbox1=#1
  \setbox0=\vbox\bgroup
    \advance\hsize by -\wd1 \advance\hsize by -2em}%
\def\textrightof#1:{%
  \setbox0=#1
  \setbox1=\vbox\bgroup
    \advance\hsize by -\wd0 \advance\hsize by -2em}%
\def\endtext{%
  \egroup
  \hbox to \hsize{\valign{\vfil##\vfil\cr%
\box0\cr%
\noalign{\hss}\box1\cr}}}%
%
\def\frameit#1#2#3{\hbox{\vrule width#1\vbox{%
  \hrule height#1\vskip#2\hbox{\hskip#2\vbox{#3}\hskip#2}%
        \vskip#2\hrule height#1}\vrule width#1}}%
\def\boxit#1{\frameit{0.4pt}{0pt}{#1}}%
\catcode`\@=12 
%
 \psfordvips   

\dofigure{
$$\psboxto(15 true cm;0cm){fig1.eps}$$
}
{Figure \figctrldgdw.}{ (left) Time traces of the half squared
amplitudes of $\pefl$ and $\phifl$, respectively labelled by
$A_p$ and $A_\phi$, in the drift wave model (right) and the
interchange model (right), showing saturation.  Only in the drift wave
model are the time traces for $A_p$ and $A_\phi$ similar.  
}

\dofigure{
$$\psboxto(15 true cm;0cm){fig2.eps}$$
}
{Figure \figcontrol.}{  (left to right) Drive and transfer
spectra, and cross coherence and phase shifts between $\pefl$ and
$\phifl$,  for the drift wave (top row) and interchange (bottom row)
models.
In the drift wave case the sink spectrum (labelled `C') is relatively
flat, and the transfer (`tr') is also due to $\dpl\Jfl$ and is positive
at short wavelength and negative at long wavelength, while in the
interchange case it is due to $\kappacv(\pefl)$ which follows the source
spectrum (`pe'). 
The drift wave case shows strong cross
coherence and a narrow phase shift distribution closer to zero than to
$\pi/2$.  The interchange case shows dominance by the longest
wavelengths, no cross coherence, and phase shifts near $\pi/2$ due to
the strong driving and weak coupling.  
The phase shift distributions contours are $0.3$,
$0.5$, and $0.8$ times the maximum, and for the cross coherence they are 
$0.37$, $0.5$, and $0.8$ times the maximum.
}

\dofigure{
$$\psboxto(15 true cm;0cm){fig3.eps}$$
}
{Figure \figbdmode.}{  (left) Transport scaling of drift wave and
interchange turbulence in toroidal geometry, from the DALF3 and reduced
resistive MHD models labelled `DW' and `BM', respectively.
The drift wave cases show the clearest scaling with 
collisionality at low $C=2.55\nu$.  At asymptotically large $C$
the trends will merge, but that limit is not reached.  (right) Half
squared amplitude ratio (including only $k_y\ne 0$)
for the two sets of cases.  Due to the adiabatic response,
$\phifl$ tracks $\pefl$ for drift wave turbulence, but in the MHD
model of the interchange cases $\pefl$ is unaffected by the Alfv\'en
dynamics, and so instead of $\phifl$ being forced towards $\pefl$ it is
forced towards zero.  The extra points marked with squares for DW
($\nu=1,2,5,10$) are with double resolution in the drift plane.
}

\dofigure{
$$\psboxto(15 true cm;0cm){fig4.eps}$$
}
{Figure \figplcoher.}{  Cross coherence between $\pefl$ and
$\phifl$, for drift wave (top row) and interchange (bottom row)
turbulence, for 
$\nu=5$, $10$, and $20$ (left to right), where $C=2.55\nu$ and
$\nu_B=C\wcv$ as defined in Eq.~(\eqresbal).  Compare
with the results in Section~\seccontrol\ for drift wave and interchange
turbulence.  The DALF3 model results in drift 
wave mode structure even for larger $\nu_B$, while the MHD model
always shows interchange mode structure. 
Contours are as in Fig.~\figcontrol.
}

\dofigure{
$$\psboxto(15 true cm;0cm){fig5.eps}$$
}
{Figure \figphasdist.}{  Phase shift distributions of $\pefl$
ahead of $\phifl$ at each $k_y$, 
for drift wave (top row) and interchange (bottom row)
turbulence, for 
$\nu=5$, $10$, and $20$ (left to right), where $C=2.55\nu$ and
$\nu_B=C\wcv$ as defined in Eq.~(\eqresbal).  Compare
with the results in Section~\seccontrol\ for drift wave and interchange
turbulence.  The DALF3 model results in drift 
wave mode structure for $\nu<10$, while the MHD model
always shows interchange mode structure.  
The transition to resistive ballooning turbulence in the DALF3 model
occurs in the longest wavelengths, $k_y\rs<0.1$.  
Contours are as in Fig.~\figcontrol.  }

\dofigure{
$$\psboxto(15 true cm;0cm){fig6.eps}$$
}
{Figure \figwtrans.}{  Spectra of the rms transfer dynamical
levels for each $k_y$ in the spectrum, comparing the
sizes of $\phifl\dpl\Jfl$ (`j'), $\phifl\kappacv(\pefl)$ (`k'), and 
$\phifl\vedl\vorfl$ (`e'),
for drift wave (top row) and interchange (bottom row) turbulence, for
$\nu=5$, $10$, and $20$ (left to right), where $C=2.55\nu$ and
$\nu_B=C\wcv$.
In drift wave turbulence the transfer through the current is
too large to be accounted for by the curvature, and at all wavelengths
the vorticity nonlinearity is balanced only by the linear time
derivative.  Nonlinear vorticity dynamics is generally much stronger
when the adiabatic coupling mechanism $\pefl\fromto\Jfl$ is present,
leading to a self consistent situation in which both mechanisms catalyse
each other. 
}

\dofigure{
$$\psboxto(15 true cm;0cm){fig7.eps}$$
}
{Figure \figpldwa.}{  Amplitude spectra of $\pefl$,
$\phifl$,and  $\vorfl$, respectively labelled by `pe', `phi', and `vor',
for drift wave (top row) and interchange (bottom row) turbulence, for
$\nu=5$, $10$, and $20$ (left to right), where $C=2.55\nu$ and
$\nu_B=C\wcv$. 
With only the density present
one cannot distinguish the mode structure or dynamics, but if $\phifl$
is also present the amplitude ratio, shown in Fig.~\figbdmode, is
decisive, as the spectrum of $\phifl$ follows that of $\pefl$ only for
drift wave turbulence.
The transition to resistive ballooning turbulence in the DALF3 model
($\phifl\gg\pefl$) occurs in the longest wavelengths, $k_y\rs<0.1$.  
}

\dofigure{
$$\psboxto(15 true cm;0cm){fig8.eps}$$
}
{Figure \figpldwb.}{  Spectra of the ExB gradient drive (`pe'),
dissipation (`C'), and the magnetic flutter drive (`mag'),
for drift wave (top row) and interchange (bottom row) turbulence, for
$\nu=5$, $10$, and $20$ (left to right), where $C=2.55\nu$ and
$\nu_B=C\wcv$. 
For drift wave turbulence the shorter wavelengths contribute more to the
energetic drive and hence the ExB transport.  Compare the positions of
the energy containing range (Fig.~\figpldwa), the energy producing range
(this figure) and the vorticity catalysing range (Fig.~\figwtrans).
These show that for drift wave turbulence the entire spectrum acts as a
single unit.  
}

\dofigure{
$$\psboxto(15 true cm;0cm){fig9.eps}$$
}
{Figure \figplfls.}{  Mean squared amplitude envelopes 
($k_y\ne 0$ only)
showing parallel structure of $\phifl$, $\pefl$,
and $\hefl=\pefl-\phifl$, respectively labelled by `phi', `pe', and  `he',
for drift wave (top row) and interchange (bottom row) turbulence, 
for $\nu=5$, $10$, and $20$ (left to right), where $C=2.55\nu$ and
$\nu_B=C\wcv$.
Drift wave mode structure is
exemplary for $\nu<10$ in the DALF3 model, with $\hefl$ flatter and
smaller than either $\pefl$ or $\phifl$. 
}

\dofigure{
$$\psboxto(15 true cm;0cm){fig10.eps}$$
}
{Figure \figplbal.}{  Amplitude envelope of the ExB transport
(`pe'), 
for drift wave (top row) and interchange (bottom row) turbulence, for
$\nu=5$, $10$, and $20$ (left to right), where $C=2.55\nu$ and
$\nu_B=C\wcv$.
The magnetic flutter transport is negligible on this scale.
The ballooning in the
transport becomes somewhat more pronounced in the transition to interchange
character for $\nu>10$, but for all cases the ballooning is much
stronger for the MHD model than for DALF3.  
}


\dofigure{
$$\psboxto(15 true cm;0cm){fig11.eps}$$
}
{Figure \figdgdwl.}{ Time evolution of the DALF3 case with
$\nu=10$ (hence $C=25.5$ and $\nu_B=1.25$) out of a linear initial state
at small amplitude, run to $t=1582$.
(top left) Half squared amplitudes of $\phifl$ (denoted $A_\phi$) and 
$\pefl$ ($A_p$).  The larger amplitude
departures of $A_\phi$ from $A_p$ reflect zonal flow activity in the fully
developed turbulence.
(top right) The transport caused by the turbulence.
(bottom left) Growth rate ($\Gamma_T$), gradient drive rate ($\Gamma_+$), 
and total dissipation rate ($\Gamma_E$). 
(bottom right) Vorticity (rms) compared to $\Gamma_+$ and $\Gamma_T$.
The linear mode leads to overshoot,
initial saturation ($t\approx 200$)
is reflected in the first drop of $\Gamma_T$ to zero,
and then the nonlinear mode structure takes over shortly thereafter,
with full development with robust transport and zonal flow activity
established after $t=400$.  The turbulence not only saturates but
changes character when the rms vorticity overcomes the linear growth
rate ($\gamma_L$, equal to $\Gamma_T$ in the linear stage).
Nonlinear saturation of the linear instability would obtain if
$\Omega_{{\rm rms}}$ were comparable to $\gamma_L$, but the situation with 
$\Omega_{{\rm rms}}\gg\gamma_L$ indicates complete supersession of the
instability by turbulence which has its own dynamics.
}

\dofigure{
$$\psboxto(15 true cm;0cm){fig12.eps}$$
}
{Figure \figplctrl.}{  Saturation and initial transition to
turbulence,
as seen in the spatial morphology of $\phifl$ and $\pefl$
in the linear growth stage
to $t=190$, and the initial saturation stage after $t=200$.
The transition between linear and nonlinear mode structure is most clear
in the disappearance of $x$-direction flows remniscent of bouyant plumes.
This represents a transition away from interchange dominated dynamics to
a more isotropic turbulence as the vorticity nonlinearity replaces the
interchange forcing as the principal mechanism supporting finite
parallel currents (nonadiabatic electron dynamics).
(Positive/negative values are indicated by solid/dashed lines.
Only $1/4$ of the computational domain in the $y$-direction is shown.
The exact moment of saturation is $t=201$, and the growth rate becomes
positive again at $t=212$.)
}

\dofigure{
$$\psboxto(15 true cm;0cm){fig13.eps}$$
}
{Figure \figplctrlb.}{  Transition to fully developed turbulence,
as seen in the spatial morphology of $\phifl$ and $\pefl$ through the
stage of nonlinear structure adjustment, and then in the stage of
statistical saturation in which the nonlinear growth rate fluctuates
near zero.  As the saturated state finds itself, the scale of motion
increases, and until $t=345$ the nonlinear growth rate is positive.
Zonal flows emerge after about $t=400$ and reach statistical equilibrium
after about $t=600$.  The zonal flows are part of the nonlinear mode
structure, but by the time they emerge the interchange driven flows of
the linear stage are long gone.  The correlation time is about 6 in
these units.
(Positive/negative values are indicated by solid/dashed lines.
Only $1/4$ of the computational domain in the $y$-direction is shown.
The exact moment of saturation is $t=201$, and the growth rate becomes
positive again at $t=212$, and negative again at $t=345$.)
}

\dofigure{
$$\psboxto(15 true cm;0cm){fig14.eps}$$
}
{Figure \figwtransl.}{  Saturation and transition to turbulence,
as seen in the dynamical transfer spectra for the ExB vorticity, for the
linear stage to about $t=190$, the saturation stage around $t=200$, and
the nonlinear structure adjustment stage after about $t=210$.
Compare with Fig.~\figwtrans.
Initially the eigenmode is controlled by interchange forcing, with the
linear polarisation drift negligible (cf.\ Eq.~\eqresponse).  But as
the amplitude becomes finite the vorticity nonlinearity, the same one
which causes the drift wave nonlinear instability, emerges to become the
principal agent supporting nonadiabatic electron dynamics.  The
transition is extremely rapid, taking place within about one correlation
time for the fully developed turbulence.
(The exact moment of saturation is $t=201$, and the growth rate becomes
positive again at $t=212$, and negative again at $t=345$.)
}

\dofigure{
$$\psboxto(15 true cm;0cm){fig15.eps}$$
}
{Figure \figplflsl.}{  Saturation and transition to turbulence,
as seen in the parallel envelope structure ($k_y\ne 0$ only), in the 
linear stage to about $t=190$, the saturation stage around $t=200$, and
the nonlinear structure adjustment stage after about $t=210$.
Compare with Fig.~\figplfls.
The principal signature of the transition is the emergence of $\phifl$
supported by the vorticity nonlinearity.  The interchange flows of the
linear stage are eliminated by the turbulent vorticity, and the
adiabatic response causes $\phifl$ to track $\pefl$.
The degree of asymmetry is also reduced, especially in $\hefl$.
This transition between linear ($\hefl\sim\pefl\gg\phifl$)
and nonlinear ($\phifl\sim\pefl\gg\hefl$)
mode structure is almost as rapid as in Fig.~\figwtransl.  
(The exact moment of saturation is $t=201$, and the growth rate becomes
positive again at $t=212$, and negative again at $t=345$.)
}

\dofigure{
$$\psboxto(15 true cm;0cm){fig16.eps}$$
}
{Figure \figphasel.}{  Phase shift distributions of $\pefl$
ahead of $\phifl$ at each $k_y$, 
for the linear stage, averaged over $50<t<150$, 
through saturation, averaged over $200<t<244$, 
and for the stage of fully developed turbulence, averaged over $502<t<611$.
Compare with Fig.~\figphasdist.
The turbulence emerges to supersede the linear structure with its own,
due to the fact that the rms vorticity level of the turbulence is larger
than the linear growth rate of the original instability.  The linear
instabilities in the range $0.3<k_y\rs<1.0$ have no role in the
turbulence.  Only in the fully developed stage does the interchange
dynamics for $k_y\rs<0.1$ emerge to make this case with $\nu_B=1.25$
the transitional one between turbulence of the drift wave and
interchange type.
}

\dofigure{
$$\psboxto(15 true cm;0cm){fig17.eps}$$
}
{Figure \figwtransll.}{Dynamical transfer spectra for the ExB vorticity,
for the linear stage, averaged over $50<t<150$, 
through saturation, averaged over $200<t<244$, 
and for the stage of fully developed turbulence, averaged over $502<t<611$,
plotted against $\kpp$ rather than $k_y$, showing the scale of motion
rather than the wavelength in the drift direction.  The linear
instability is dominantly in the same range which is later dominated by
the turbulence, making the linear interchange dominated mode
irrelevant.  When interchange effects do enter, as in this transitional
case with $\nu_B=1.25$, they do so at larger scale where they more
easily overcome the native vorticity of the drift wave turbulence.
}


\dofigure{
$$\psboxto(15 true cm;0cm){fig18.eps}$$
}
{Figure \figtorsat.}{  Saturation mechanism in drift wave
turbulence in toroidal geometry.  The run with $\nu=2$
(hence $C=5.1$ and $\nu_B=0.25$) taken to $t=1000$ (solid curves)
is restarted from $t=500$ (dashed curves) with the
ExB vorticity nonlinearity either left out (`novor') or with all ExB
nonlinearities except the vorticity one left out (`voronly').
Without the vorticity nonlinearity, the linear drive 
is balanced by mixing via $\vedl\pefl$ and saturation occurs.  Without
the pressure nonlinearity, the vorticity is vigorously scattered via
$\vedl\vorfl$ with little dissipative effect, and this nonlinear
excitation continues indefinitely without saturation.
}


\dofigure{
$$\psboxto(15 true cm;0cm){fig19.eps}$$
}
{Figure \figbdmodeti.}{  Transport scaling (left two frames) and
amplitude ratios (right) of drift wave (solid lines) and
resistive MHD (dashed lines) turbulence in toroidal geometry under the
DALFTI model.  Both ion and electron heat transport are shown as a pair;
in each case the ion transport curve is the one lying slightly higher in
the pair.
The MHD model, which neglects the drift wave coupling
terms between $\pefl$ and $\Jfl$, is insensitive to collisionality
($C=2.55\nu$) because although $\phifl$ is too small, $\pifl$ is too
large, compared to the drift wave model.  The transition to the MHD
regime at a lower $\nu>3$ is assisted by the warm ion interchange 
physics, since $\tifl$ does not feel the adiabatic response.  The extra
points marked with squares for DW ($\nu=1,2,5,10$)
are with double resolution in the
drift plane.  Compare with Fig.~\figbdmode.
}

\dofigure{
$$\psboxto(15 true cm;0cm){fig20.eps}$$
}
{Figure \figplcoherti.}{  Cross coherence between $\nefl$ and
$\phifl$, for drift wave (top row) and interchange (bottom row)
turbulence, for 
$\nu=2$, $5$, and $10$ (left to right), where $C=2.55\nu$ and
$\nu_B=C\wcv$ as defined in Eq.~(\eqresbal).  
The turbulence makes the transition
from drift wave to resistive ballooning mode structure for $\nu_B$ of
about $0.5$ (here, $\nu=5$).
Compare to Fig.~\figplcoher.  
Interchange turbulence with
warm ions is very violent at small scales due to the effects of
gyroviscosity, reducing the timestep and shortening the runs.
}

\dofigure{
$$\psboxto(15 true cm;0cm){fig21.eps}$$
}
{Figure \figphasdistti.}{  Phase shift distributions of $\nefl$
ahead of $\phifl$ at each $k_y$, 
for drift wave (top row) and interchange (bottom row)
turbulence, for 
$\nu=2$, $5$, and $10$ (left to right), where $C=2.55\nu$ and
$\nu_B=C\wcv$ as defined in Eq.~(\eqresbal).  
The turbulence makes the transition
from drift wave to resistive ballooning mode structure for $\nu_B$ of
about $0.5$.  Compare to Fig.~\figphasdist.
The tendency of the phase shift to go to $-\pi$ at small scales is the
signature of the nonlinearity in the gyroviscosity which conserves
energy but not vorticity for warm ions, producing the effects seen in
Fig.~\figplcoherti. 
As in the DALF3 model, the transition to resistive ballooning turbulence
in the DALF3 model occurs in the longest wavelengths, $k_y\rs<0.1$.  
}

\dofigure{
$$\psboxto(15 true cm;0cm){fig22.eps}$$
}
{Figure \figplflsti.}{  Mean squared amplitude envelopes 
($k_y\ne 0$ only)
showing parallel structure of $\phifl$, $\nefl$, $\tefl$, and $\tifl$,
respectively labelled by `phi', `ne', `Te', and  `Ti',
for drift wave (top row) and interchange (bottom row) turbulence, 
for $\nu=2$, $5$, and $10$ (left to right), where $C=2.55\nu$ and
$\nu_B=C\wcv$ as defined in Eq.~(\eqresbal).
The turbulence makes the transition
from drift wave to resistive ballooning mode structure for $\nu_B$ of
about $0.5$, with additional effects due to the contribution of $\pifl$
to the vorticity, as described in the text.  
Note $\pefl=\nefl+\tefl$
and $\pifl=\tau_i\nefl+\tifl$ in this model.
}

\dofigure{
$$\psboxto(15 true cm;0cm){fig23.eps}$$
}
{Figure \figtransbetas.}{  Transport scaling as a function of the
ideal ballooning parameter $\alpha_M$, for the DALF3 (`3') and
DALFTI (`I') models, expressed as diffusivities, with $\chi_{e,i}$
including the convective contributions.
The collisionality was $C=2.55$, in the drift wave
regime.  The transport is given in physical units ($\msqsec$,
assuming deuterium
ions, and $B=2.5\tesla$, and $\Lpp=4.2\cm$), compensating
for the effect of varying $\bhat$ on the normalisation scale
$\rs^2c_s/\Lpp$.  The ion temperature assists the transition to
ideal MHD in the measure and for the same reason as for the resistive
ballooning cases: more total pressure gradient, and $\tifl>\tefl$.
But the ideal ballooning threshold, now between $0.2$ and $0.6$,
is lowered by a factor of at least two relative to the linear analysis.
The experimentally interesting range is actually $\alpha_M>0.2$,
corresponding to $\bhat>1$ for the DALFTI model.
}

\dofigure{
$$\psboxto(15 true cm;0cm){fig24.eps}$$
}
{Figure \figplcoherbetas.}{  Cross coherence between $\nefl$
and $\phifl$, 
for the DALFTI model at $C=2.55$ hence $\nu_B=0.25$, for various
$\alpha_M$ in the range of the sharp transport rise, noting
$\alpha_M=0.2\bhat$.  The turbulence makes the transition 
from drift wave to ideal ballooning mode structure in this parameter
range.  Compare to Fig.~\figplcoherti.
}

\dofigure{
$$\psboxto(15 true cm;0cm){fig25.eps}$$
}
{Figure \figphasdistbetas.}{  Phase shift distributions of $\nefl$
ahead of $\phifl$ at each $k_y$, 
for the DALFTI model at $C=2.55$ hence $\nu_B=0.25$, for various
$\alpha_M$ in the range of the sharp transport rise, noting
$\alpha_M=0.2\bhat$.  The turbulence makes the transition 
from drift wave to ideal ballooning mode structure in this parameter
range.  Compare to Fig.~\figphasdistti.
Here as well, the transition to resistive ballooning turbulence in the
DALF3 model occurs in the longest wavelengths, $k_y\rs<0.1$.  
}

\dofigure{
$$\psboxto(15 true cm;0cm){fig26.eps}$$
}
{Figure \figplflsbetas.}{  Mean squared amplitude envelopes 
($k_y\ne 0$ only)
showing parallel structure of $\phifl$, $\nefl$, $\tefl$, and $\tifl$,
respectively labelled by `phi', `ne', `Te', and  `Ti',
for the DALFTI model at $C=2.55$ hence $\nu_B=0.25$, for various
$\alpha_M$ in the range of the sharp transport rise, noting
$\alpha_M=0.2\bhat$.  The turbulence makes the transition 
from drift wave to ideal ballooning mode structure in this parameter
range.  Compare to Fig.~\figplflsti.
Note $\pefl=\nefl+\tefl$ and $\pifl=\tau_i\nefl+\tifl$ in this model.
}


\end